\documentclass{article}

\usepackage[final,nonatbib]{neurips_2019}

\usepackage[utf8]{inputenc} 
\usepackage[T1]{fontenc}    
\usepackage{hyperref}       
\usepackage{url}            
\usepackage{booktabs}       
\usepackage{amsfonts}       
\usepackage{nicefrac}       
\usepackage{microtype}      

\usepackage{amsmath}
\usepackage{graphicx}
\usepackage{caption}
\usepackage{subcaption}
\usepackage{algorithm}
\usepackage[noend]{algpseudocode}
\usepackage{float}

\DeclareMathOperator*{\argmax}{argmax}

\title{Channel Decomposition into Painting Actions}

\author{%
  Shih-Chieh Su\thanks{In Workshop of Knowledge Representation and Reasoning. Demo and code are available and updated at \url{https://github.com/jessysu/cpia}.} \\
  Microsoft\\
  San Diego, CA 92130 \\
  \texttt{jessysu@gmail.com} \\
}

\begin{document}

\maketitle

\begin{abstract}
    This work presents a method to decompose a convolution layer of the deep neural network into painting actions. The pre-trained knowledge in the appointed operation layer is used to guide the neural painter. To behave like the human painter, the actions are driven by the cost simulating the hand movement, the paint color change, the stroke shape and the stroking style. To help planning, the Mask R-CNN is applied to detect the object areas and decide the painting order. The proposed painting system introduces a variety of extensions in artistic styles, based on the chosen parameters. Further experiments are performed to evaluate the channel penetration and the channel sensitivity on the strokes.

\end{abstract}

\section{Introduction}

For years, the convolutional neural networks have been digesting the visual world and serving a wide range of applications. Neural style transfer is one of the most popular applications. Soon after the pioneering work by Gatys \textit{et al.}~\cite{gatys2016image}, the feed-forward network incorporating convolutional layers has been introduced to perform near realtime style transfer~\cite{johnson2016perceptual, ulyanov2016texture}. In recent work~\cite{li2017universal, jing2018stroke}, the stroke and the attention factors have been considered. Although fast, current style transfer work generates the whole transferred frame in one feed. This leaves the audience wondering that, in which stroke order the neural painter would paint the art that can lead to the style transfer output.

On the other hand, the stroke composure process has been studied in the literature. The steps to paint the image (or to write the letters) is typically referred to as stroke-based rendering (or inverse graphics). To learn the painting behavior without pairing stroke-wise training data, reinforcement learning is applied to help stroke planning, in recent work like SPIRAL\cite{ganin2019synthesizing}, StrokeNet\cite{zheng2019strokenet} and LearningToPaint\cite{huang2019learning}. While the strokes can typically be arranged in the coarse-to-fine order as the painting architecture of the designed, the stroke shape and stroke order may differ from that of the human painter.

Decomposing the target into reasonable amount of stroking actions of reasonable amount of stroke shapes remains challenging. Which parts of the target needs to be painted, in which kind of artistic style? How does the human painter plan and compose the paint with strokes? For different painter to paint the same object or the same scene, how different would their approaches be? We try to find clues from the generator network.

This work presents a strategy to decompose the channel response of the generator networks into stroke actions, called the \textit{channel stroke}. The channel stroke considers the burden of the human painter in changing paint brushes and changing colors. Leveraging the channel depth of the generator networks, the proposed strategy strokes through the same channels continuously over the regions with high receptive field response.

\begin{figure}[!t]
    \centering
    \begin{subfigure}[b]{\textwidth}
        \includegraphics[width=0.16\textwidth]{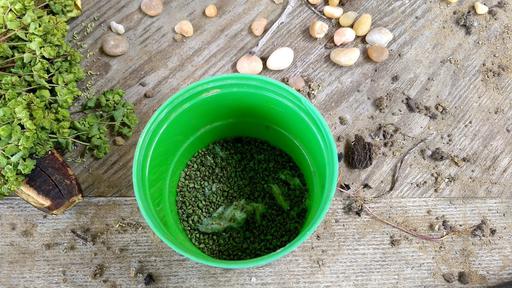}
        \hfill
        \includegraphics[width=0.09\textwidth]{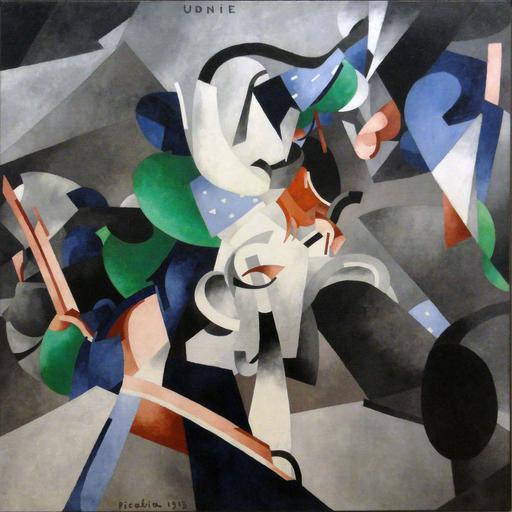}
        \includegraphics[width=0.09\textwidth]{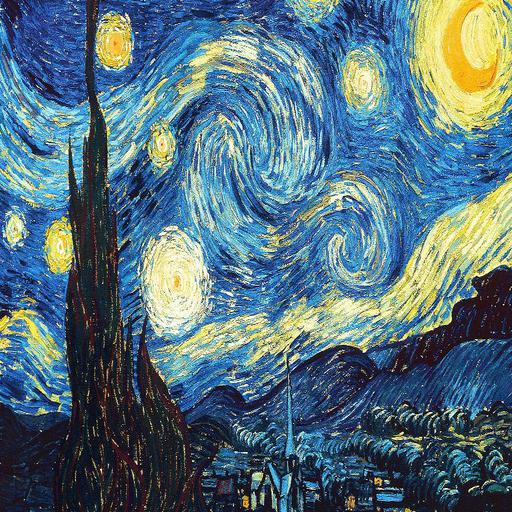}
        \includegraphics[width=0.09\textwidth]{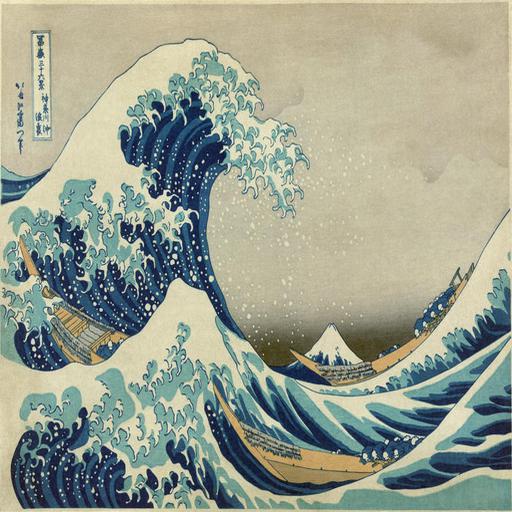}
        \hfill
        \includegraphics[width=0.16\textwidth]{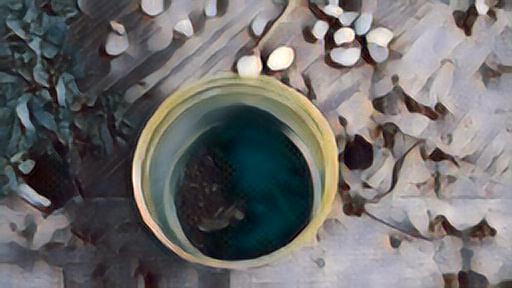}
        \includegraphics[width=0.16\textwidth]{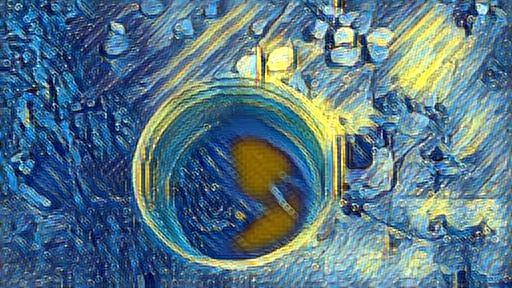}
        \includegraphics[width=0.16\textwidth]{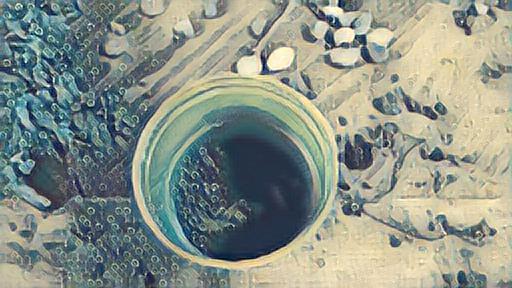}
    \end{subfigure}
    \begin{subfigure}[b]{0.16\textwidth}
        \centering
        \includegraphics[width=\textwidth]{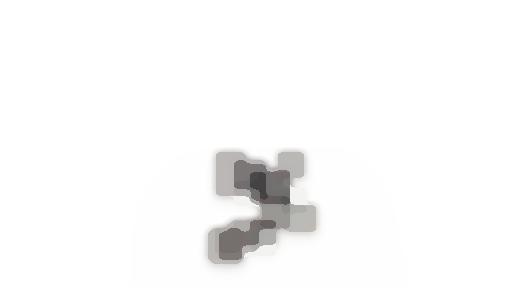}
        \includegraphics[width=\textwidth]{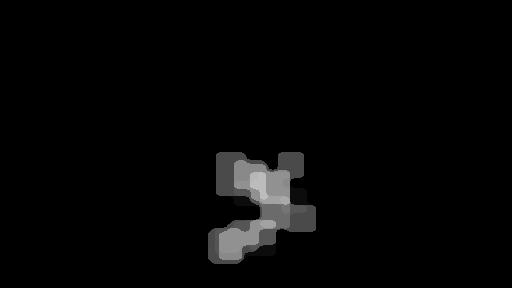}
        \includegraphics[width=\textwidth]{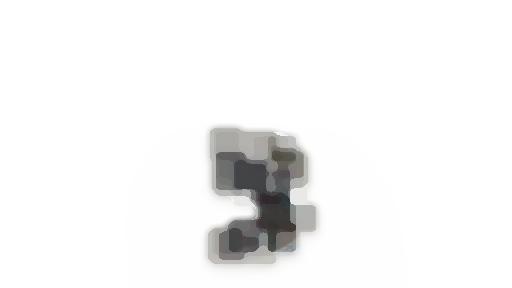}
        \includegraphics[width=\textwidth]{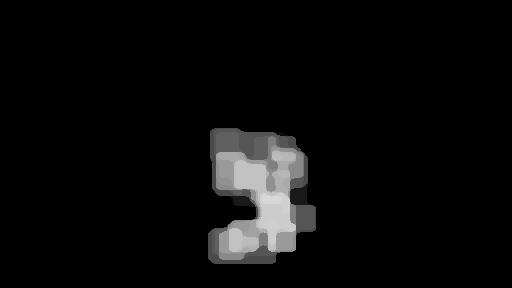}
        \includegraphics[width=\textwidth]{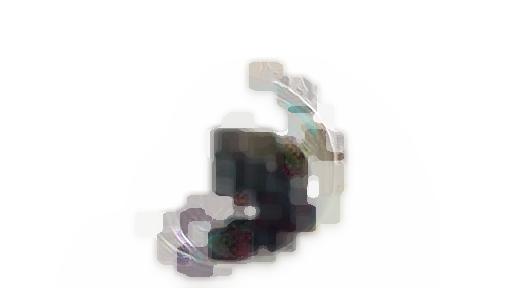}
        \includegraphics[width=\textwidth]{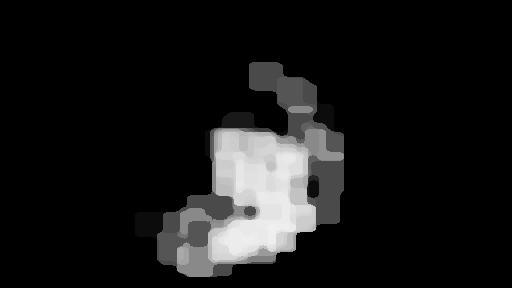}
        \includegraphics[width=\textwidth]{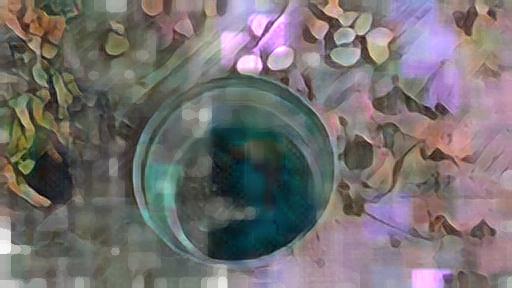}
        \caption{big, $l=9$}
    \end{subfigure}
    \begin{subfigure}[b]{0.16\textwidth}
        \includegraphics[width=\textwidth]{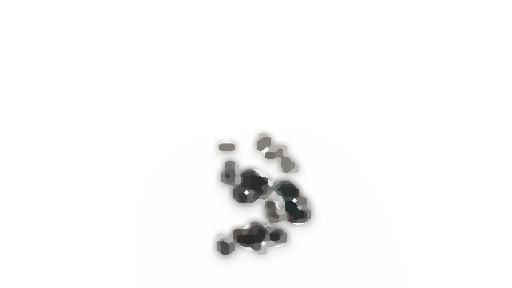}
        \includegraphics[width=\textwidth]{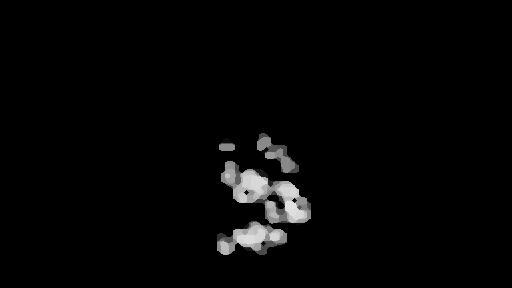}
        \includegraphics[width=\textwidth]{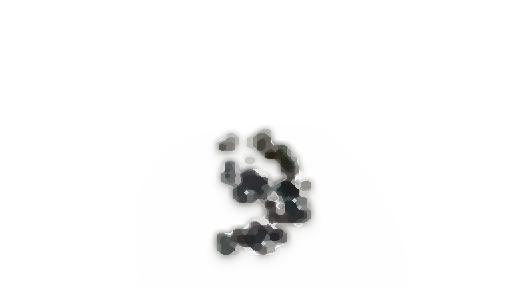}
        \includegraphics[width=\textwidth]{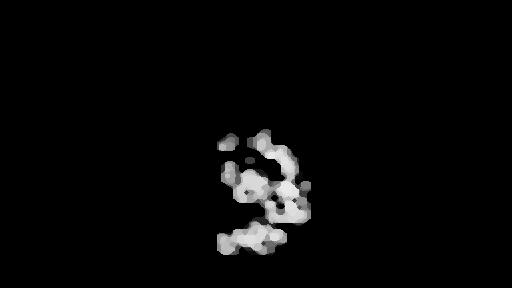}
        \includegraphics[width=\textwidth]{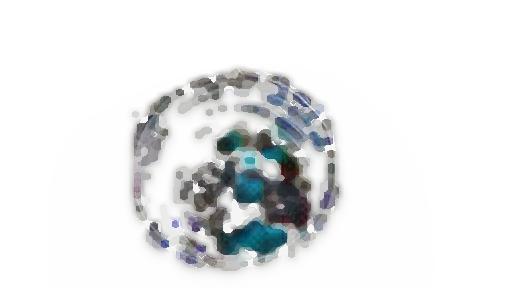}
        \includegraphics[width=\textwidth]{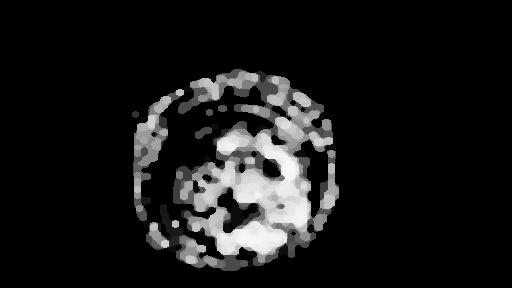}
        \includegraphics[width=\textwidth]{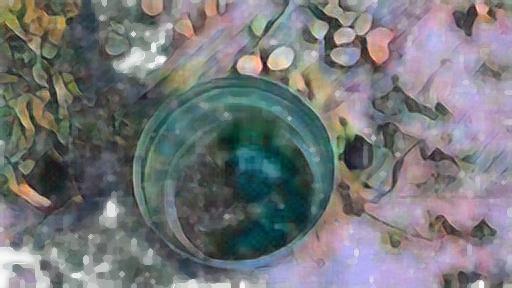}
        \caption{small, $l=9$}
    \end{subfigure}
    \begin{subfigure}[b]{0.16\textwidth}
        \includegraphics[width=\textwidth]{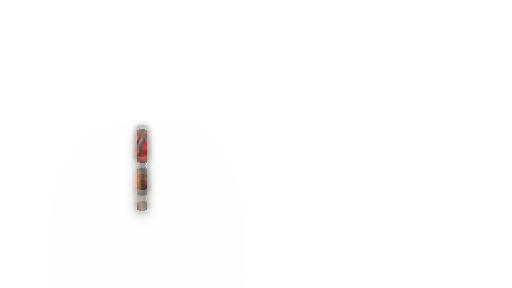}
        \includegraphics[width=\textwidth]{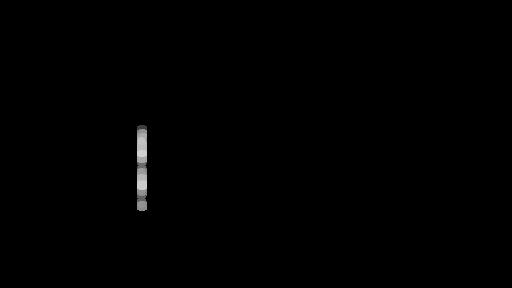}
        \includegraphics[width=\textwidth]{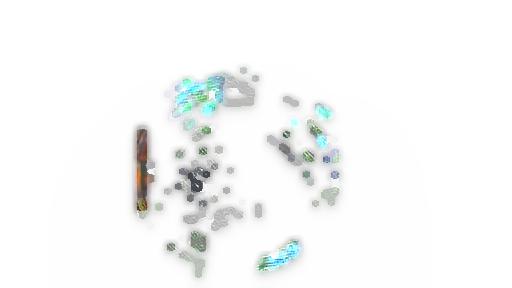}
        \includegraphics[width=\textwidth]{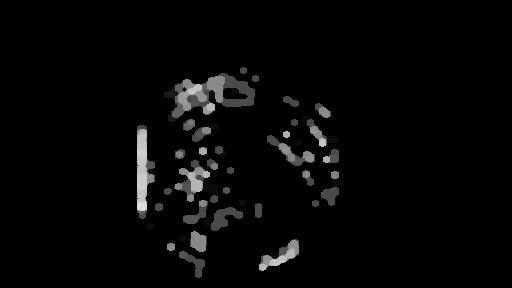}
        \includegraphics[width=\textwidth]{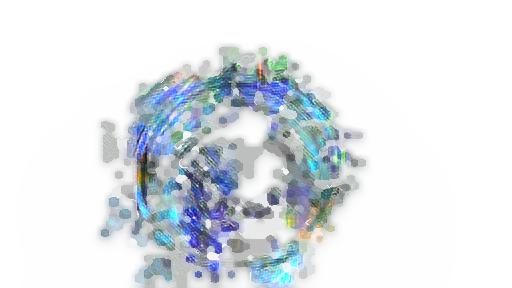}
        \includegraphics[width=\textwidth]{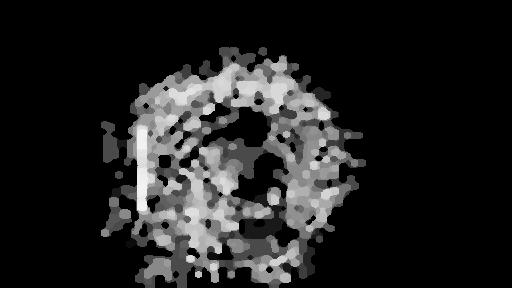}
        \includegraphics[width=\textwidth]{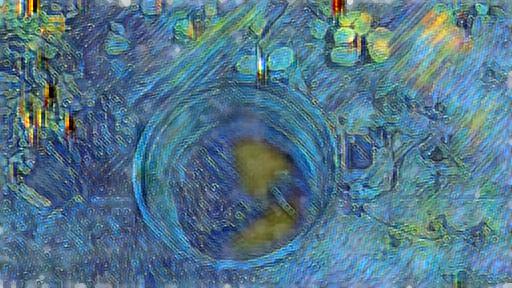}
        \caption{small, $l=9$}
    \end{subfigure}
    \begin{subfigure}[b]{0.16\textwidth}
        \includegraphics[width=\textwidth]{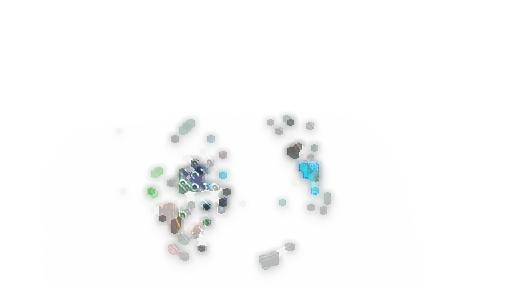}
        \includegraphics[width=\textwidth]{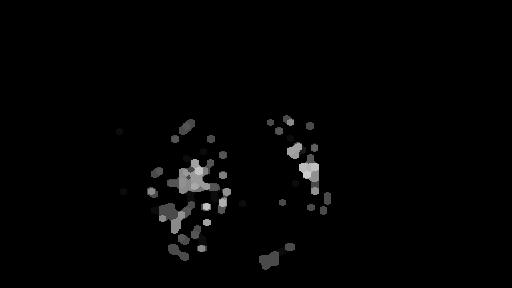}
        \includegraphics[width=\textwidth]{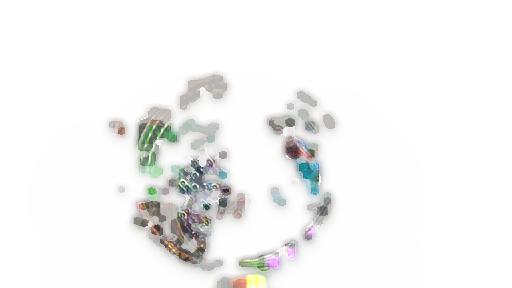}
        \includegraphics[width=\textwidth]{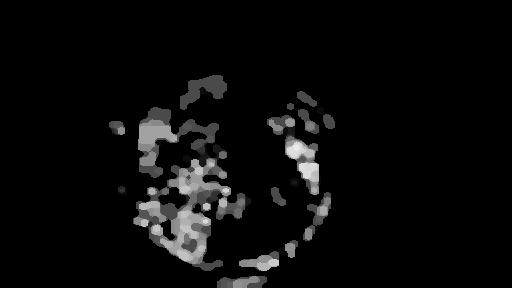}
        \includegraphics[width=\textwidth]{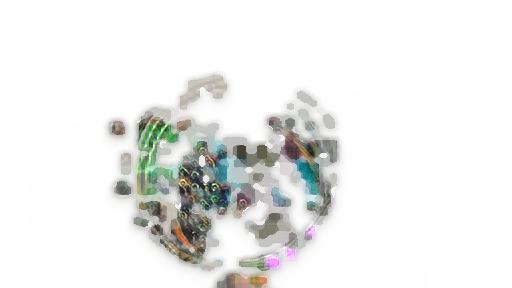}
        \includegraphics[width=\textwidth]{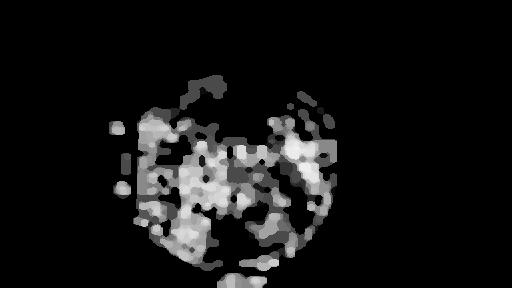}
        \includegraphics[width=\textwidth]{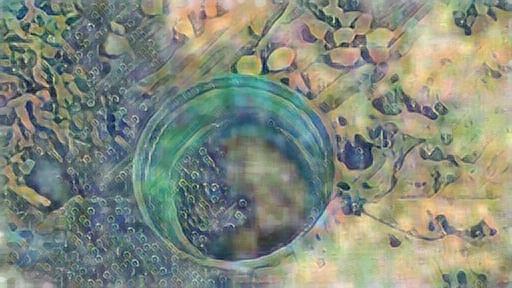}
        \caption{small, $l=9$}
    \end{subfigure}
    \begin{subfigure}[b]{0.16\textwidth}
        \includegraphics[width=\textwidth]{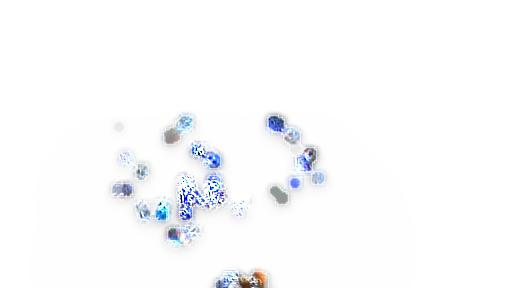}
        \includegraphics[width=\textwidth]{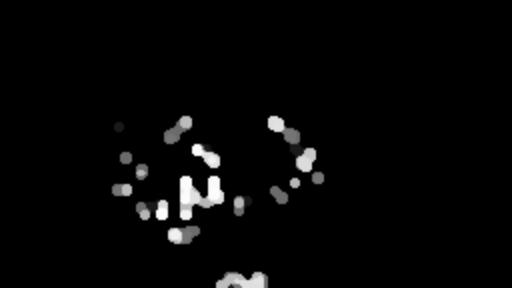}
        \includegraphics[width=\textwidth]{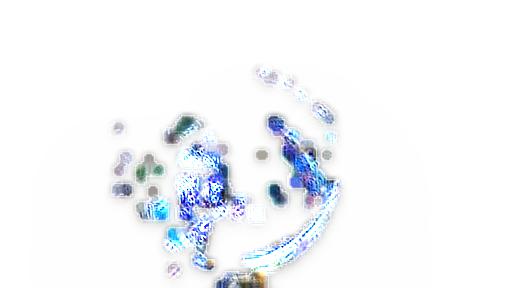}
        \includegraphics[width=\textwidth]{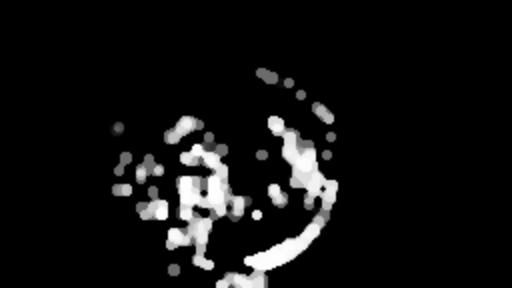}
        \includegraphics[width=\textwidth]{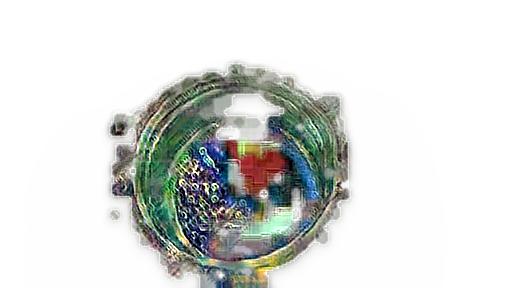}
        \includegraphics[width=\textwidth]{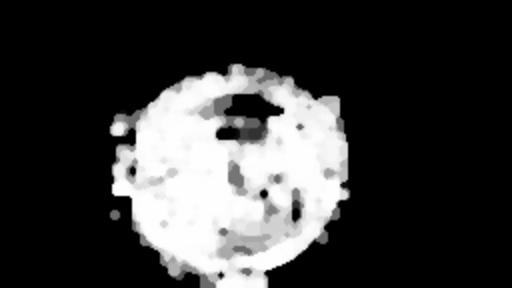}
        \includegraphics[width=\textwidth]{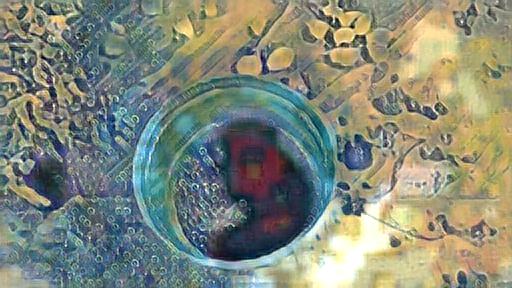}
        \caption{small, $l=8$}
    \end{subfigure}
    \begin{subfigure}[b]{0.16\textwidth}
        \includegraphics[width=\textwidth]{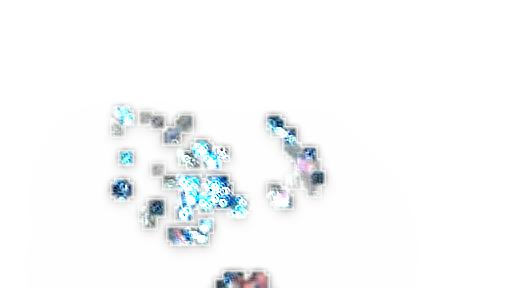}
        \includegraphics[width=\textwidth]{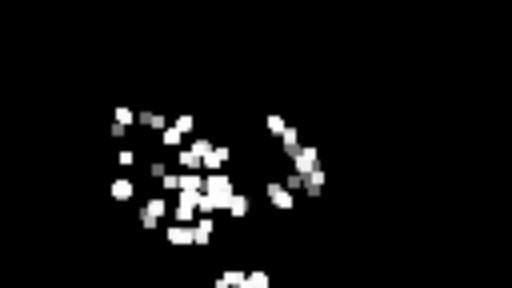}
        \includegraphics[width=\textwidth]{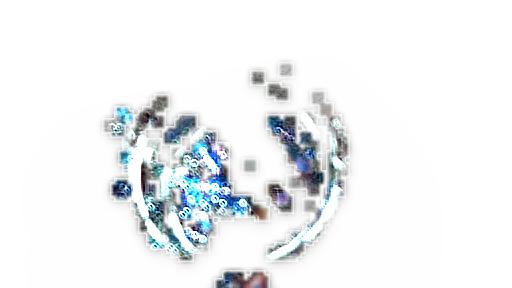}
        \includegraphics[width=\textwidth]{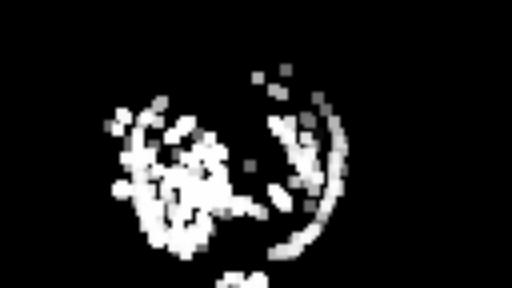}
        \includegraphics[width=\textwidth]{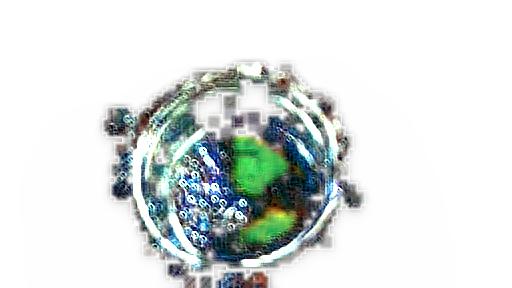}
        \includegraphics[width=\textwidth]{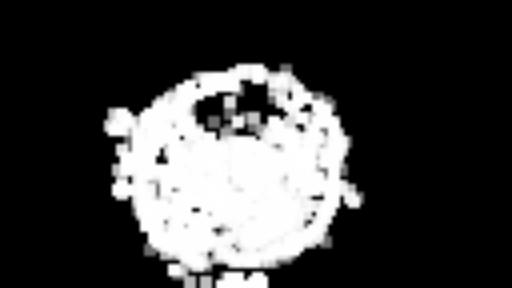}
        \includegraphics[width=\textwidth]{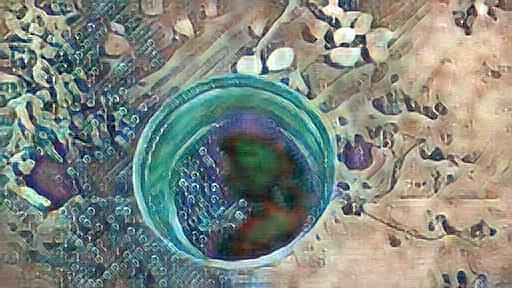}
        \caption{small, $l=7$}
    \end{subfigure}
    
    \caption{CPIA over the style transfer transformer of the MSG\cite{zhang2018multi}: $l$ indicates the operation layer and big/small means the stroke size (see $\mathcal{N}$ in Section~\ref{subsec:cs}). The top row consists of the content image, the style images (from left to right, \textit{Udnie}, \textit{Starry Night} and \textit{the Great Wave Off Kanagawa}), and original style transfer outputs. The remaining rows are the intermediate outputs, the intermediate stroke maps, and the final outputs. (a) and (b) are styled by \textit{Udnie}. (c) is styled by \textit{Starry Night}. (d-f) are styled by \textit{the Great Wave Off Kanagawa}. MSG has 13 layers where layers 2-9 are the operable bottleneck layers.}
    \label{fig:PIA_steps}
\end{figure}

Experiments are performed over the generator in the GANs and the transformer network of the style transfer. The layer in any of these networks decides the stroke-able space $(c,h,w)$ for the channel stroke. The cost in actions in the stroke-able space then quantifies the burden of the action and makes decision on either continuing stroke, change color, or stop painting. Depending on the learned knowledge in the pre-trained CNN layers, the channel stroking location and style varies. When applied toward the style transfer network, the stroke style varies on top of the neural style (Fig.~\ref{fig:PIA_steps}).

Mask R-CNN\cite{he2017mask} is used to help the neural painter plan via understanding what it is painting. With the knowledge of recognized objects, the neural painter then focus in painting the object regions one by one. The plan thus covers what to paint, whether the background is painted, and to what detail each region is painted, where the stroke detail is already parameterized in the channel stroke above. The tune-able planning helps to put appropriate focus over different regions. The whole system is called \textit{channel painter in action}, CPIA (Fig.~\ref{fig:PIA_block}).

The paper details the method on channel decomposition and rendering over limited amount of channels. Qualitative results are presented with quantified channel coverage at the operation layer. Based on the existing pre-trained networks, the proposed CPIA provides: 1. stroke composure actions, 2. additional tune-able artistic outcome, 3. controllable brush shape and movement. It is unsupervised and without additional training data.

\section{Related Work}

\textbf{Generative neural networks} provides the tensor space for channel decomposition. The back-propagation concept of the autoencoders has been introduced in 1980's~\cite{rumelhart1985learning,ballard1987modular}, making the neurons learn on the errors between the generated target and the real target. The infrastructural improvement on parallel computing later leads to two popular generative approaches: the variational autoencoders (VAEs\cite{kingma2013auto}) and the generative adversarial networks (GANs\cite{goodfellow2014generative}). The VAEs uses a framework of probabilistic graphical models to generate the output by maximizing the lower bound of the likelihood of the data. While the GANs leverages a discriminative network to judge and improve the output of the generative network. After the adoption of the deep convolutional nets (DCGAN\cite{radford2016unsupervised}), the task-oriented GANs have been applied to image-to-image translation (Pix2Pix\cite{isola2017image}, CycleGAN\cite{zhu2017unpaired}, GDWCT\cite{cho2019image}), concept-to-image translation (GAWWN\cite{reed2016learning}, PG$^2$\cite{ma2017pose}, StyleGAN\cite{karras2019style}) and text-to-image translation (StackGAN\cite{zhang2017stackgan}, BigGAN\cite{brock2019}), among other domain-specific GANs\cite{jin2017towards, chen2018cartoongan, wang2018high}. In this paper, the BigGAN is used to generate the CPIA painting targets from keywords, as in Fig.~\ref{fig:PIA_steps}.

\textbf{Neural style transfer} is one major domain that CPIA can be applied. In the work by Gatys \textit{et al.}~\cite{gatys2016image}, the authors formulate the style transfer cost as a combination of the content loss and the style loss. The loss is measured over the pre-trained VGGnet\cite{simonyan2014very}, from the generated image to both the content image and the style image. The transformer networks with deep convolutional layers are introduced in~\cite{johnson2016perceptual, ulyanov2016texture} to speed up the style transfer - the whole transformer is trained on a particular style. Then comes the transformer attempting to learn multiple styles in one single network, such as~\cite{dumoulin2017learned, zhang2018multi}. In the following sections, the transformer of MSG\cite{zhang2018multi} is decomposed into the CPIA actions.

\textbf{Stroke-based rendering}, or inverse graphic, without the training stroke sequence is challenging. To deal without the training stroke sequence, a discriminative network guides the distributed reinforcement learners to make meaningful progress in SPIRAL\cite{ganin2019synthesizing}. The computation cost is high for the deep reinforcement learners with large and continuous action space. That can be mitigated by creating a differentiable environment, like the ones in WorldModels\cite{ha2018recurrent}, PlaNet\cite{hafner2019learning} and StrokeNet\cite{zheng2019strokenet}. The ongoing research has delved into various stroking agents that generate very different output styles. For cartoon-like stroking, LearningToPaint\cite{huang2019learning} efficiently generates the simple strokes to compose the complex image. On the other hand, the NeuralPainter\cite{nakano2019neural} abstracts and recreates the image into a sketch-like output.

\section{Methods}

Let $\Lambda^{(l)}(Y)$ denote the layer operation of the $l^{th}$ layer on its input $Y$. The generator network of $L$ layers can be expressed as 
\begin{equation} \label{eq:gff}
\begin{array}{ll}
Y^{(l)} = \Lambda^{(l)} \Big(Y^{(l-1)}\Big) , & \forall l \in \{1,2,\cdots,L\},
\end{array}
\end{equation}
where $Y^{(0)}$ is the input of the network and $Y^{(L)}$ is the output. The operations and corresponding weights in $\Lambda^{(l)}$ were trained with or without the input $Y^{(0)}$.

Extending the forward path of the neural network to allow additional layer operations $\Phi^{(l)}$ leads to
\begin{equation} \label{eq:gffe}
\begin{array}{ll}
Y^{(l)} = \Phi^{(l)} \bigg( \Lambda^{(l)} \Big(Y^{(l-1)}\Big) \bigg) , & \forall l \in \{1,2,\cdots,L\},
\end{array}
\end{equation}

\begin{figure}[!t]
     \centering
     \begin{subfigure}[b]{0.32\textwidth}
         \centering
         \includegraphics[width=\textwidth]{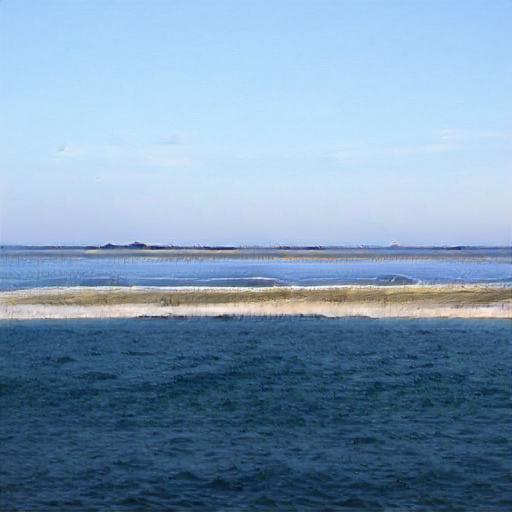}
         \caption{raw image generated from \cite{brock2019}}
         \label{fig:GDP_intro_1}
     \end{subfigure}
     \hfill
     \begin{subfigure}[b]{0.32\textwidth}
         \centering
         \includegraphics[width=\textwidth]{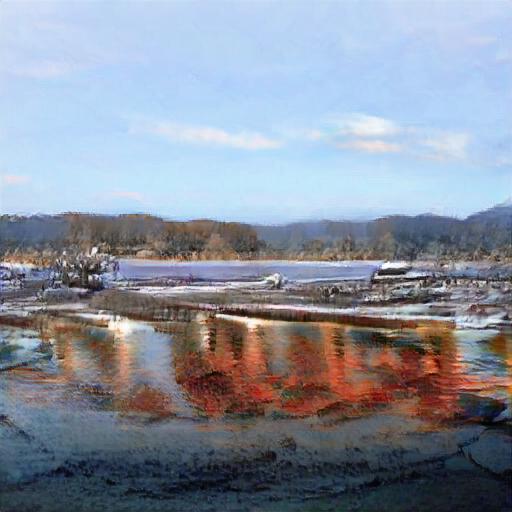}
         \caption{channel flush $l=3,\tau=512$}
         \label{fig:GDP_intro_2}
     \end{subfigure}
     \hfill
     \begin{subfigure}[b]{0.32\textwidth}
         \centering
         \includegraphics[width=\textwidth]{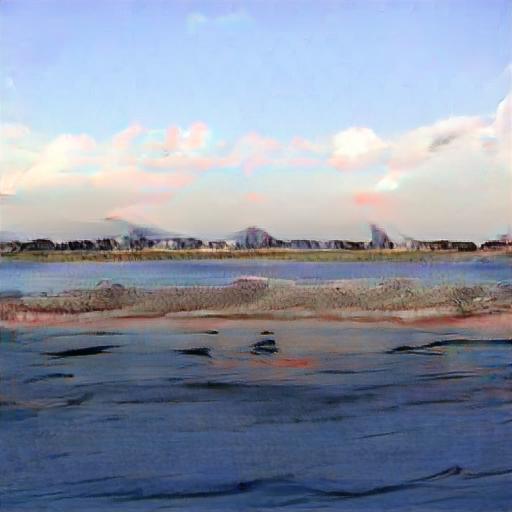}
         \caption{channel flush $l=6,\tau=512$}
         \label{fig:GDP_intro_3}
     \end{subfigure}
     \begin{subfigure}[b]{0.32\textwidth}
         \centering
         \includegraphics[width=\textwidth]{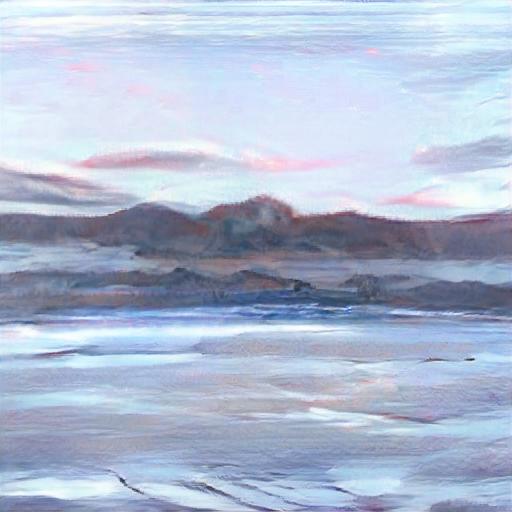}
         \caption{channel stroke $l=4,\tau=256$}
         \label{fig:GDP_intro_4}
     \end{subfigure}
     \hfill
     \begin{subfigure}[b]{0.32\textwidth}
         \centering
         \includegraphics[width=\textwidth]{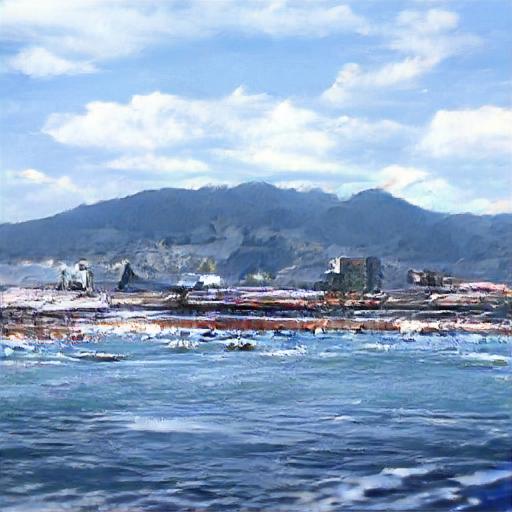}
         \caption{channel stroke $l=3,\tau=512$}
         \label{fig:GDP_intro_5}
     \end{subfigure}
     \hfill
     \begin{subfigure}[b]{0.32\textwidth}
         \centering
         \includegraphics[width=\textwidth]{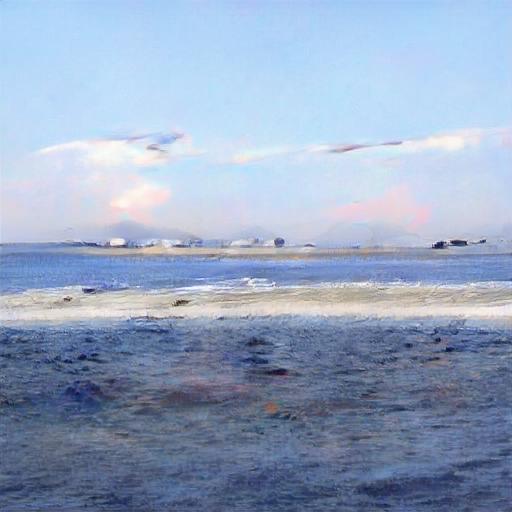}
         \caption{channel stroke $l=6,\tau=512$}
         \label{fig:GDP_intro_6}
     \end{subfigure}
        \caption{Channel decomposition of the BigGAN~\cite{brock2019}: (a) is generated using the keyword "seashore". From (a), we use the channel flush in Section~\ref{subsec:cf} to generate (b) and (c), and the channel stroke in Section~\ref{subsec:cs} to generate (d-f). BigGAN has 15 layers and 14 of them are generative bottleneck blocks.}
        \label{fig:GDP_intro}
\end{figure}

\subsection{Channel Flush} \label{subsec:cf}

Next, we provide implementations of the layer operations $\Phi^{(l)}$. One key finding in our experiments is that, for the intermediate layer images $Y^{(1)}\cdots Y^{(L-1)}$, the decomposed representation preserves the spatial information of the output image $Y^{(L)}$, while the detail at each location can be truncated and represented by the high response channel(s). By only showing $\tau$ channels out of the total $C$ channels at layer $L$, we force the later layers to respond only on the top $\tau$ channels at each location of $L$. This observation shed light on the following operation
\begin{equation} \label{eq:cf}
\begin{array}{l}
\Phi^{(l)} \Big(Y\Big) = M\Big(Y\Big) \odot Y  \\
and \\
M_{c,h,w}(Y, \tau) = \left\{
\begin{array}{ll}
1 & \text{if } \Big\|Y_{(c,h,w)} < Y_{(c',h,w)}\Big\| \leq \tau, \forall c' \in \{1, \cdots, C\} \\
\\
0 & \text{otherwise,}
\end{array}
\right.
\end{array}
\end{equation}
where $\odot$ is the Hadamard product, $\|\cdot\|$ is the cardinality of a set, and $\tau$ is the channel limit out of the $C$ channels at layer $l$ per location $h,w$. The tensor $M$ masks the original layer output $Y$ and picks only the top $\tau$ channels of $Y$ at each location $h,w$ for rendering into later layers.

The channel flush provides a way to focus the image render on high response channels. On any spatial location of the operating layer, the lower response channels are muted. The number of channels to choose from also matters. In CNNs \cite{lecun2004learning, krizhevsky2012imagenet}, the depth (channels) is traded with the breadth (spatial size). As a consequence, the operation layer of the channel flush is better in the middle of the network - avoiding the last layers which have low channel variety, and the initial layers which have low spatial resolution.

The result of the channel decomposition is shown in Fig.~\ref{fig:GDP_intro}. We start with an image generated from the BigGAN\cite{brock2019}, then apply the channel flush and the channel stroke, which is described next.

\subsection{Channel Stroke} \label{subsec:cs}

Sometimes the channel flush incurs unnecessary discontinuities over the output image. To deal with this issue, we extend the $\Phi^{(l)} \big(\cdot\big)$ in Eq.~\ref{eq:cf} into an operation set of channel strokes at layer $l$. Let $(C,H,W)$ denote the channel depth, height and width of its output $Y^{(l)}$. We define $\mathcal{N}(c,h,w;m)\in \mathbb{N}^3$ as the set of neighborhood pixels $(c',h',w')$ near pixel $(c,h,w)$, where $Y_{c',h',w'} \geq m\cdot Y_{c,h,w}$ for all $(c',h',w')$ in $\mathcal{N}(c,h,w;m)$. The quantifier $m$ is a real number from $(0,1)$, which means the sensitivity of the stroke. When $m$ is close to one, the stroke sensitivity is high. Thus the channel stroke can only turn on the neighboring pixels with highly similar response as the stroke pixel. The simplest case of $\mathcal{N}(c,h,w;m)$ is a square box centered at $(c,h,w)$ with each side of $2z+1$ pixels on channel $c$. In this case, the parameter $z$ is the stroke size.

The channel stroke algorithm (Alg.\ref{alg:cs}) updates the mask tensor $M$ in $\mathbb{Z}_2^{C\times H\times W}$ and the cost tensor $G$ in $\mathbb{R}^{C\times H\times W}$, on each of its iteration. The mask $M$ then filters the layer $l$ to response $Y$ in $\mathbb{R}^{C\times H\times W}$. The stopping criterion $\mathcal{S}$ terminates the procedure when current response $Y_{c,h,w}$ is lower than a fraction of $Y_{max}$. Other possible choices for $\mathcal{S}$ include number of strokes and the fraction of painted locations.

\begin{algorithm}
  \caption{Channel Stroke}\label{alg:cs}
  \begin{algorithmic}[1]
    \Procedure{ChannelStroke}{$\mathcal{S},\mathcal{N},m,\tau$}
    \State mask $M\gets 0$
    \State cost $G\gets 0$
    \While{true}
    \State choose pixel $(c,h,w) \gets \argmax\limits_{c,h,w}(1-G) \odot Y$ conditioned on 
      \begin{equation*}
      M_{c,h,w}=0 \text{ and } \sum_{c'=1}^{C}M_{c',h,w} < \tau
      \end{equation*}
      \label{alg:cs_pick}
    \If{stopping criterion $\mathcal{S}$ met}
      \State break
    \EndIf
    \State extend stroke $M_{c',h',w'}\gets 1$ for all $(c',h',w')$ in $\mathcal{N}(c,h,w;m)$ \label{alg:cs_ext}
    \State update $G$ according to the chosen pixel $(c,h,w)$ \label{alg:cs_cost}
    \EndWhile
    \EndProcedure
  \end{algorithmic}
\end{algorithm}

The intuition of channel stroke is that, for a human painter to paint, one needs to select a color of paint, the painting brush, and the pattern to paint. Once these items are selected, the painter can stroke on the canvas and then extend the stroke over a certain region. In Alg.\ref{alg:cs}, the color of paint and pattern to paint are controlled by channel $c$, which is chosen in Step~\ref{alg:cs_pick} according the current most responsive pixel $(c,h,w)$. The neighborhood $\mathcal{N}(c,h,w;m)$ decides the stroke shape, which can be related to the painting brush of the human painter.

At the end of each stroke, the human painter can either continue to use the same color to paint other area on the canvas, or switch to other color. To be effective, it is desirable to keep using the brush of current color as much as possible on the same level of painting detail. We factor this behavior into the cost $G$ at the stroke pixel $(c,h,w)$. The channel stroke will continue the stroke into nearby stroke pixel at the same channel. Note that the neighborhood extension in Step~\ref{alg:cs_ext} is about the shape of the stroke, while the stroke continuation in Step~\ref{alg:cs_cost} models the behavior of the human painter in switching color.

\begin{figure}[!t]
     \centering
     \begin{subfigure}[]{\textwidth}
         \centering 
         \hspace*{\fill}
         \includegraphics[width=0.15\textwidth]{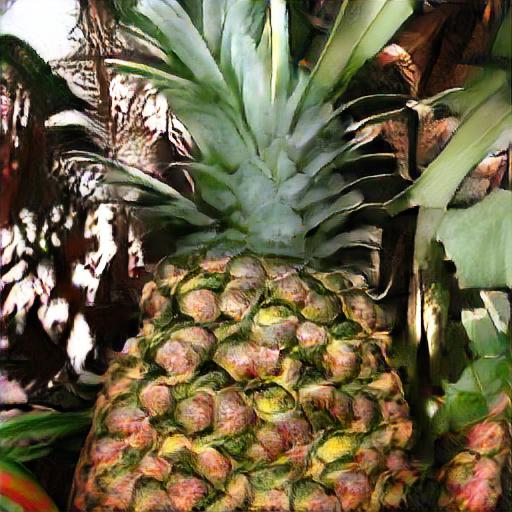}
         \hspace{7em} \includegraphics[width=0.15\textwidth]{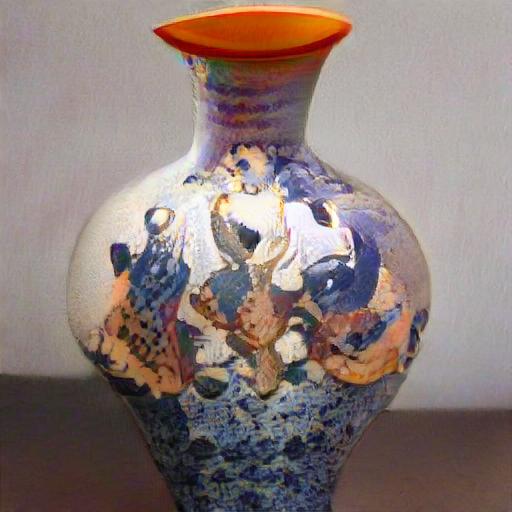}
         \hspace{7em} \includegraphics[width=0.15\textwidth]{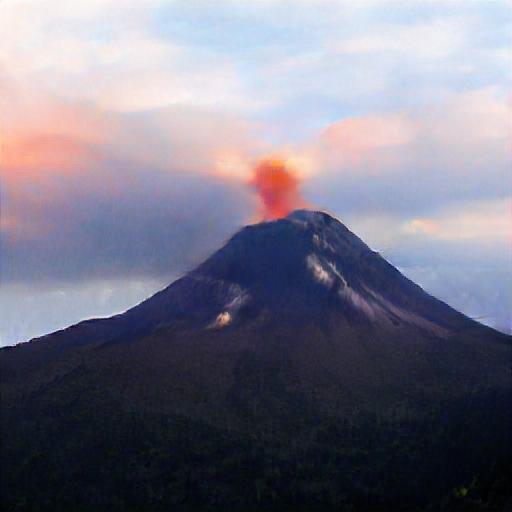}
         \hspace*{\fill}
     \end{subfigure}

     \begin{subfigure}[]{\textwidth}
         \centering
         \includegraphics[width=0.16\textwidth]{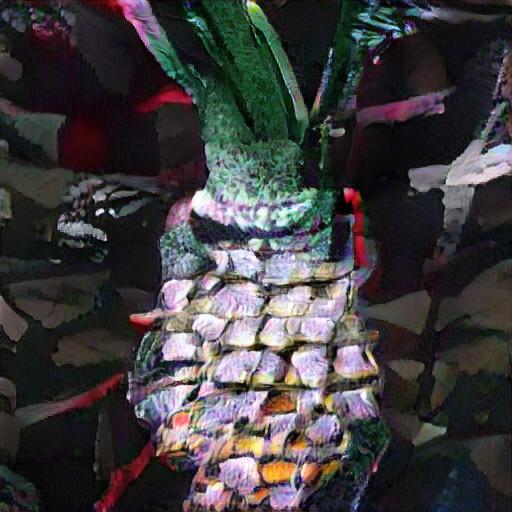}
         \hspace{-6.85em} \includegraphics[width=0.04\textwidth]{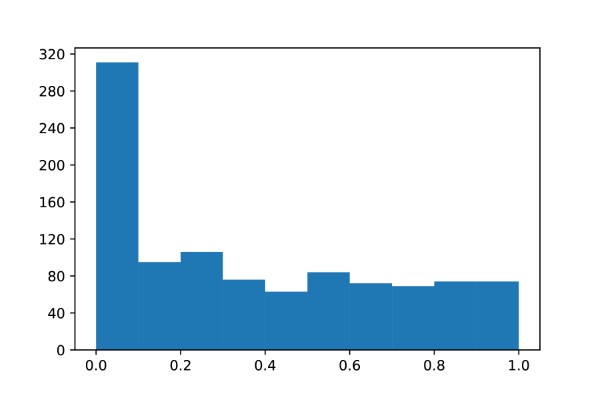}
         \hspace{4.5em} \includegraphics[width=0.16\textwidth]{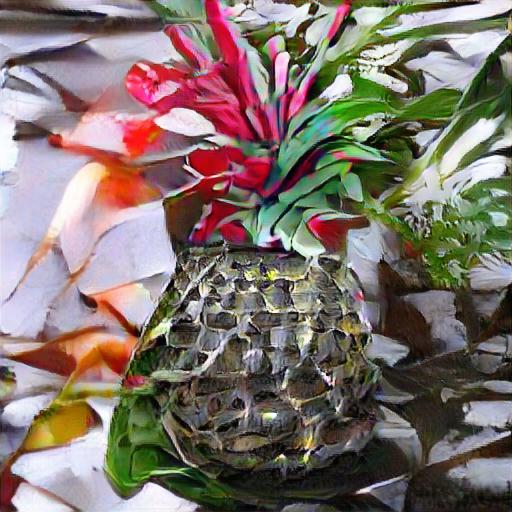}
         \hspace{-6.85em} \includegraphics[width=0.04\textwidth]{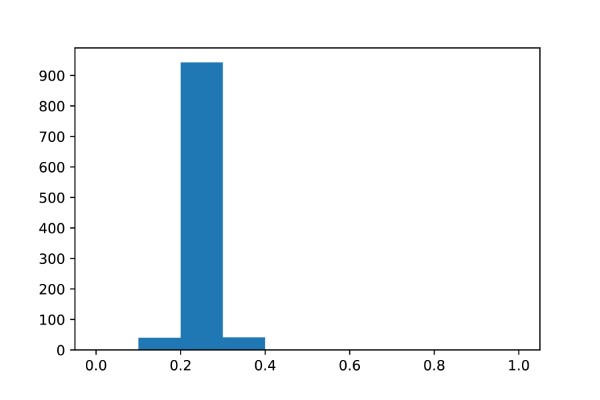}
         \hspace{4.5em}\hfill
         \includegraphics[width=0.16\textwidth]{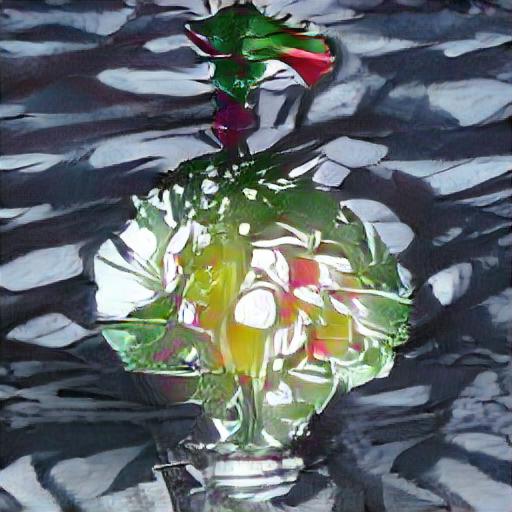}
         \hspace{-6.85em} \includegraphics[width=0.04\textwidth]{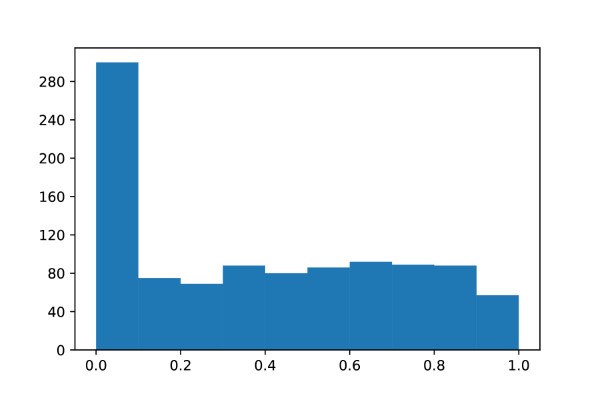}
         \hspace{4.5em} \includegraphics[width=0.16\textwidth]{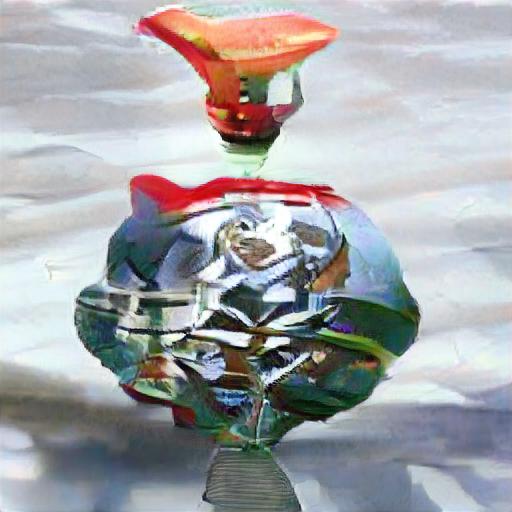}
         \hspace{-6.85em} \includegraphics[width=0.04\textwidth]{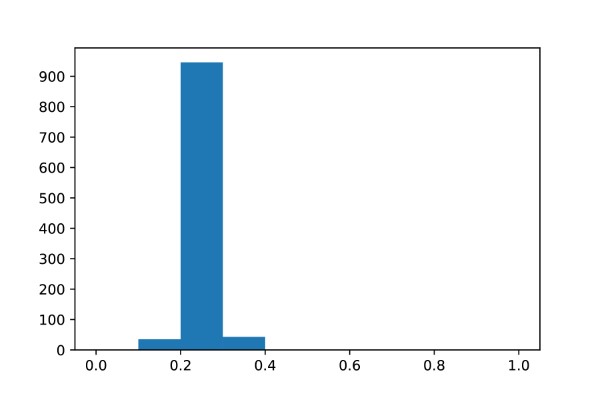}
         \hspace{4.5em}\hfill
         \includegraphics[width=0.16\textwidth]{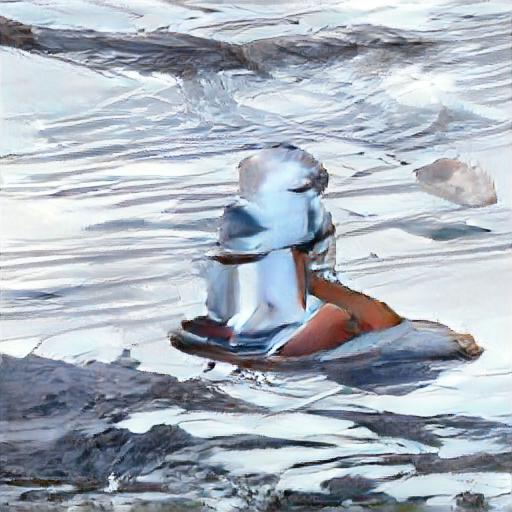}
         \hspace{-6.85em} \includegraphics[width=0.04\textwidth]{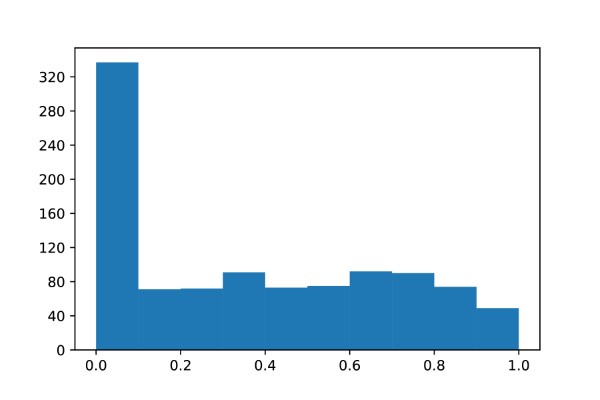}
         \hspace{4.7em}\includegraphics[width=0.16\textwidth]{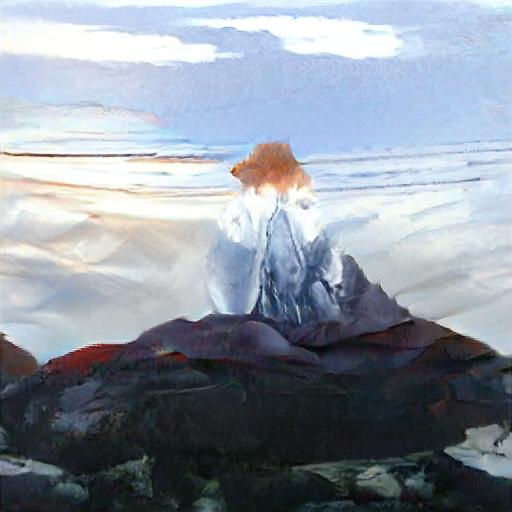}
         \hspace{-6.85em} \includegraphics[width=0.04\textwidth]{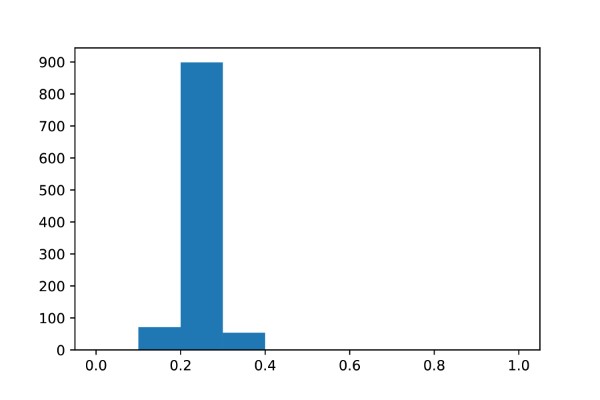}
         \hspace{4.6em}
     \end{subfigure}

     \begin{subfigure}[]{\textwidth}
         \centering
         \includegraphics[width=0.16\textwidth]{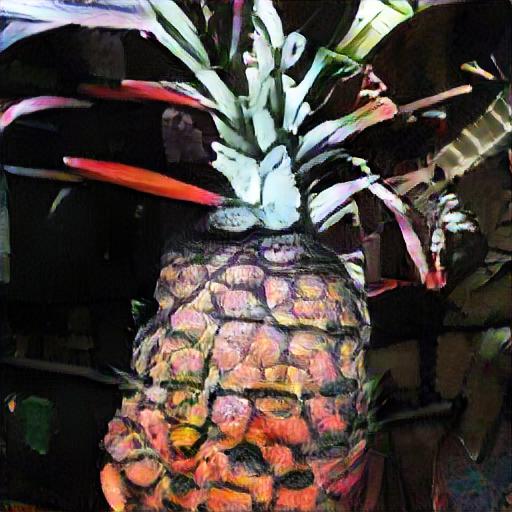}
         \hspace{-6.85em} \includegraphics[width=0.04\textwidth]{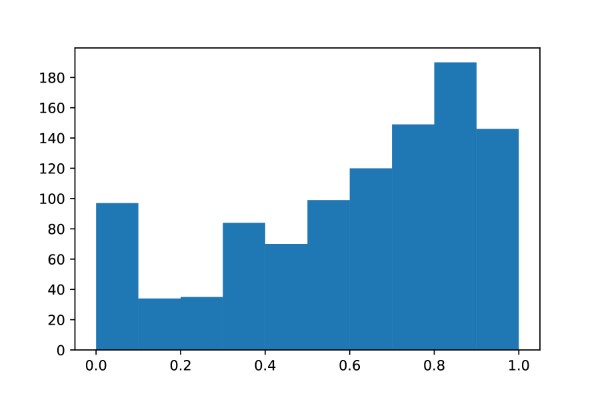}
         \hspace{4.5em} \includegraphics[width=0.16\textwidth]{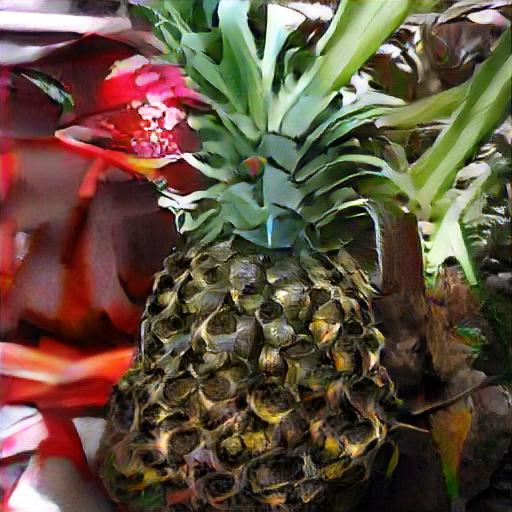}
         \hspace{-6.85em} \includegraphics[width=0.04\textwidth]{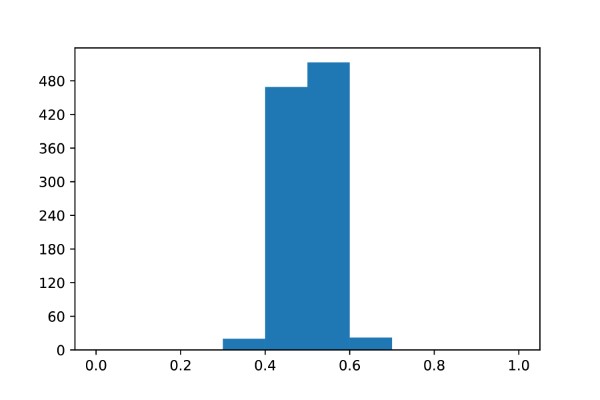}
         \hspace{4.5em}\hfill
         \includegraphics[width=0.16\textwidth]{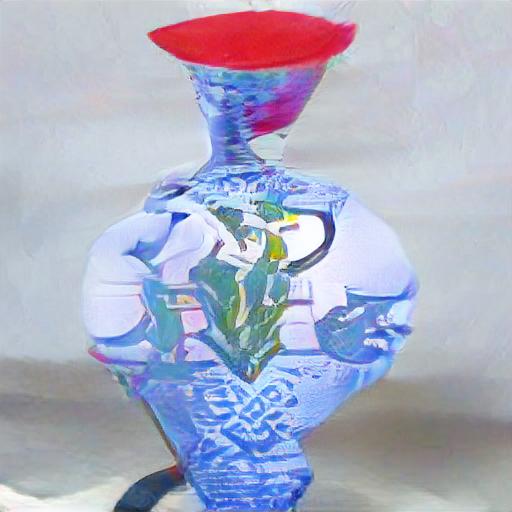}
         \hspace{-6.85em} \includegraphics[width=0.04\textwidth]{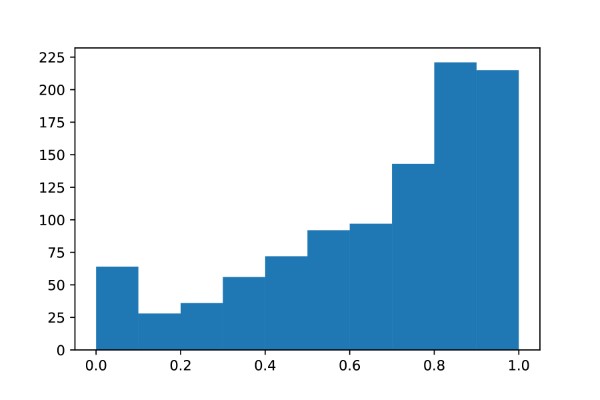}
         \hspace{4.5em} \includegraphics[width=0.16\textwidth]{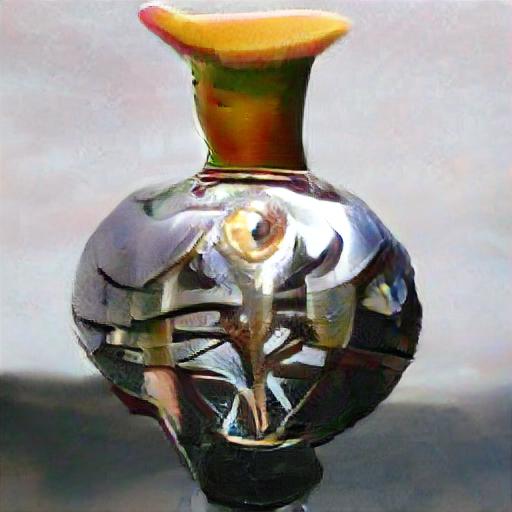}
         \hspace{-6.85em} \includegraphics[width=0.04\textwidth]{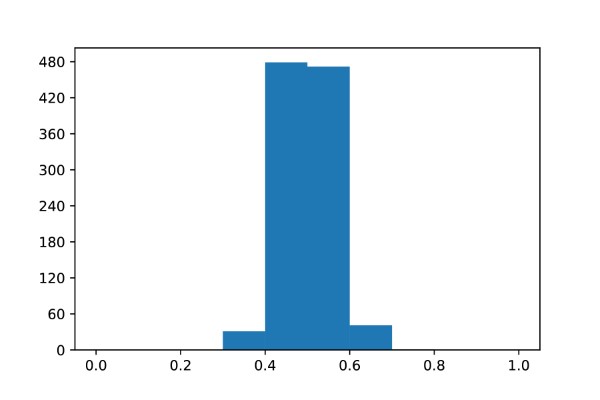}
         \hspace{4.5em}\hfill
         \includegraphics[width=0.16\textwidth]{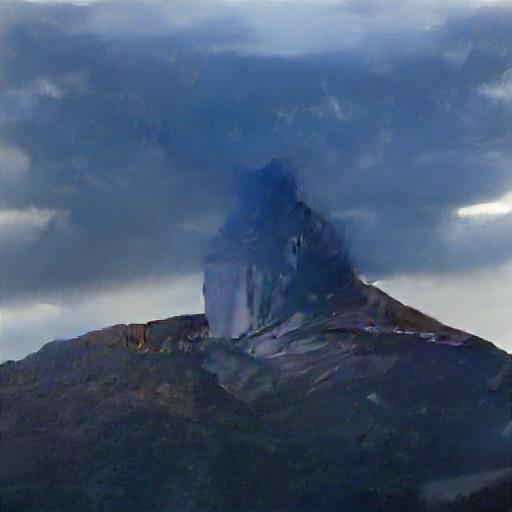}
         \hspace{-6.85em} \includegraphics[width=0.04\textwidth]{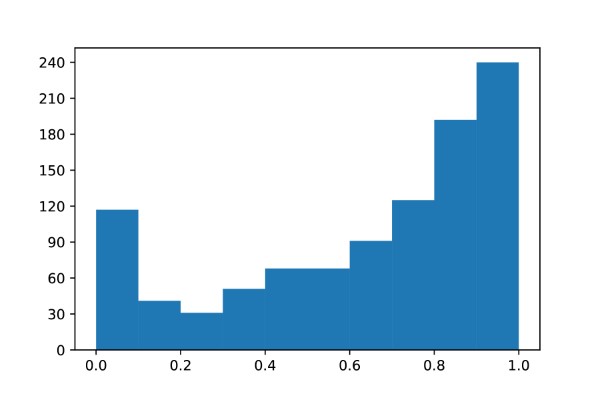}
         \hspace{4.5em} \includegraphics[width=0.16\textwidth]{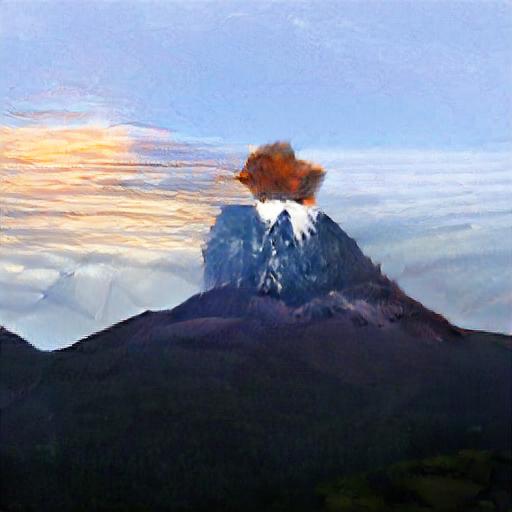}
         \hspace{-6.85em} \includegraphics[width=0.04\textwidth]{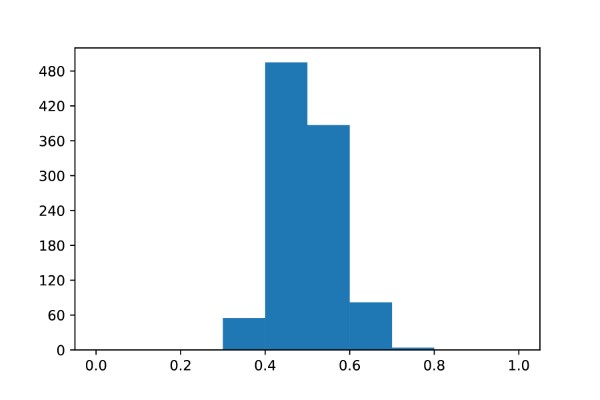}
         \hspace{4.6em}
     \end{subfigure}

     \begin{subfigure}[]{\textwidth}
         \centering
         \includegraphics[width=0.16\textwidth]{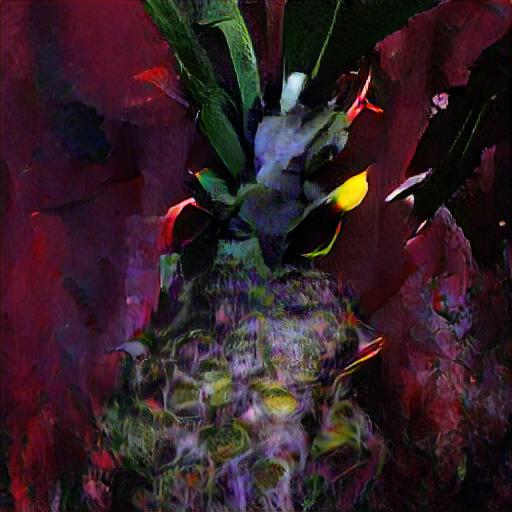}
         \hspace{-6.85em} \includegraphics[width=0.04\textwidth]{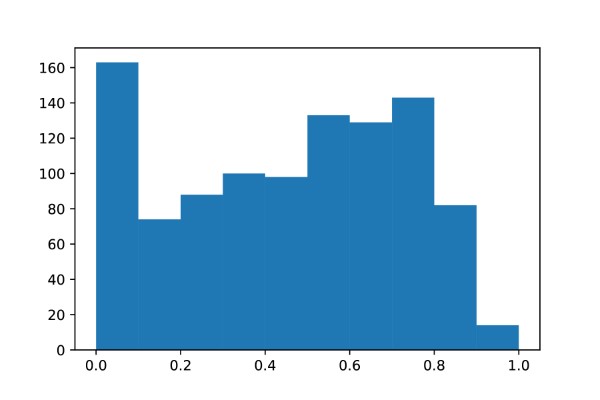}
         \hspace{4.5em} \includegraphics[width=0.16\textwidth]{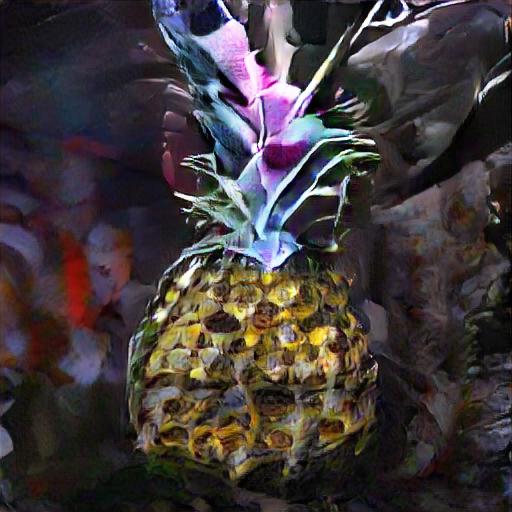}
         \hspace{-6.85em} \includegraphics[width=0.04\textwidth]{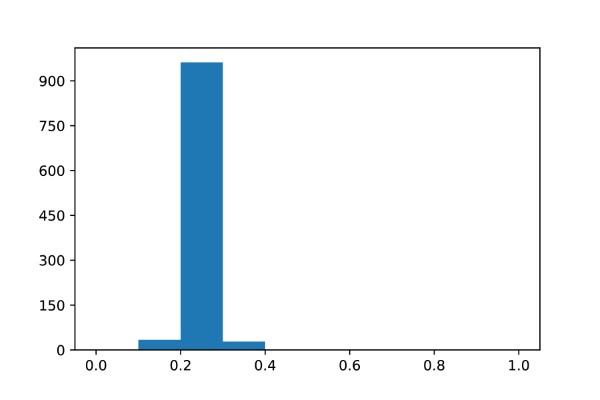}
         \hspace{4.5em}\hfill
         \includegraphics[width=0.16\textwidth]{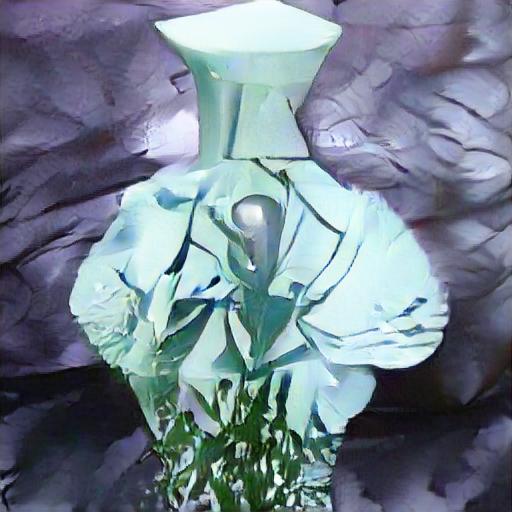}
         \hspace{-6.85em} \includegraphics[width=0.04\textwidth]{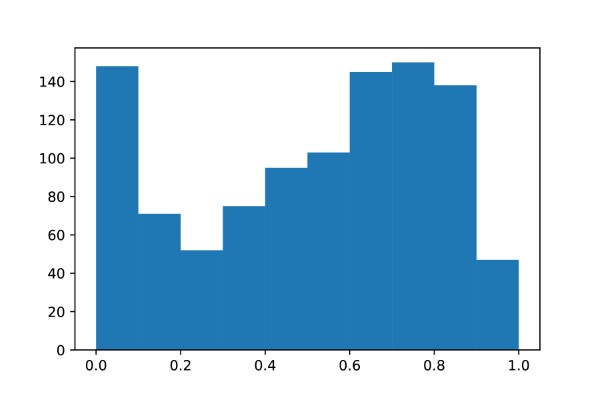}
         \hspace{4.5em} \includegraphics[width=0.16\textwidth]{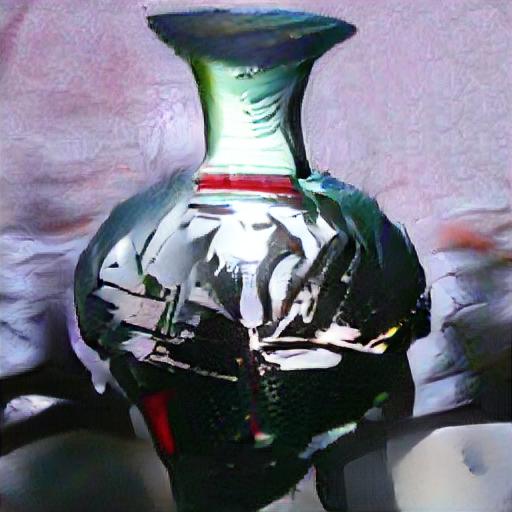}
         \hspace{-6.85em} \includegraphics[width=0.04\textwidth]{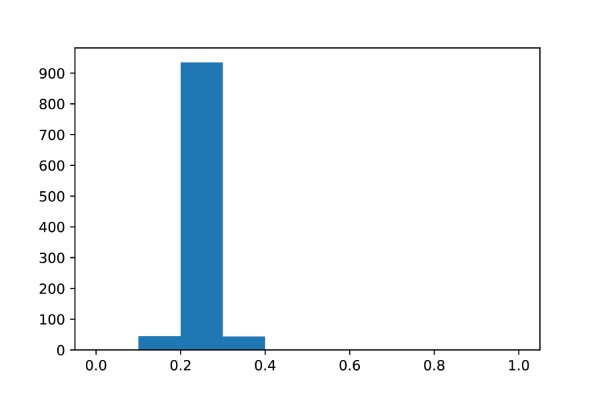}
         \hspace{4.5em}\hfill
         \includegraphics[width=0.16\textwidth]{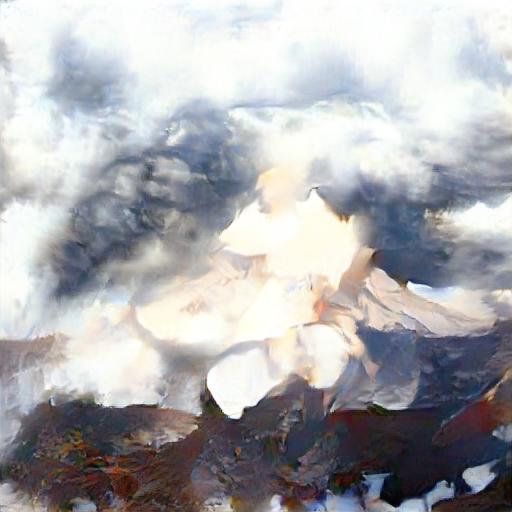}
         \hspace{-6.85em} \includegraphics[width=0.04\textwidth]{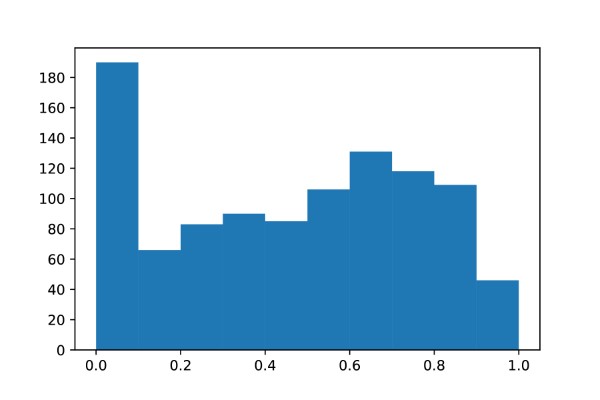}
         \hspace{4.5em} \includegraphics[width=0.16\textwidth]{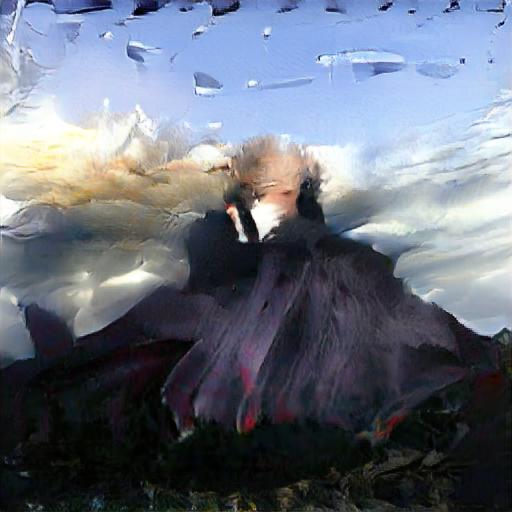}
         \hspace{-6.85em} \includegraphics[width=0.04\textwidth]{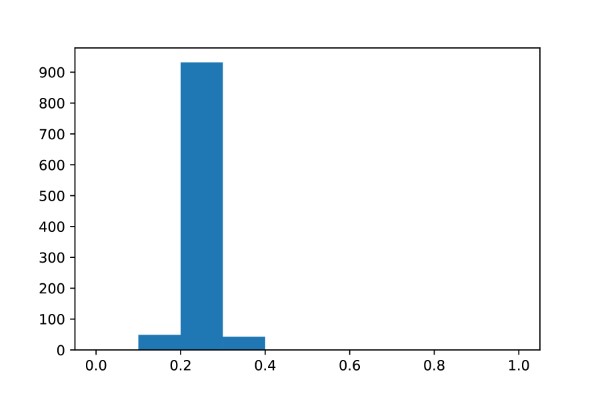}
         \hspace{4.6em}
     \end{subfigure}

     \begin{subfigure}[]{\textwidth}
         \centering
         \includegraphics[width=0.16\textwidth]{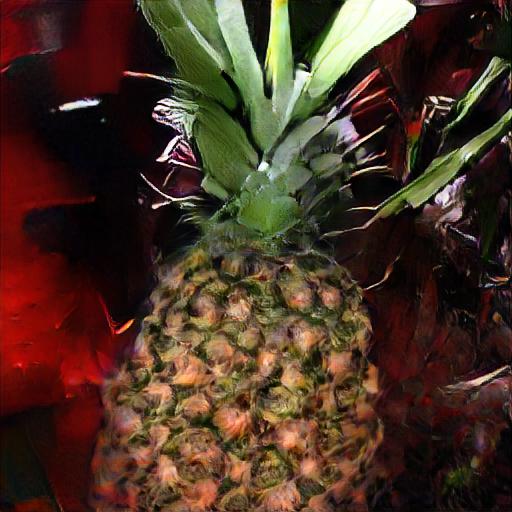}
         \hspace{-6.85em} \includegraphics[width=0.04\textwidth]{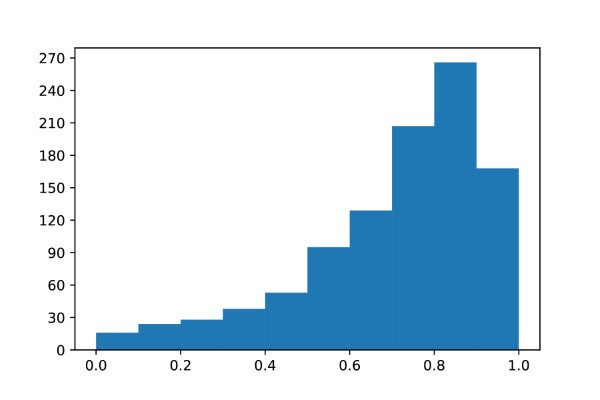}
         \hspace{4.5em} \includegraphics[width=0.16\textwidth]{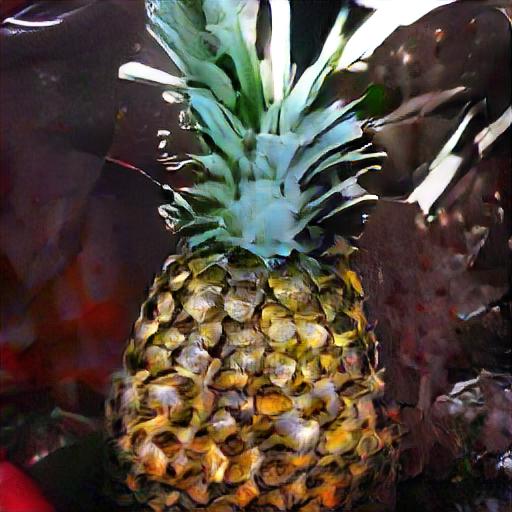}
         \hspace{-6.85em} \includegraphics[width=0.04\textwidth]{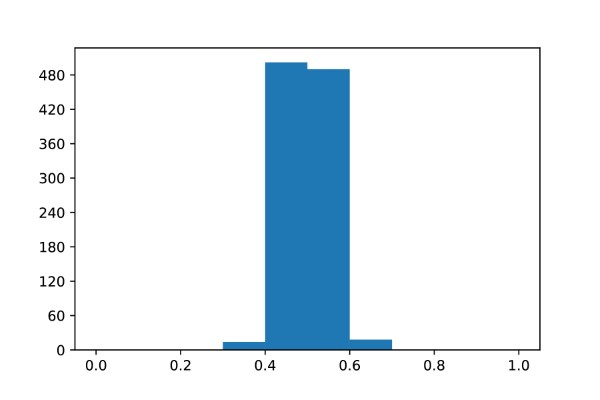}
         \hspace{4.5em}\hfill
         \includegraphics[width=0.16\textwidth]{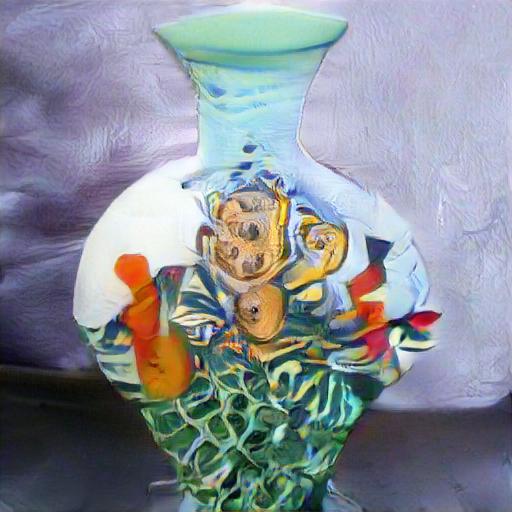}
         \hspace{-6.85em} \includegraphics[width=0.04\textwidth]{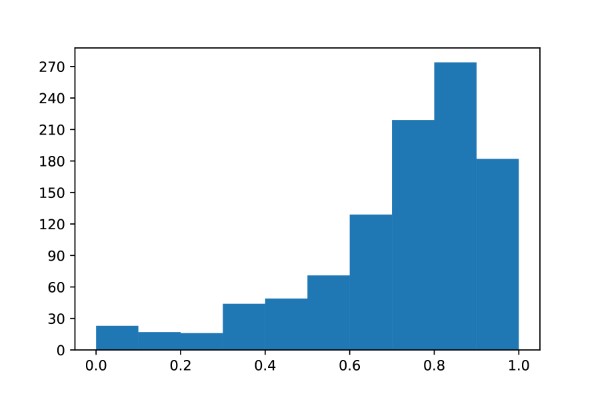}
         \hspace{4.5em} \includegraphics[width=0.16\textwidth]{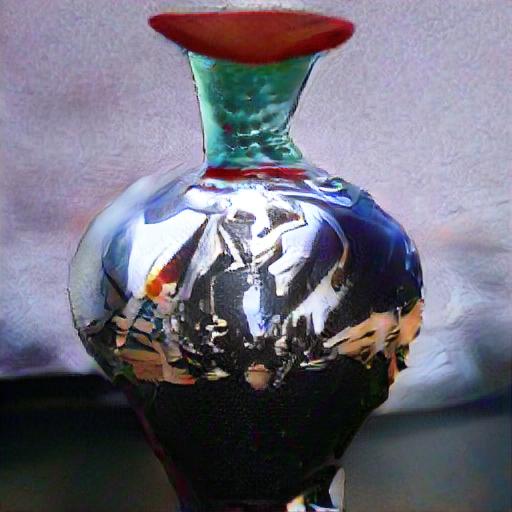}
         \hspace{-6.85em} \includegraphics[width=0.04\textwidth]{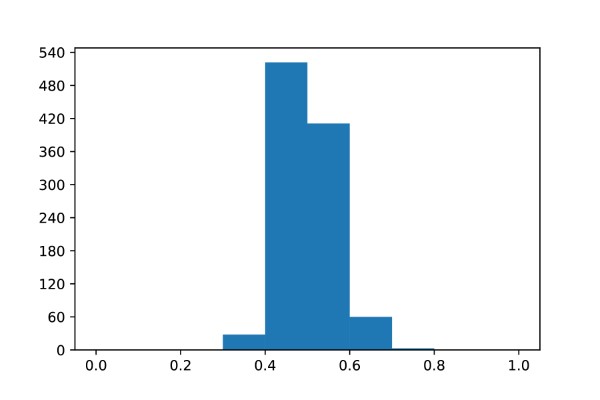}
         \hspace{4.5em}\hfill
         \includegraphics[width=0.16\textwidth]{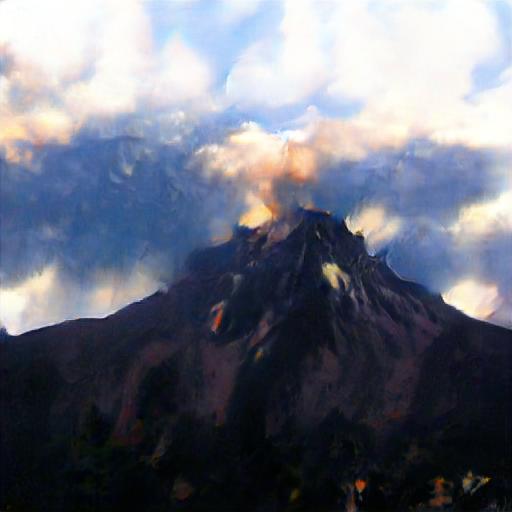}
         \hspace{-6.85em} \includegraphics[width=0.04\textwidth]{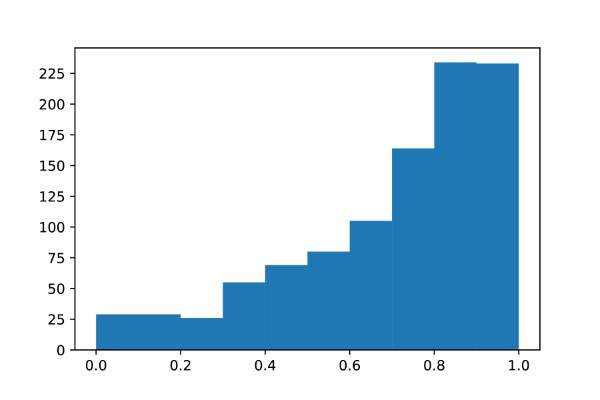}
         \hspace{4.5em} \includegraphics[width=0.16\textwidth]{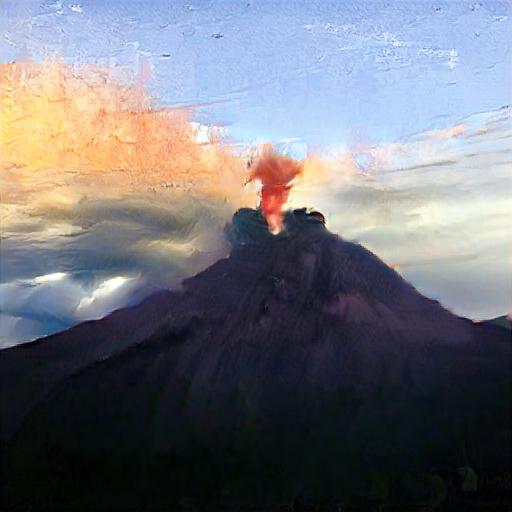}
         \hspace{-6.85em} \includegraphics[width=0.04\textwidth]{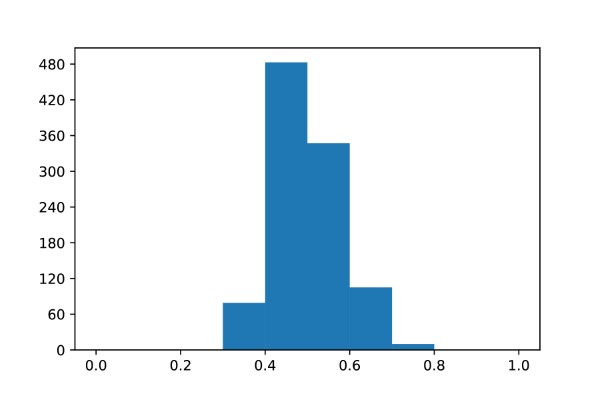}
         \hspace{4.6em}
     \end{subfigure}

    \begin{subfigure}[]{\textwidth}
        \centering
        \includegraphics[width=0.16\textwidth]{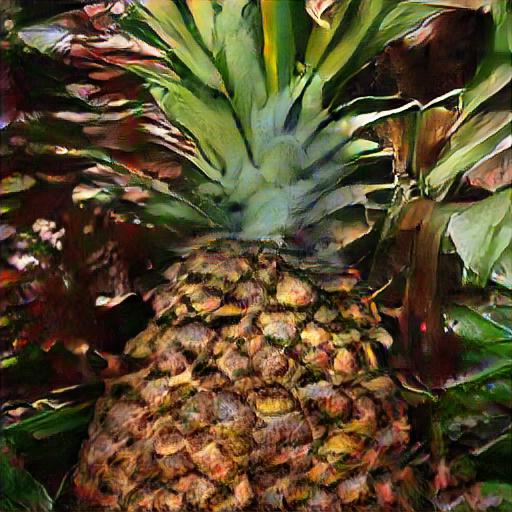}
        \hspace{-6.85em} \includegraphics[width=0.04\textwidth]{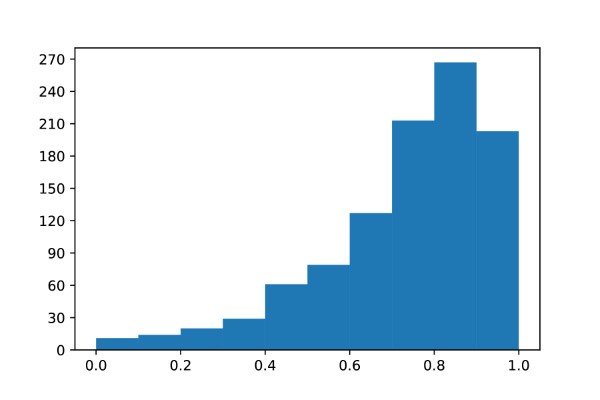}
        \hspace{4.5em} \includegraphics[width=0.16\textwidth]{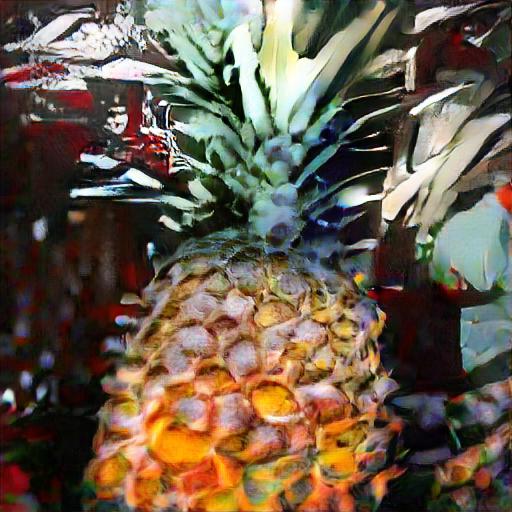}
        \hspace{-6.85em} \includegraphics[width=0.04\textwidth]{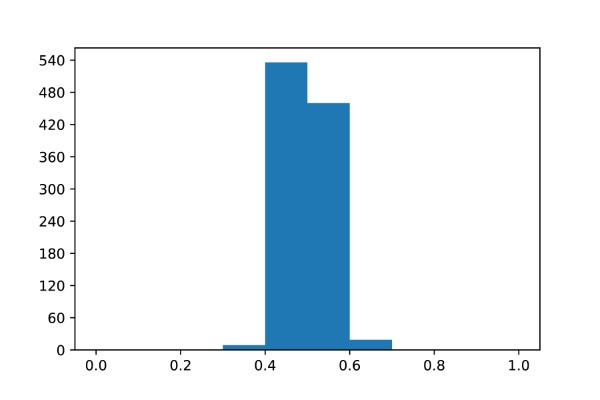}
        \hspace{4.5em}\hfill
        \includegraphics[width=0.16\textwidth]{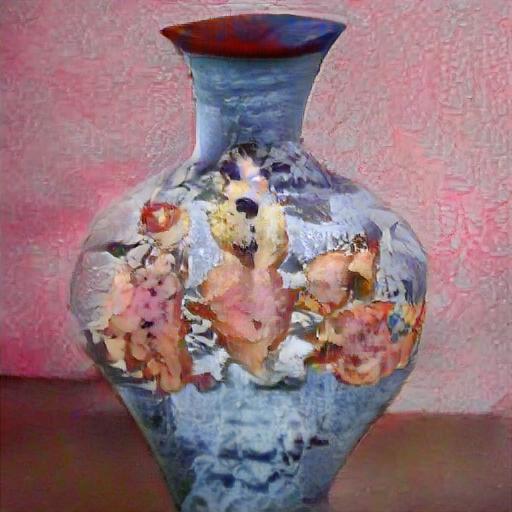}
        \hspace{-6.85em} \includegraphics[width=0.04\textwidth]{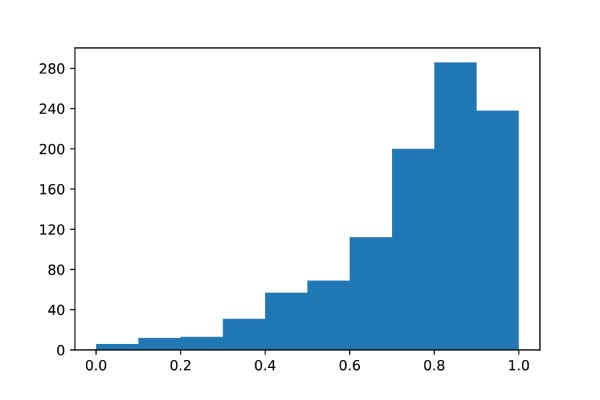}
        \hspace{4.5em} \includegraphics[width=0.16\textwidth]{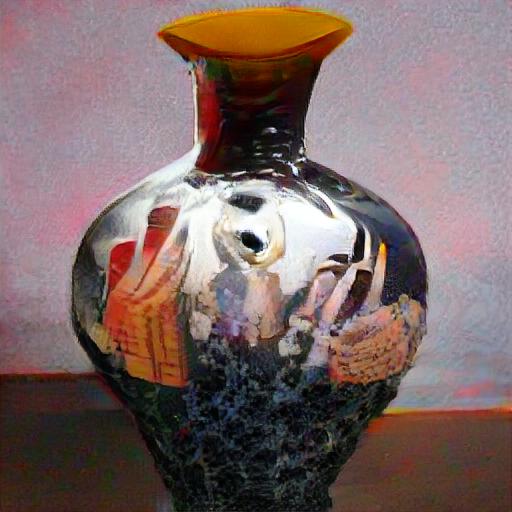}
        \hspace{-6.85em} \includegraphics[width=0.04\textwidth]{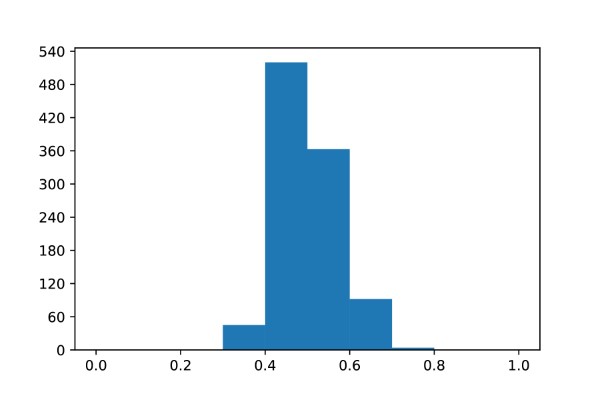}
        \hspace{4.5em}\hfill
        \includegraphics[width=0.16\textwidth]{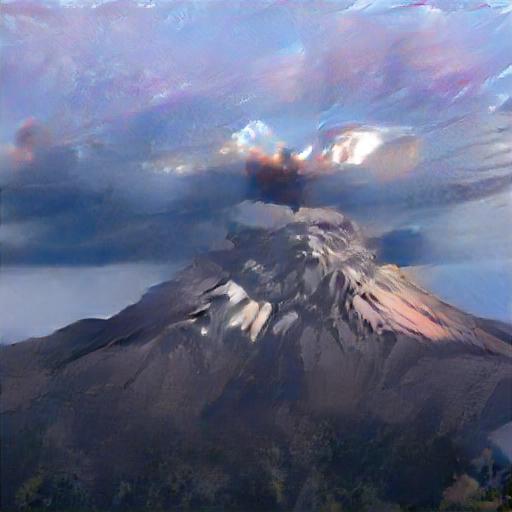}
        \hspace{-6.85em} \includegraphics[width=0.04\textwidth]{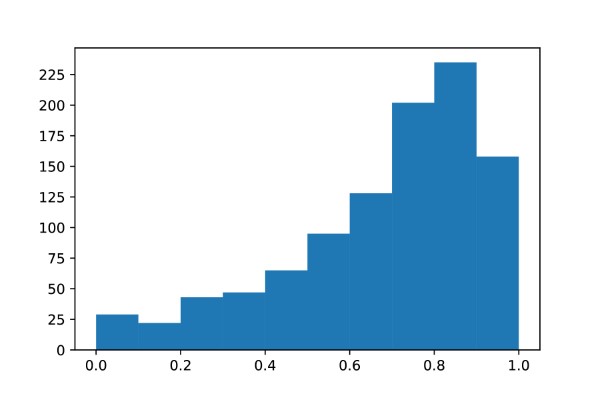}
        \hspace{4.5em} \includegraphics[width=0.16\textwidth]{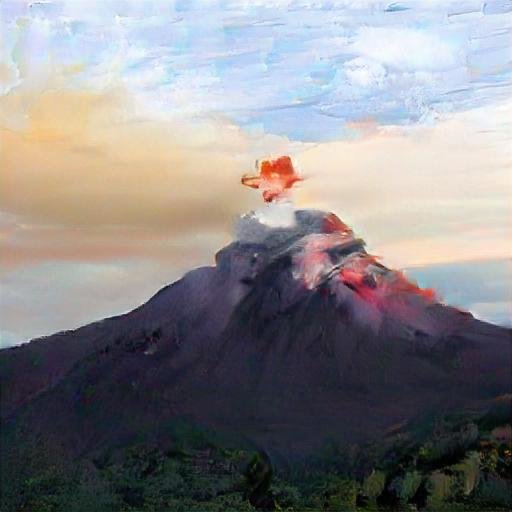}
        \hspace{-6.85em} \includegraphics[width=0.04\textwidth]{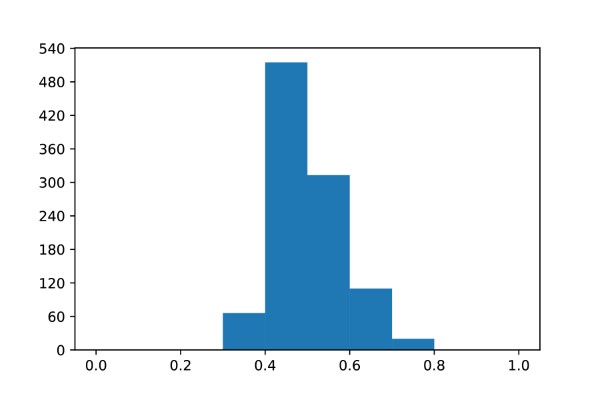}
        \hspace{4.6em}
    \end{subfigure}

    \begin{subfigure}[]{\textwidth}
        \centering
        \includegraphics[width=0.16\textwidth]{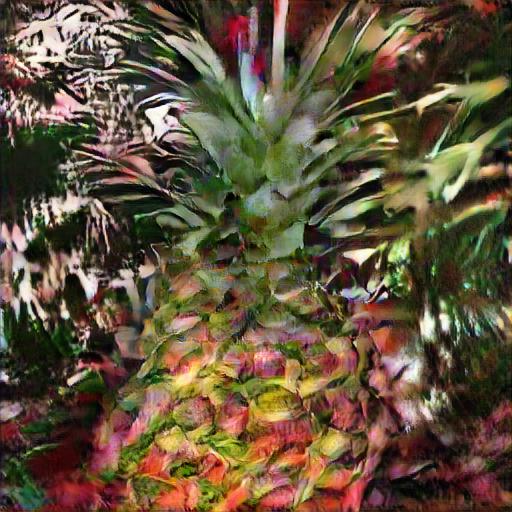}
        \hspace{-6.85em} \includegraphics[width=0.04\textwidth]{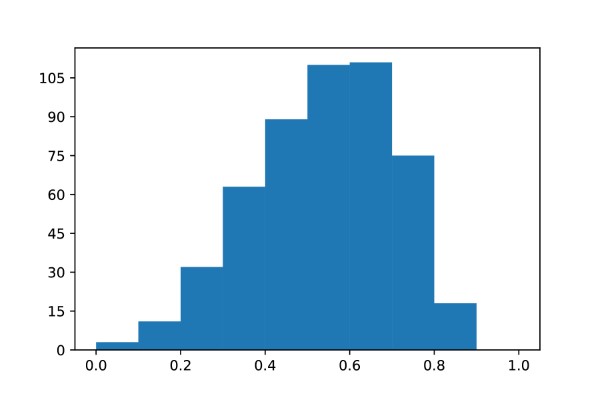}
        \hspace{4.5em} \includegraphics[width=0.16\textwidth]{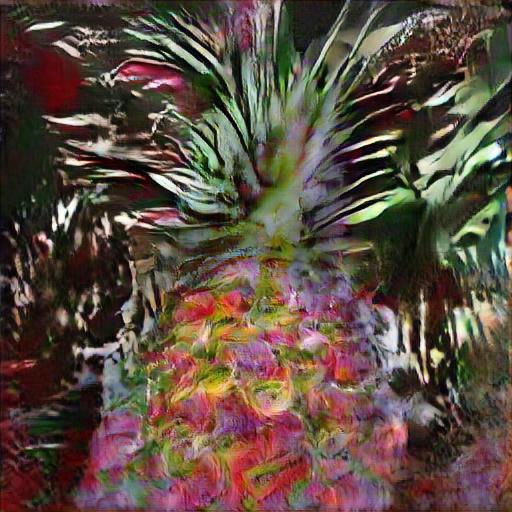}
        \hspace{-6.85em} \includegraphics[width=0.04\textwidth]{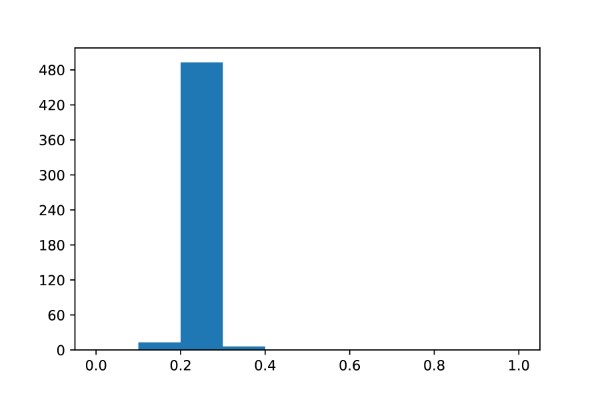}
        \hspace{4.5em}\hfill
        \includegraphics[width=0.16\textwidth]{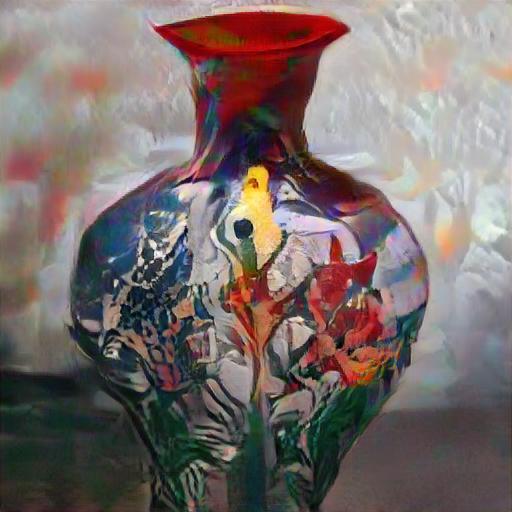}
        \hspace{-6.85em} \includegraphics[width=0.04\textwidth]{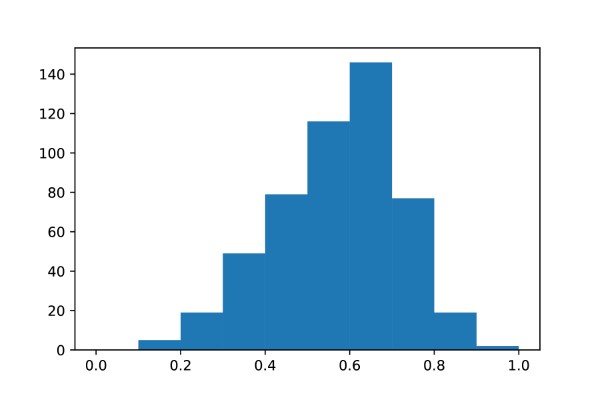}
        \hspace{4.5em} \includegraphics[width=0.16\textwidth]{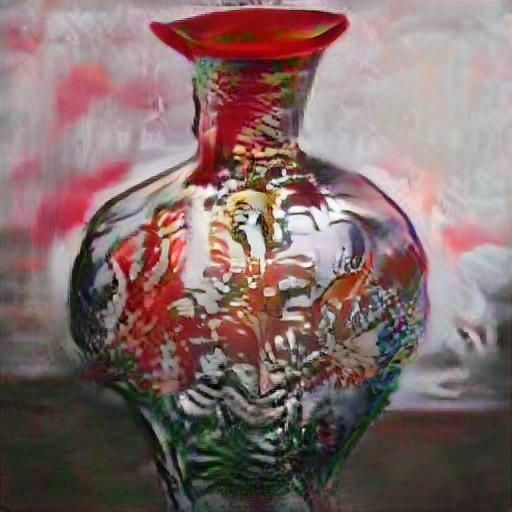}
        \hspace{-6.85em} \includegraphics[width=0.04\textwidth]{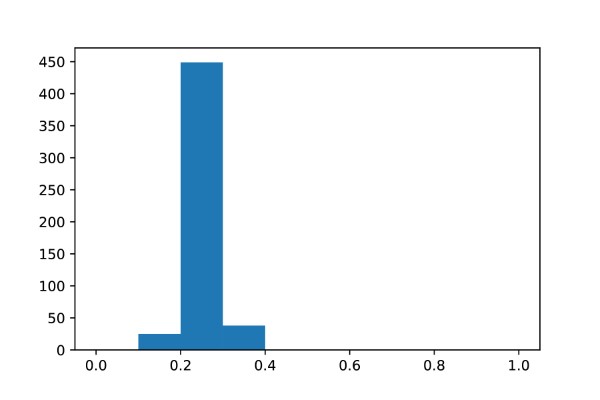}
        \hspace{4.5em}\hfill
        \includegraphics[width=0.16\textwidth]{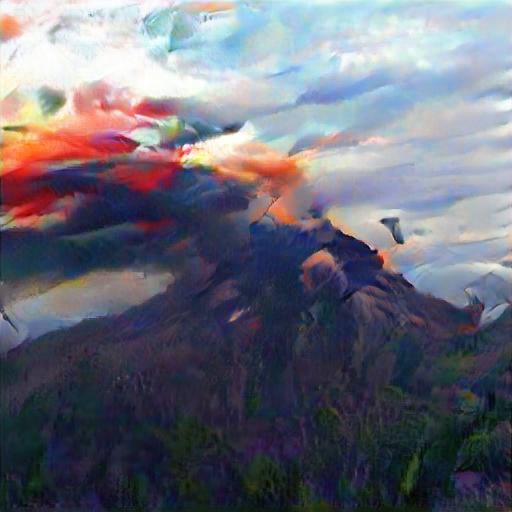}
        \hspace{-6.85em} \includegraphics[width=0.04\textwidth]{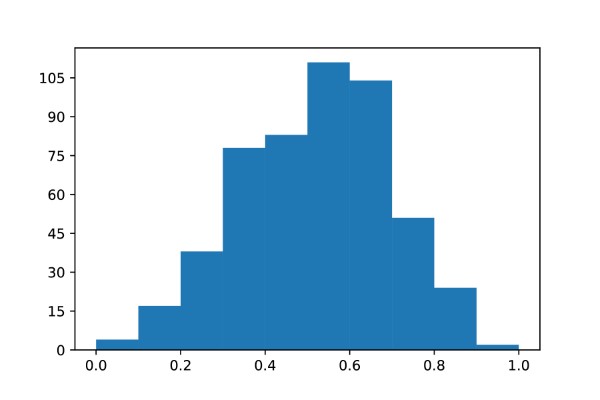}
        \hspace{4.5em} \includegraphics[width=0.16\textwidth]{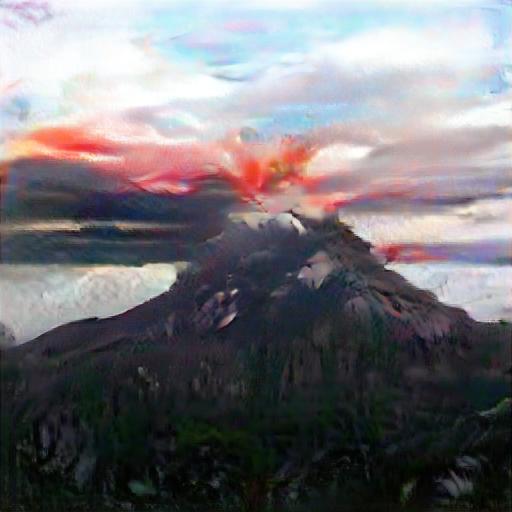}
        \hspace{-6.85em} \includegraphics[width=0.04\textwidth]{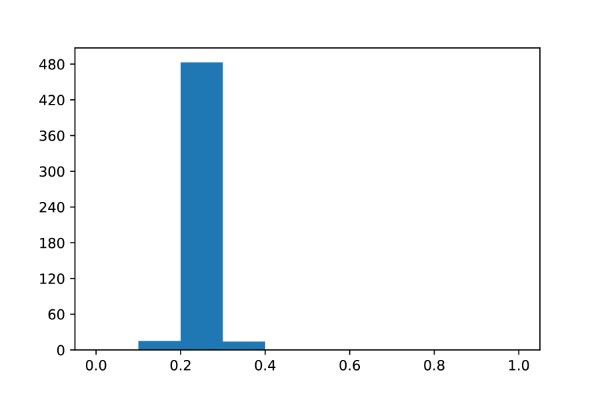}
        \hspace{4.6em}
    \end{subfigure}
    \caption{Comparison between the channel stroke (left of pair) and channel flush (right of pair). Histogram of channel coverage floats over each image. From top to bottom, the results in each row are (1) BigGAN~\cite{brock2019} from keywords "pineapple", "vase" and "volcano" respectively (2) $l=4,\tau=256$ (3) $l=4,\tau=512$ (4) $l=5,\tau=256$ (5) $l=5,\tau=512$ (6) $l=6,\tau=512$ (7) $l=7,\tau=128$}
    \label{fig:GDP_stroke}
\end{figure}

\subsection{Action Cost} \label{subsec:ac}

We first consider the channel cost, which reflects the cost in changing the brush color for human. Let $g_c$ be the constant cost accounting to changing channel. The channel cost tensor $J$ is
\begin{equation} \label{eq:cc}
J_{c',h',w'}(c) = \left\{
\begin{array}{ll}
g_c & \text{if } c'=c \text{, where } c \text{ is the current stroking channel}  \\
\\
0 & \text{otherwise}
\end{array}
\right.
\end{equation}

The next cost factor is the stroke movement. Naturally, the stroke continues into its nearby region. Therefore, the movement cost increases as the distance from the current stroke location increases. The movement cost $K$ is
\begin{equation} \label{eq:mc}
K_{c',h',w'}(h,w;\sigma) = 1 - e^{-\frac{(h'-h)^2+(w'-w)^2}{2\sigma^2}},
\end{equation}
where $\sigma$ is the standard deviation of the Gaussian kernel centered at the current stroking location $h,w$.

The overall cost is then the Hadamard product of the individual cost components, $G = J\odot K$. This cost is being updated in every channel stroke iteration at Alg.\ref{alg:cs} Step~\ref{alg:cs_cost}. Further stroke behavior modeling may incorporate location cost for top-to-bottom and left-to-right handwriting behavior. It is also possible to add the directional cost to make the stroke continue in the same direction and attain certain artistic feel. Here we focus on quantifying the burden in between continuing with the current brush or changing color. The cost $G$ provides the next stroking location and channel. 

In Fig.\ref{fig:GDP_stroke}, we compare the results from channel flush and channel stroke over different operation layers $l$ and different channel limits $\tau$. We also provide a histogram of channel coverage for each outcome image. For an arbitrary channel, the coverage means the fraction of locations having their mask $M$ turned on. The results from the channel flush have more concentrated coverage compared to those from the channel stroke. Because channel stroke extends the stroke into its neighborhood and continues onto the nearby stroke-able region, some channels tend to have higher coverage then others. That causes in the diverged channel coverage.

With neighborhood extension and stroke continuation, the channel stroke overcomes the occasional discontinuity issue in the channel flush, while keeping the artistic look from blocking out the low response channels at each $h,w$ location. When the channel limit $\tau$ becomes closer to the number of channels at the operation layer, the output image becomes closer to the original output without channel operations.

\begin{figure}[!t]
    \centering
    \includegraphics[width=0.8\textwidth]{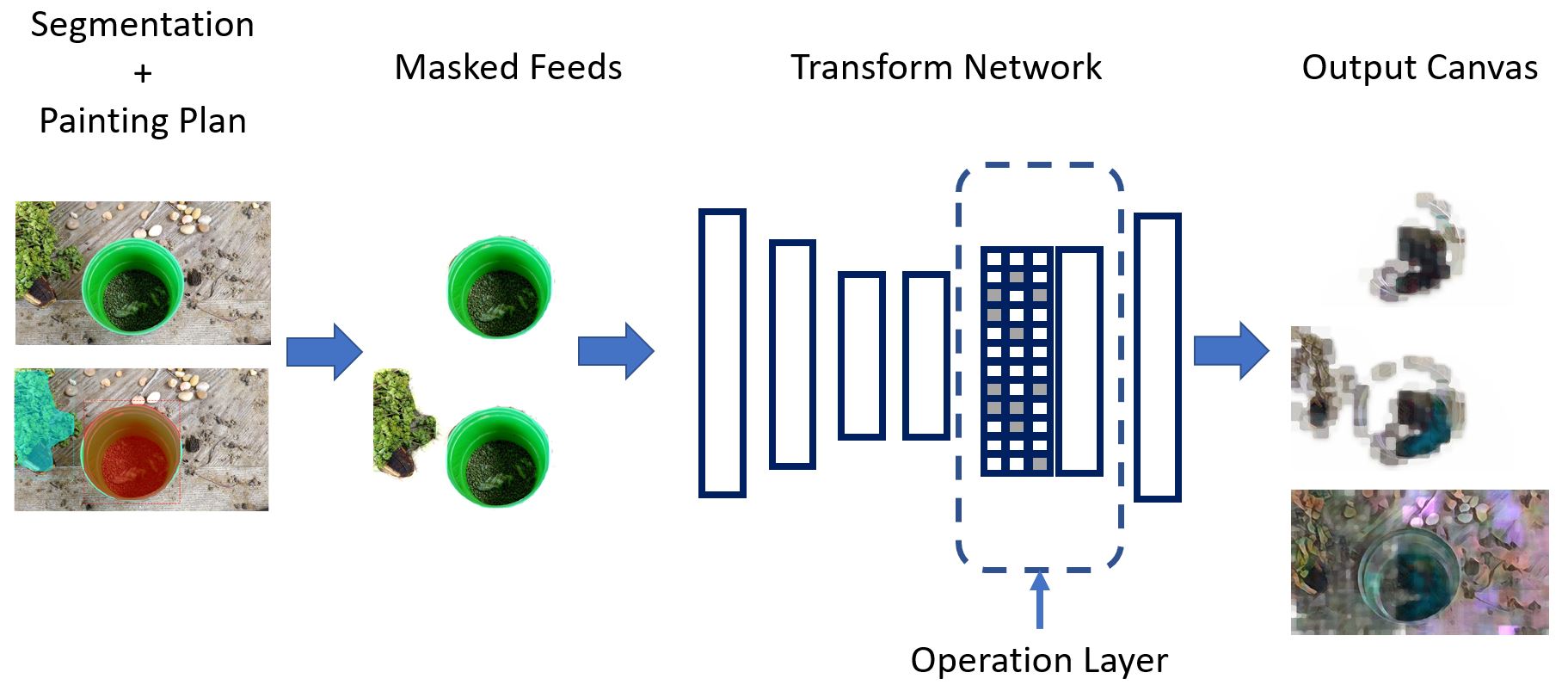}
    \caption{Block diagram for applying the CPIA over generator network}
    \label{fig:PIA_block}
\end{figure}

\begin{figure}[!t]
    \centering
    \begin{subfigure}[b]{\textwidth}
        \includegraphics[width=0.057\textwidth]{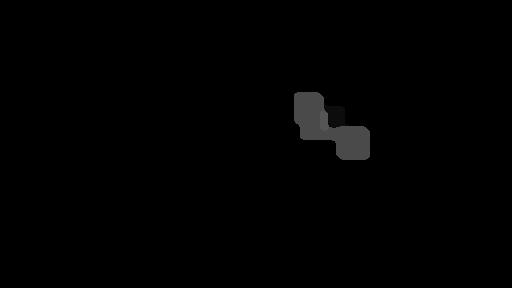}
        \includegraphics[width=0.057\textwidth]{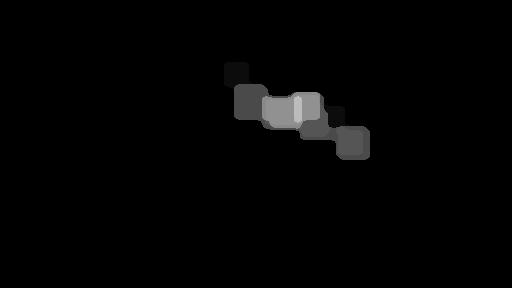}
        \includegraphics[width=0.057\textwidth]{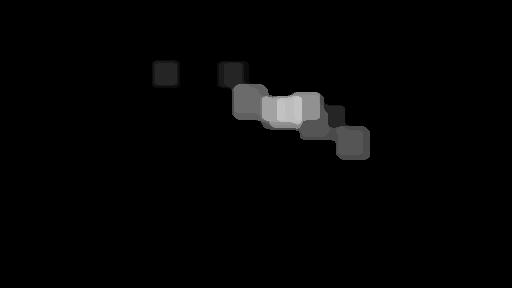}
        \includegraphics[width=0.057\textwidth]{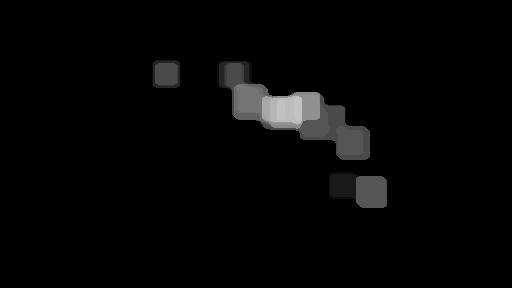}
        \includegraphics[width=0.057\textwidth]{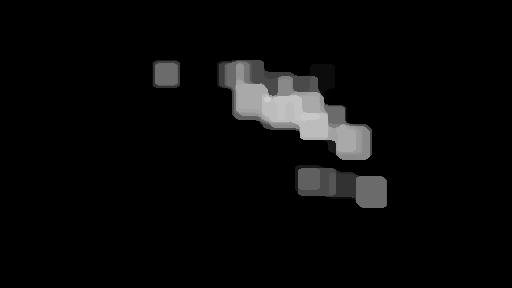}
        \includegraphics[width=0.057\textwidth]{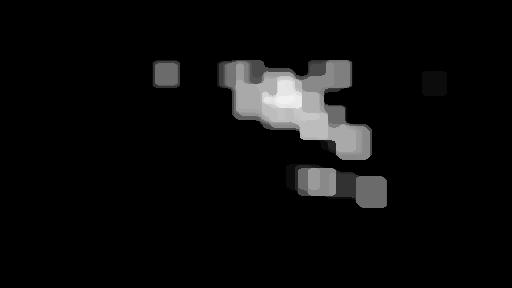}
        \includegraphics[width=0.057\textwidth]{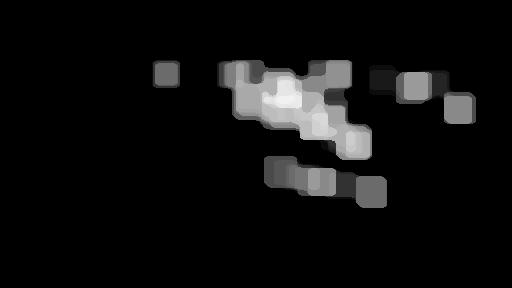}
        \includegraphics[width=0.057\textwidth]{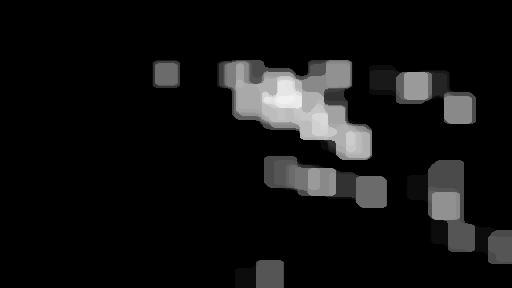}
        \includegraphics[width=0.057\textwidth]{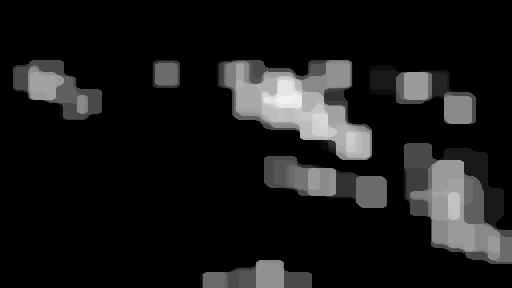}
        \includegraphics[width=0.057\textwidth]{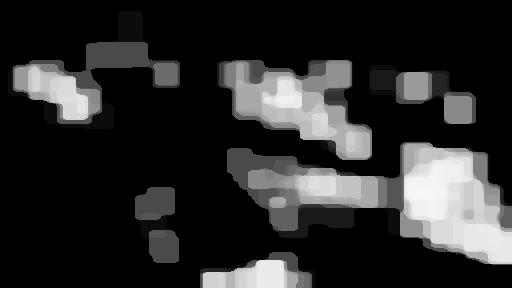}
        \includegraphics[width=0.057\textwidth]{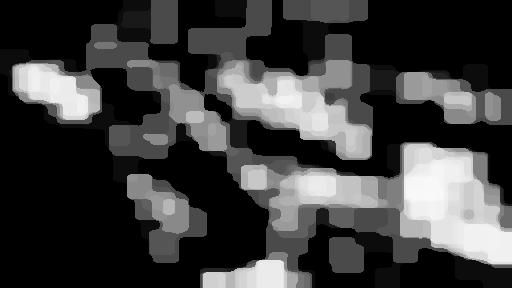}
        \includegraphics[width=0.057\textwidth]{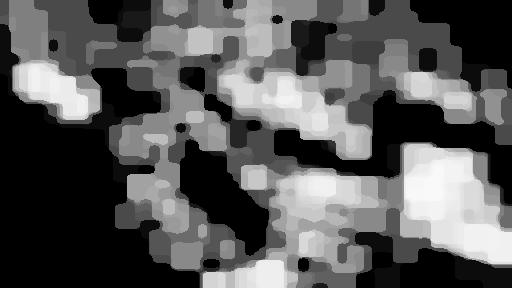}
        \includegraphics[width=0.057\textwidth]{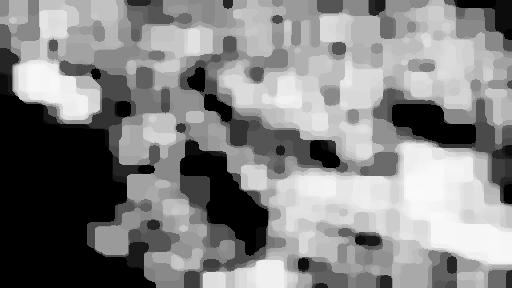}
        \includegraphics[width=0.057\textwidth]{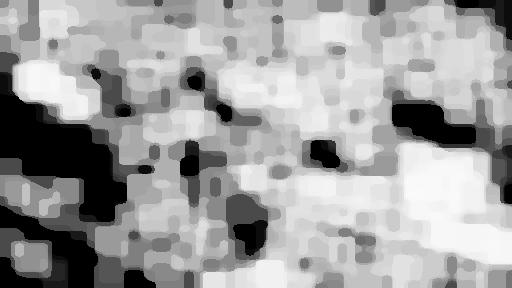}
        \includegraphics[width=0.057\textwidth]{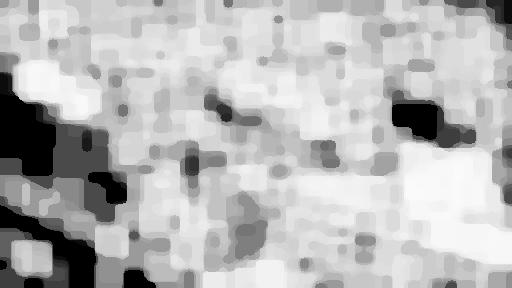}
        \includegraphics[width=0.057\textwidth]{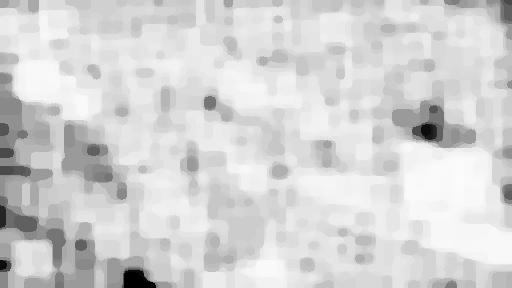}
    \end{subfigure}
    \begin{subfigure}[b]{\textwidth}
        \centering
        \includegraphics[width=0.24\textwidth]{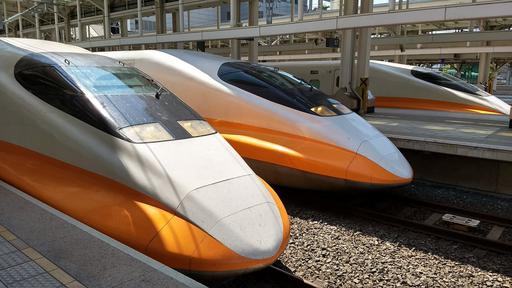}
        \includegraphics[width=0.24\textwidth]{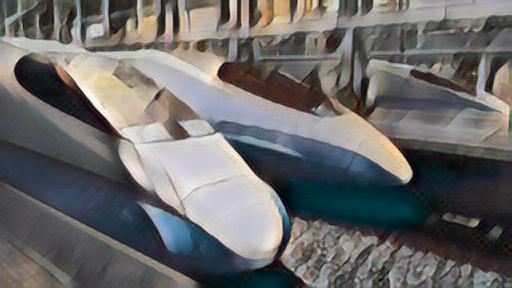}
        \includegraphics[width=0.24\textwidth]{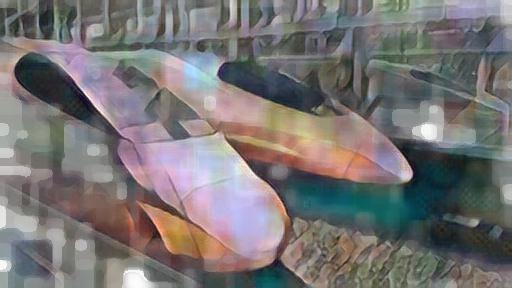}
        \includegraphics[width=0.24\textwidth]{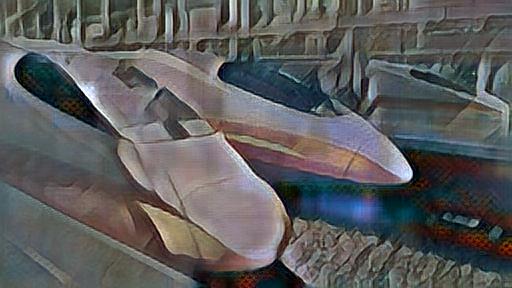}
    \end{subfigure}
    \begin{subfigure}[b]{\textwidth}
        \includegraphics[width=0.057\textwidth]{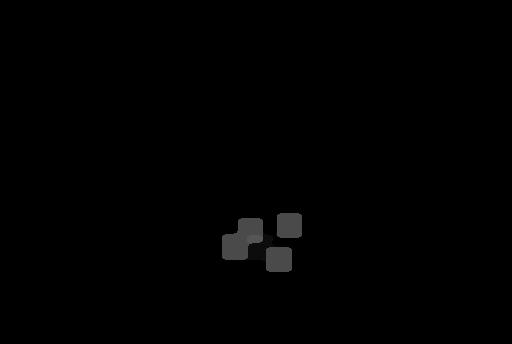}
        \includegraphics[width=0.057\textwidth]{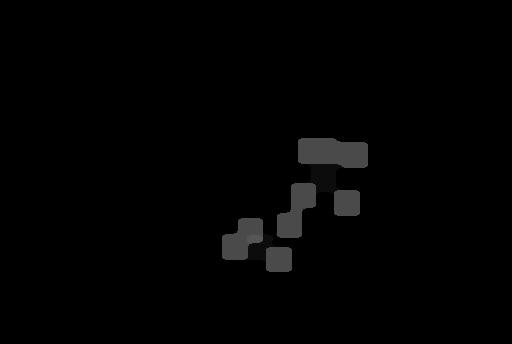}
        \includegraphics[width=0.057\textwidth]{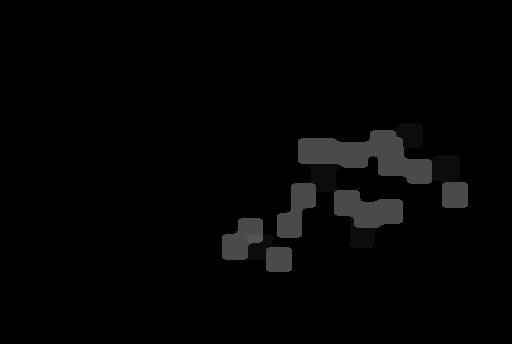}
        \includegraphics[width=0.057\textwidth]{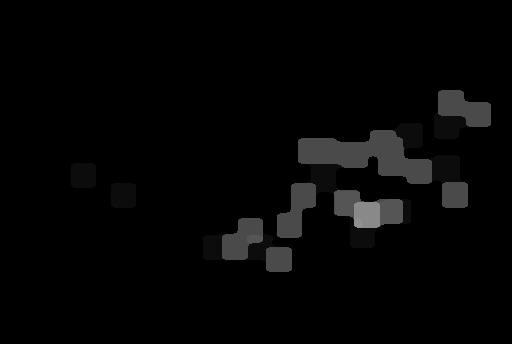}
        \includegraphics[width=0.057\textwidth]{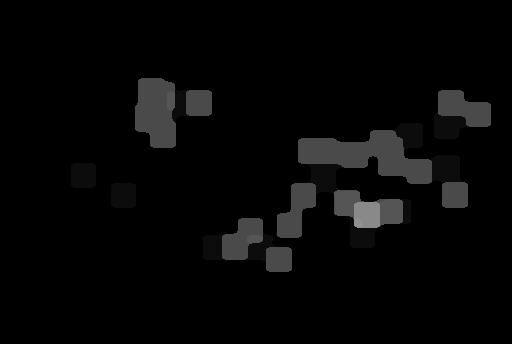}
        \includegraphics[width=0.057\textwidth]{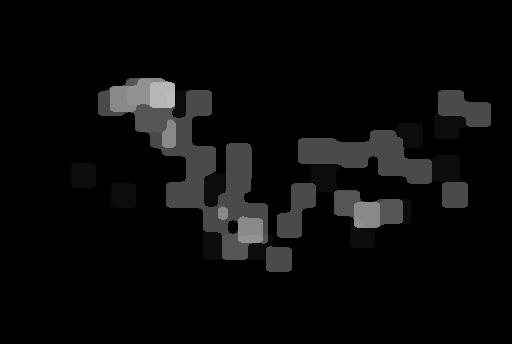}
        \includegraphics[width=0.057\textwidth]{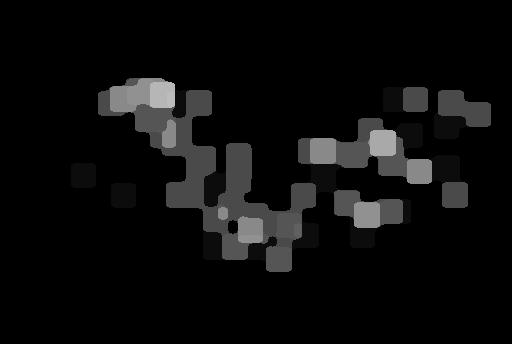}
        \includegraphics[width=0.057\textwidth]{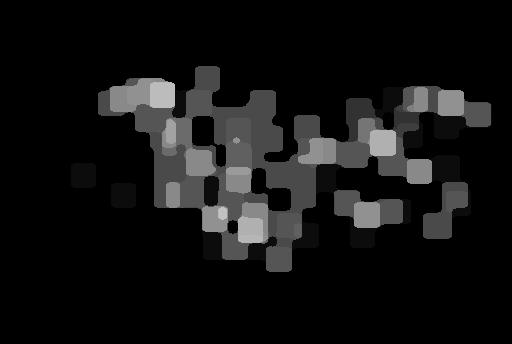}
        \includegraphics[width=0.057\textwidth]{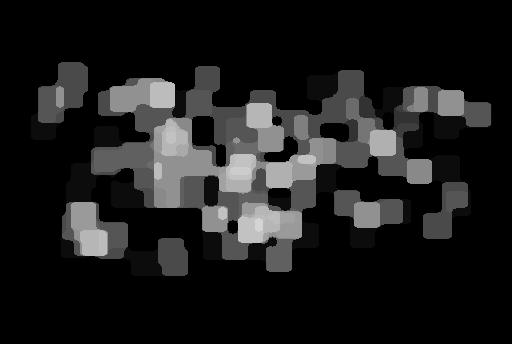}
        \includegraphics[width=0.057\textwidth]{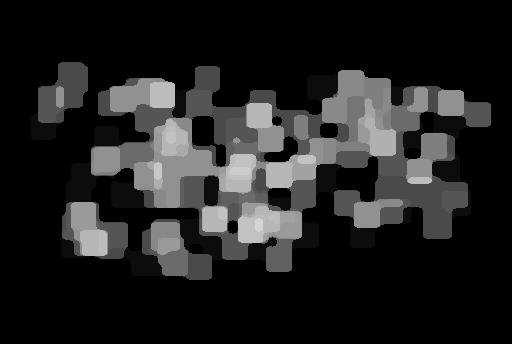}
        \includegraphics[width=0.057\textwidth]{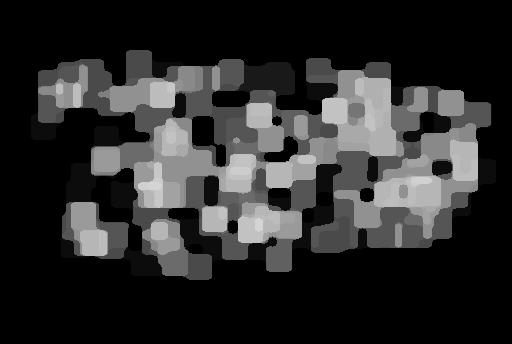}
        \includegraphics[width=0.057\textwidth]{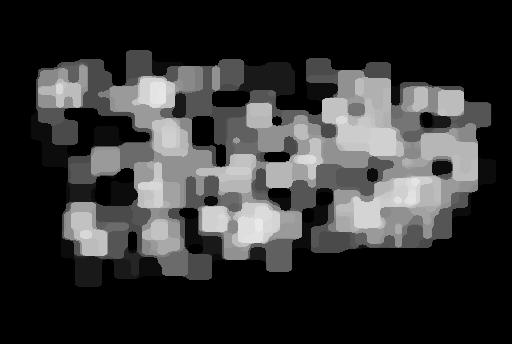}
        \includegraphics[width=0.057\textwidth]{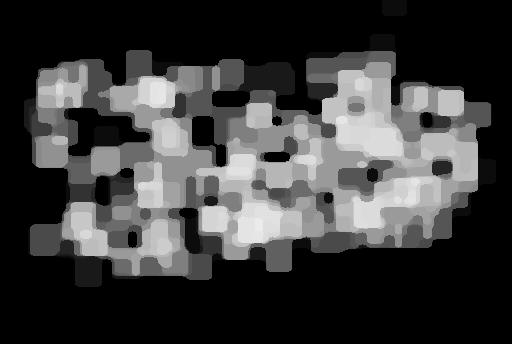}
        \includegraphics[width=0.057\textwidth]{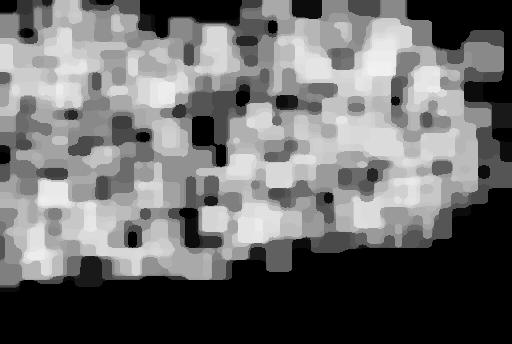}
        \includegraphics[width=0.057\textwidth]{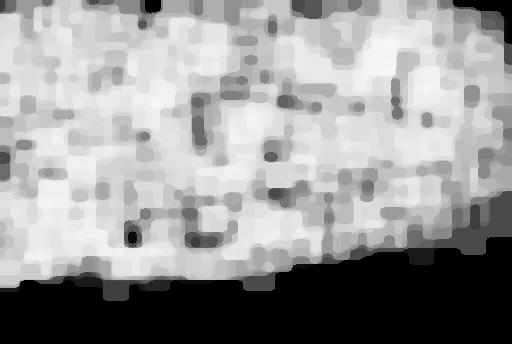}
        \includegraphics[width=0.057\textwidth]{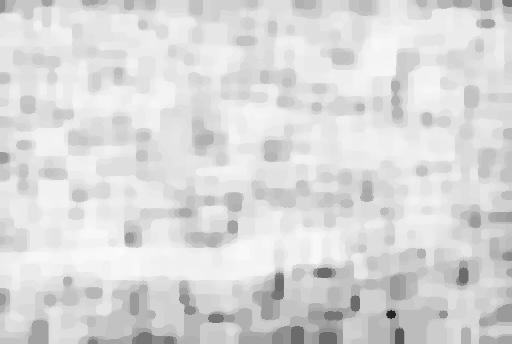}
    \end{subfigure}
    \begin{subfigure}[b]{\textwidth}
        \centering
        \includegraphics[width=0.24\textwidth]{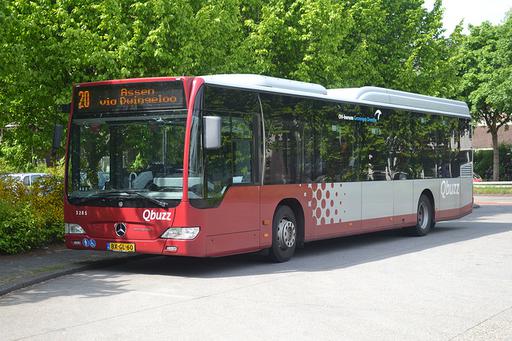}
        \includegraphics[width=0.24\textwidth]{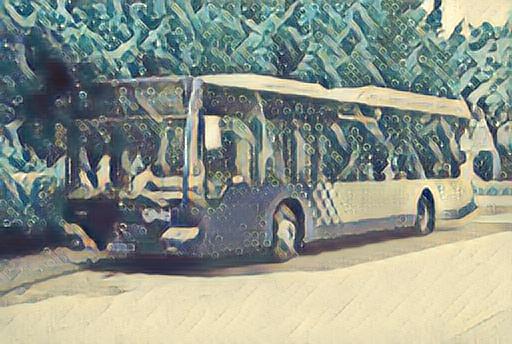}
        \includegraphics[width=0.24\textwidth]{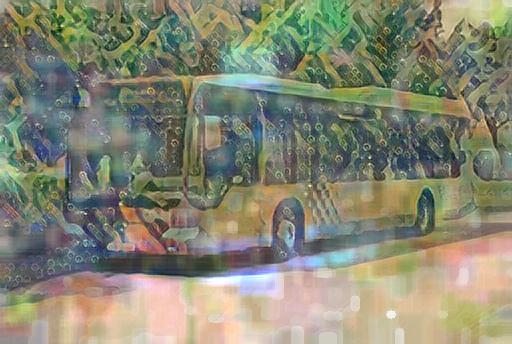}
        \includegraphics[width=0.24\textwidth]{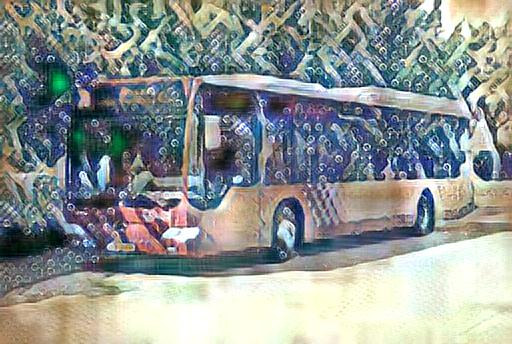}
    \end{subfigure}
    
    \begin{subfigure}[b]{\textwidth}
        \includegraphics[width=0.057\textwidth]{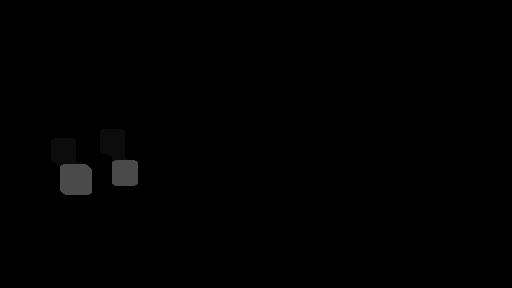}
        \includegraphics[width=0.057\textwidth]{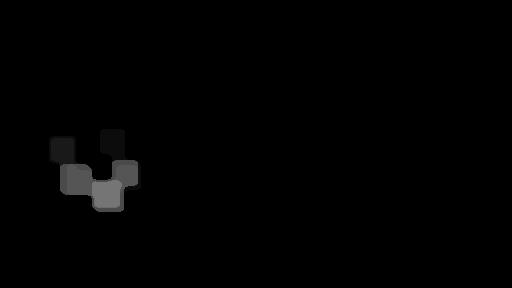}
        \includegraphics[width=0.057\textwidth]{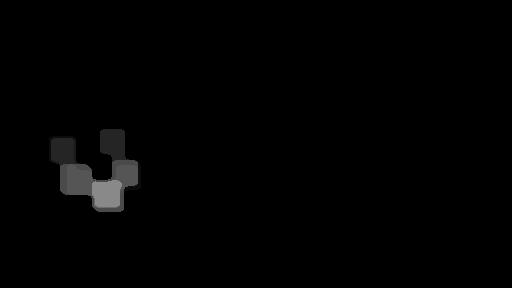}
        \includegraphics[width=0.057\textwidth]{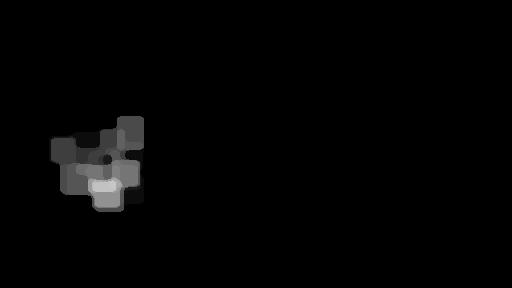}
        \includegraphics[width=0.057\textwidth]{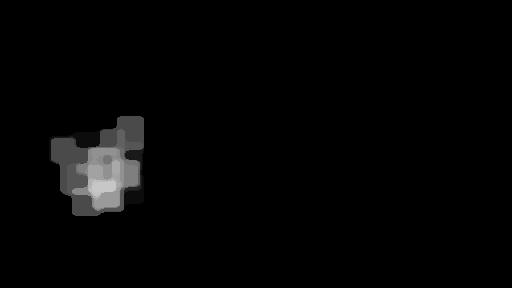}
        \includegraphics[width=0.057\textwidth]{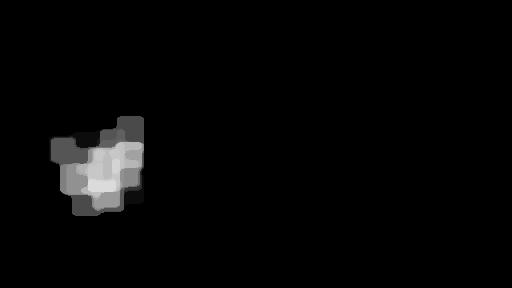}
        \includegraphics[width=0.057\textwidth]{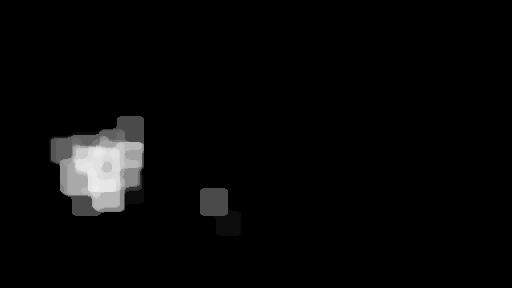}
        \includegraphics[width=0.057\textwidth]{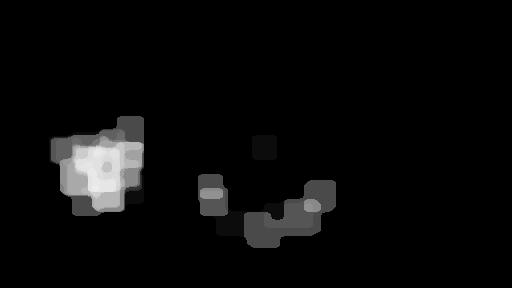}
        \includegraphics[width=0.057\textwidth]{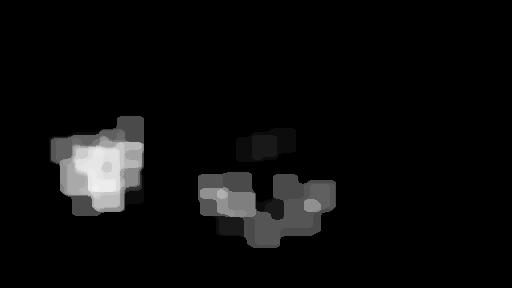}
        \includegraphics[width=0.057\textwidth]{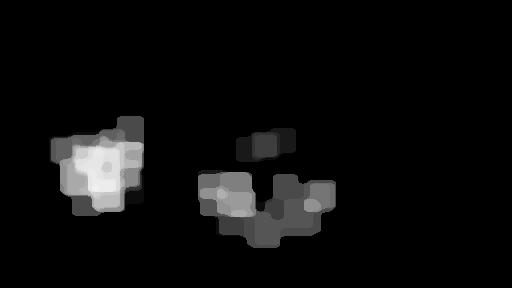}
        \includegraphics[width=0.057\textwidth]{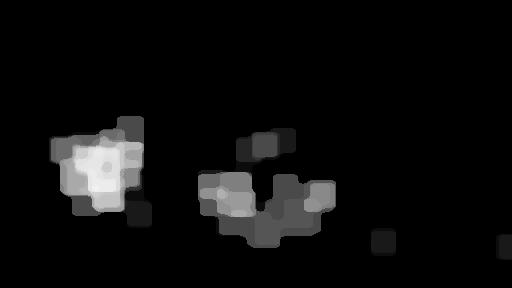}
        \includegraphics[width=0.057\textwidth]{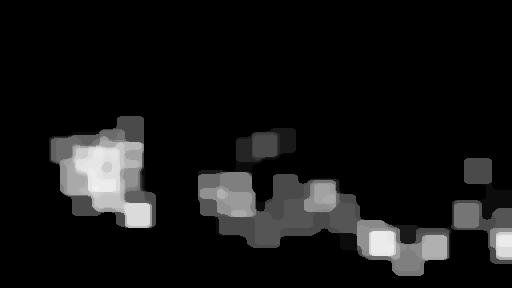}
        \includegraphics[width=0.057\textwidth]{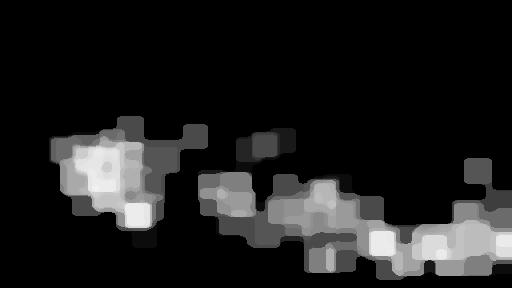}
        \includegraphics[width=0.057\textwidth]{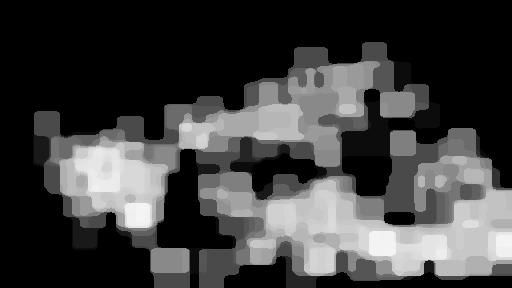}
        \includegraphics[width=0.057\textwidth]{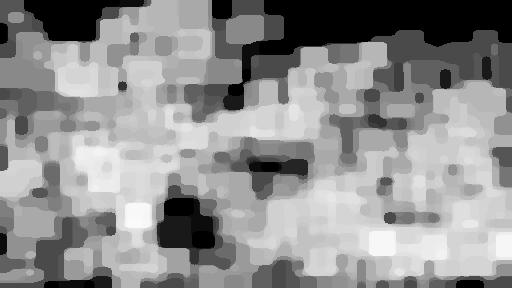}
        \includegraphics[width=0.057\textwidth]{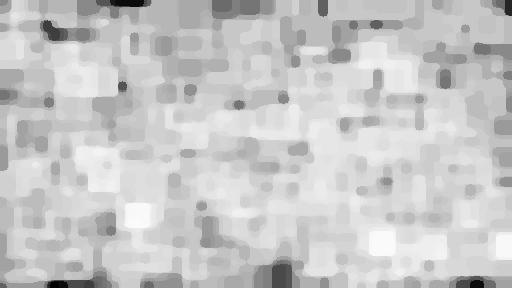}
    \end{subfigure}
    \begin{subfigure}[b]{\textwidth}
        \centering
        \includegraphics[width=0.24\textwidth]{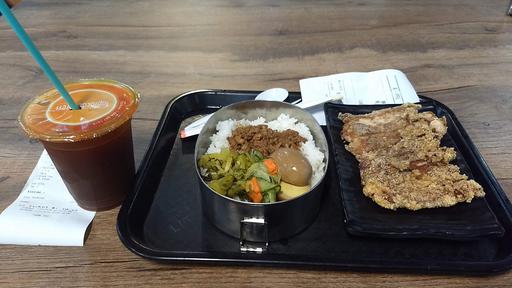}
        \includegraphics[width=0.24\textwidth]{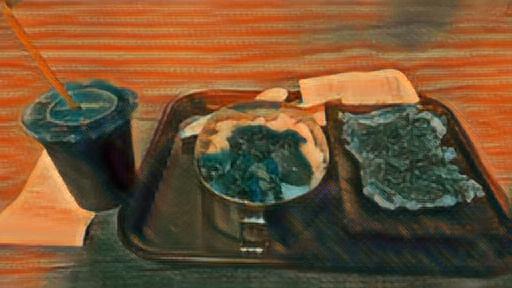}
        \includegraphics[width=0.24\textwidth]{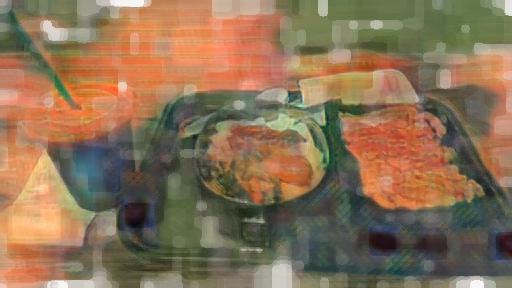}
        \includegraphics[width=0.24\textwidth]{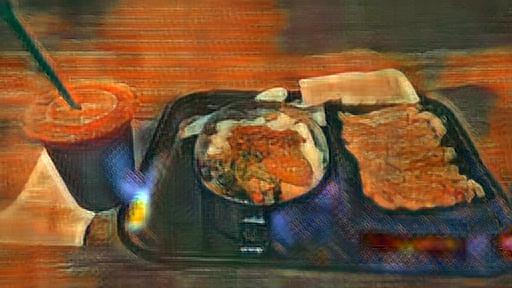}
    \end{subfigure}
    
    \begin{subfigure}[b]{\textwidth}
        \includegraphics[width=0.057\textwidth]{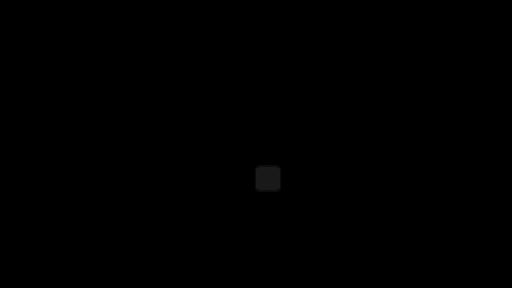}
        \includegraphics[width=0.057\textwidth]{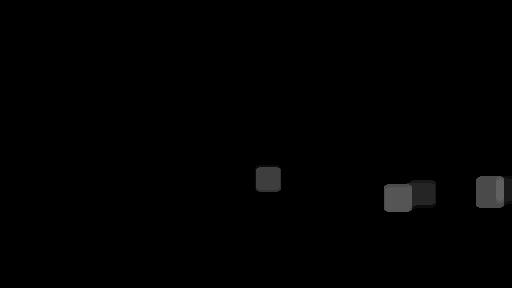}
        \includegraphics[width=0.057\textwidth]{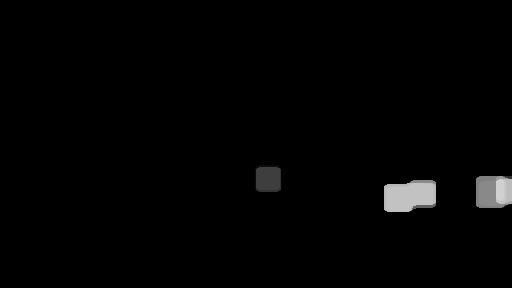}
        \includegraphics[width=0.057\textwidth]{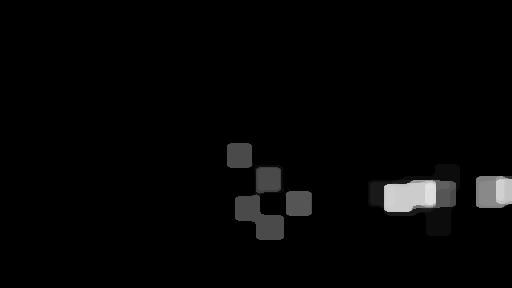}
        \includegraphics[width=0.057\textwidth]{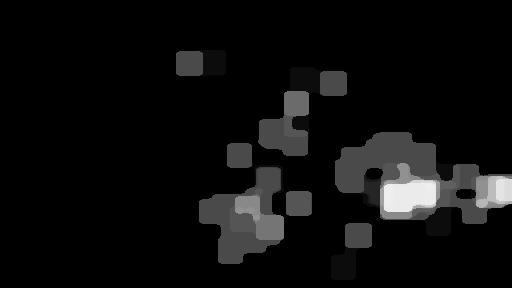}
        \includegraphics[width=0.057\textwidth]{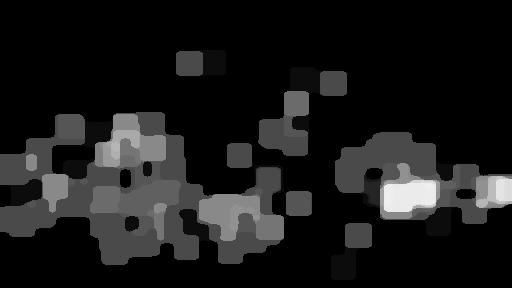}
        \includegraphics[width=0.057\textwidth]{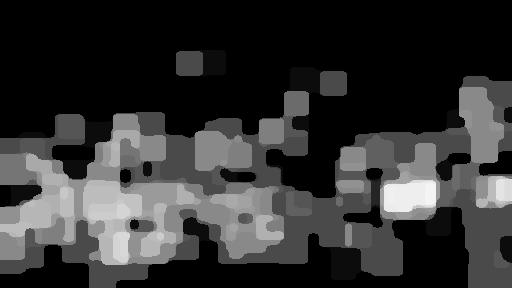}
        \includegraphics[width=0.057\textwidth]{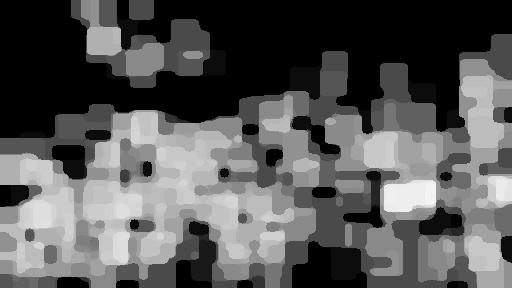}
        \includegraphics[width=0.057\textwidth]{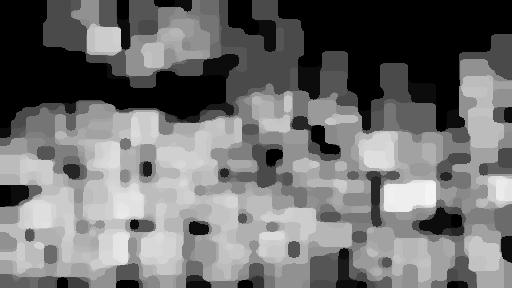}
        \includegraphics[width=0.057\textwidth]{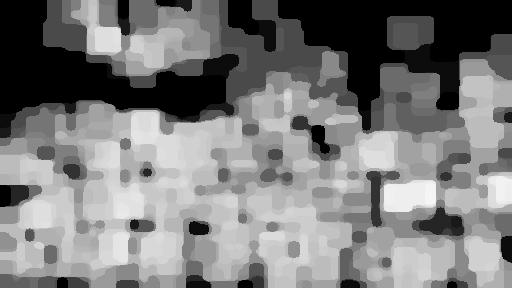}
        \includegraphics[width=0.057\textwidth]{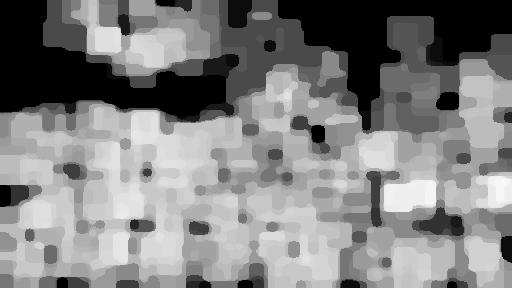}
        \includegraphics[width=0.057\textwidth]{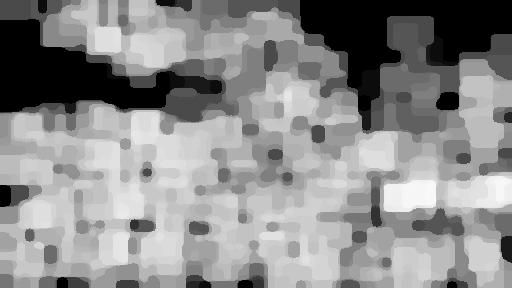}
        \includegraphics[width=0.057\textwidth]{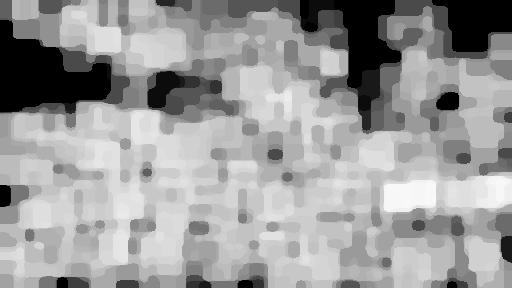}
        \includegraphics[width=0.057\textwidth]{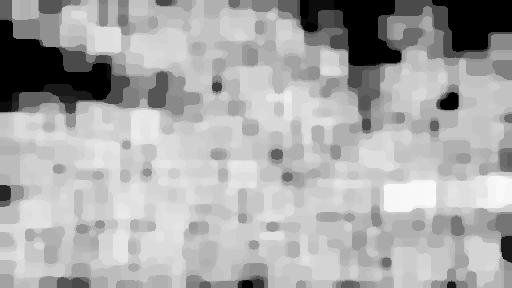}
        \includegraphics[width=0.057\textwidth]{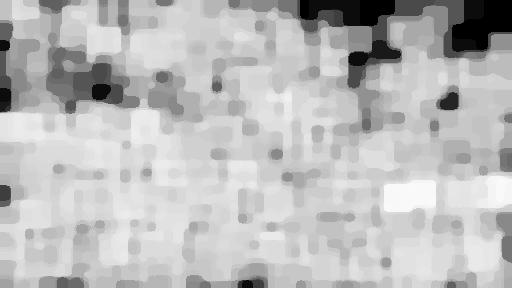}
        \includegraphics[width=0.057\textwidth]{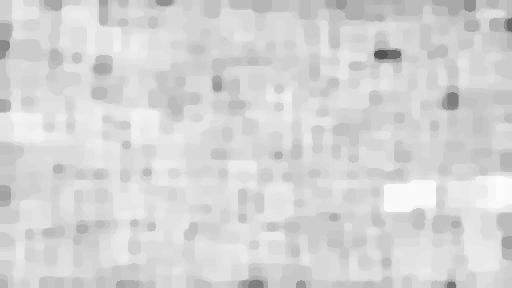}
    \end{subfigure}
    
    \begin{subfigure}[b]{0.24\textwidth}
        \includegraphics[width=\textwidth]{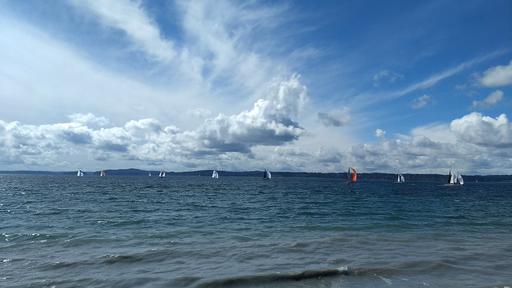}
        \caption{raw}
    \end{subfigure}
    \begin{subfigure}[b]{0.24\textwidth}
        \includegraphics[width=\textwidth]{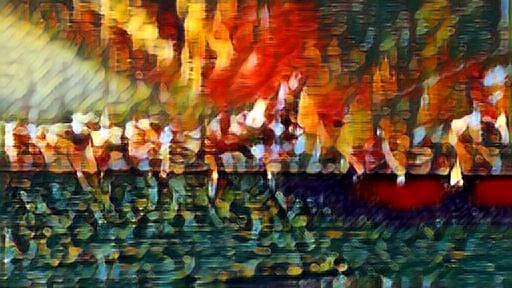}
        \caption{styled}
    \end{subfigure}
    \begin{subfigure}[b]{0.24\textwidth}
        \includegraphics[width=\textwidth]{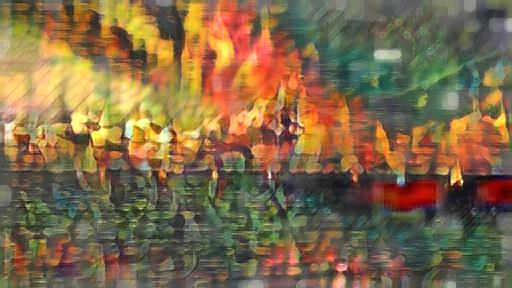}
        \caption{painted @ $l=9$}
    \end{subfigure}
    \begin{subfigure}[b]{0.24\textwidth}
        \includegraphics[width=\textwidth]{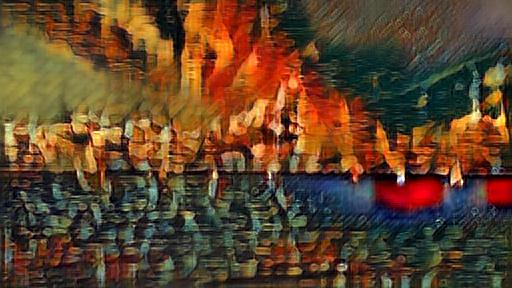}
        \caption{painted @ $l=7$}
    \end{subfigure}
    \caption{CPIA result over the style transfer network in \cite{zhang2018multi}. Images are painted with $\tau=20$ in (b) and with $\tau=128$ in (c). Each row of the stroke maps illustrates the painting actions of the (c) image in the following row.}
    \label{fig:PIA_results}
\end{figure}

\subsection{Channel Painting in Action}

Based on the channel stroke, we propose the channel painting-in-action (CPIA) framework. This framework first analyzes the input image into painting regions, and come up with a painting plan. The painting plan contains a list of step images, which are composed of certain masked regions of the input image. The step images are then sequentially fed into the pre-trained generator network. The operation layer carries out the stroke actions and paint on the output canvas. The framework is presented in Fig.~\ref{fig:PIA_block}. In the implementation of this work, we use the Mask R-CNN to mark out the objects as regions of interest (ROIs). The step images are prepared according to these ROIs.

The step images mask out the regions to ensure channel continuity on those regions. When the first step image is fed, the channel stroke algorithm is applied to paint on the masked regions of the current step image, until the stopping criterion $\mathcal{S}$ is met. The output canvas contains only strokes on the regions that have been exposed to channel stroke so far. Then the next step image is fed. The painting process continues until the list of step images is exhausted.

Because of the convolution kernels, the stroked response will propagate into its nearby regions in the next convolutional layer of the CNNs. The cascaded propagation tends to tint the whole canvas at the last layer. Therefore, the stroke maps are used as the post-generator masks to wipe out the response of the non-stroking regions at the last layer.

Note that although there can be as many as $\tau$ channels being stroked for each $h,w$ location, the painting process does not guarantee that every location has exactly $\tau$ channels stroked. In fact, at the operation layer, it is very likely that the locations with more high response channels have reached the maximum $\tau$ channels being stroked. While the less response locations have less than $\tau$ channels being stroked.

Fig.~\ref{fig:PIA_results} presents the CPIA result over the style transfer network in \cite{zhang2018multi}. The original style transfer offers the sharper result in general. On the other hand, the CPIA introduces additional artistic styling components on top of the existing style transfer. Such components are defined by the parameters $\mathcal{S}$, $\mathcal{N}$, $m$, $l$ and $\tau$. At the same time, the step images in the painting plan can also be seen as a controlling factor that decides the painting priority of each masked region.

\section{Discussion and Implementation Details} \label{sec:disc}

Working with the stroke penetration parameter, the visually plausible stroke outcome requires the proper stroke mask after the last layer. With low penetration in channels, the response can be very dimmed and need to be masked in low opacity on the background. The mask opacity increases as the location collects more stroked channels.

In current implementations, the painting area ordering is based on the object and its detection score of each ROIs from the Mask R-CNN. We can optionally add side information about the semantics to the planning - such as paint the persons at the end, or paint the large objects first.

To stop the painting at each step image feed of the CPIA, we leverage the stopping criterion $\mathcal{S}$ in Alg.\ref{alg:cs}. $\mathcal{S}$ can be a global threshold cutting of the response ratio at each ROI, or can be the stopping condition tailored toward different object types or sizes. The former is adopted in this work.

\section{Conclusion} \label{sec:con}

The proposed channel stroke strategy utilizes the knowledge learned and stored within the deep convolutional generator network. On top of that, the CPIA leverages the learned object and segmentation knowledge in frameworks like Mask R-CNN to plan the painting regions. The CPIA works with the existing generator networks and the existing image segmentation tools, without additional training data on stroking order.

Looking ahead, we seek to drive the stroking factors in a more adaptive way. The stroke size (bundled in the neighborhood $\mathcal{N}$) can vary based on the stroking location. Depending on the stroking channel $c$, the stroke penetration can possibly change in a responsive fashion. In this work, one single layer of the CNN is chosen as the operation layer. A further expansion is to investigate the multi-layer coordination of the channel stroke. The channel decomposition for the generator networks still has much to explore and the application may be beyond the artistic rendering and the step-wise painting.


\newpage

\small

\bibliographystyle{unsrt}
\bibliography{ref}

\begin{thebibliography}{10}

\bibitem{gatys2016image}
Leon~A Gatys, Alexander~S Ecker, and Matthias Bethge.
\newblock Image style transfer using convolutional neural networks.
\newblock In {\em Proceedings of the IEEE conference on computer vision and
  pattern recognition (CVPR)}, pages 2414--2423, 2016.

\bibitem{johnson2016perceptual}
Justin Johnson, Alexandre Alahi, and Li~Fei-Fei.
\newblock Perceptual losses for real-time style transfer and super-resolution.
\newblock In {\em Proceedings of the European conference on computer vision
  (ECCV)}, pages 694--711, 2016.

\bibitem{ulyanov2016texture}
Dmitry Ulyanov, Vadim Lebedev, Andrea Vedaldi, and Victor~S Lempitsky.
\newblock Texture networks: Feed-forward synthesis of textures and stylized
  images.
\newblock In {\em International Conference on Machine Learning (ICML)},
  volume~1, page~4, 2016.

\bibitem{li2017universal}
Yijun Li, Chen Fang, Jimei Yang, Zhaowen Wang, Xin Lu, and Ming-Hsuan Yang.
\newblock Universal style transfer via feature transforms.
\newblock In {\em Advances in neural information processing systems (NIPS)},
  pages 386--396, 2017.

\bibitem{jing2018stroke}
Yongcheng Jing, Yang Liu, Yezhou Yang, Zunlei Feng, Yizhou Yu, Dacheng Tao, and
  Mingli Song.
\newblock Stroke controllable fast style transfer with adaptive receptive
  fields.
\newblock In {\em Proceedings of the European Conference on Computer Vision
  (ECCV)}, pages 238--254, 2018.

\bibitem{ganin2019synthesizing}
Yaroslav Ganin, Tejas Kulkarni, Igor Babuschkin, SM~Eslami, and Oriol Vinyals.
\newblock Synthesizing programs for images using reinforced adversarial
  learning.
\newblock {\em International Conference on Learning Representations (ICLR)},
  2019.

\bibitem{zheng2019strokenet}
Ningyuan Zheng, Yifan Jiang, and Dingjiang Huang.
\newblock Strokenet: A neural painting environment.
\newblock In {\em International Conference on Learning Representations (ICLR)},
  2019.

\bibitem{huang2019learning}
Zhewei Huang, Wen Heng, and Shuchang Zhou.
\newblock Learning to paint with model-based deep reinforcement learning.
\newblock {\em arXiv preprint arXiv:1903.04411}, 2019.

\bibitem{he2017mask}
Kaiming He, Georgia Gkioxari, Piotr Doll{\'a}r, and Ross Girshick.
\newblock Mask r-cnn.
\newblock In {\em Proceedings of the IEEE international conference on computer
  vision (ICCV)}, pages 2961--2969, 2017.

\bibitem{zhang2018multi}
Hang Zhang and Kristin Dana.
\newblock Multi-style generative network for real-time transfer.
\newblock In {\em Proceedings of the European Conference on Computer Vision
  (ECCV)}, 2018.

\bibitem{rumelhart1985learning}
David~E Rumelhart, Geoffrey~E Hinton, and Ronald~J Williams.
\newblock Learning internal representations by error propagation.
\newblock Technical report, California Univ San Diego La Jolla Inst for
  Cognitive Science, 1985.

\bibitem{ballard1987modular}
Dana~H Ballard.
\newblock Modular learning in neural networks.
\newblock In {\em Proceedings of the sixth National conference on Artificial
  intelligence-Volume 1}, pages 279--284. AAAI Press, 1987.

\bibitem{kingma2013auto}
Diederik~P Kingma and Max Welling.
\newblock Auto-encoding variational bayes.
\newblock In {\em International Conference on Learning Representations (ICLR)},
  2014.

\bibitem{goodfellow2014generative}
Ian Goodfellow, Jean Pouget-Abadie, Mehdi Mirza, Bing Xu, David Warde-Farley,
  Sherjil Ozair, Aaron Courville, and Yoshua Bengio.
\newblock Generative adversarial nets.
\newblock In {\em Advances in neural information processing systems (NIPS)},
  pages 2672--2680, 2014.

\bibitem{radford2016unsupervised}
Alec Radford, Luke Metz, and Soumith Chintala.
\newblock Unsupervised representation learning with deep convolutional
  generative adversarial networks.
\newblock In {\em International Conference on Learning Representations (ICLR)},
  2016.

\bibitem{isola2017image}
Phillip Isola, Jun-Yan Zhu, Tinghui Zhou, and Alexei~A Efros.
\newblock Image-to-image translation with conditional adversarial networks.
\newblock In {\em Proceedings of the IEEE conference on computer vision and
  pattern recognition (CVPR)}, pages 1125--1134, 2017.

\bibitem{zhu2017unpaired}
Jun-Yan Zhu, Taesung Park, Phillip Isola, and Alexei~A Efros.
\newblock Unpaired image-to-image translation using cycle-consistent
  adversarial networks.
\newblock In {\em Proceedings of the IEEE international conference on computer
  vision (ICCV)}, pages 2223--2232, 2017.

\bibitem{cho2019image}
Wonwoong Cho, Sungha Choi, David~Keetae Park, Inkyu Shin, and Jaegul Choo.
\newblock Image-to-image translation via group-wise deep whitening-and-coloring
  transformation.
\newblock In {\em Proceedings of the IEEE Conference on Computer Vision and
  Pattern Recognition (CVPR)}, pages 10639--10647, 2019.

\bibitem{reed2016learning}
Scott~E Reed, Zeynep Akata, Santosh Mohan, Samuel Tenka, Bernt Schiele, and
  Honglak Lee.
\newblock Learning what and where to draw.
\newblock In {\em Advances in Neural Information Processing Systems (NIPS)},
  pages 217--225, 2016.

\bibitem{ma2017pose}
Liqian Ma, Xu~Jia, Qianru Sun, Bernt Schiele, Tinne Tuytelaars, and Luc
  Van~Gool.
\newblock Pose guided person image generation.
\newblock In {\em Advances in Neural Information Processing Systems (NIPS)},
  pages 406--416, 2017.

\bibitem{karras2019style}
Tero Karras, Samuli Laine, and Timo Aila.
\newblock A style-based generator architecture for generative adversarial
  networks.
\newblock In {\em Proceedings of the IEEE Conference on Computer Vision and
  Pattern Recognition (CVPR)}, pages 4401--4410, 2019.

\bibitem{zhang2017stackgan}
Han Zhang, Tao Xu, Hongsheng Li, Shaoting Zhang, Xiaogang Wang, Xiaolei Huang,
  and Dimitris~N Metaxas.
\newblock Stackgan: Text to photo-realistic image synthesis with stacked
  generative adversarial networks.
\newblock In {\em Proceedings of the IEEE International Conference on Computer
  Vision (CVPR)}, pages 5907--5915, 2017.

\bibitem{brock2019}
Andrew Brock, Jeff Donahue, and Karen Simonyan.
\newblock Large scale {GAN} training for high fidelity natural image synthesis.
\newblock In {\em International Conference on Learning Representations (ICLR)},
  2019.

\bibitem{jin2017towards}
Yanghua Jin, Jiakai Zhang, Minjun Li, Yingtao Tian, Huachun Zhu, and Zhihao
  Fang.
\newblock Towards the automatic anime characters creation with generative
  adversarial networks.
\newblock In {\em Advances in neural information processing systems (NIPS)},
  2017.

\bibitem{chen2018cartoongan}
Yang Chen, Yu-Kun Lai, and Yong-Jin Liu.
\newblock Cartoongan: Generative adversarial networks for photo cartoonization.
\newblock In {\em Proceedings of the IEEE Conference on Computer Vision and
  Pattern Recognition (CVPR)}, pages 9465--9474, 2018.

\bibitem{wang2018high}
Ting-Chun Wang, Ming-Yu Liu, Jun-Yan Zhu, Andrew Tao, Jan Kautz, and Bryan
  Catanzaro.
\newblock High-resolution image synthesis and semantic manipulation with
  conditional gans.
\newblock In {\em Proceedings of the IEEE conference on computer vision and
  pattern recognition (CVPR)}, pages 8798--8807, 2018.

\bibitem{simonyan2014very}
Karen Simonyan and Andrew Zisserman.
\newblock Very deep convolutional networks for large-scale image recognition.
\newblock In {\em International Conference on Learning Representations (ICLR)},
  2015.

\bibitem{dumoulin2017learned}
Vincent Dumoulin, Jonathon Shlens, and Manjunath Kudlur.
\newblock A learned representation for artistic style.
\newblock {\em International Conference on Learning Representations (ICLR)},
  2017.

\bibitem{ha2018recurrent}
David Ha and J{\"u}rgen Schmidhuber.
\newblock Recurrent world models facilitate policy evolution.
\newblock In {\em Advances in Neural Information Processing Systems (NIPS)},
  pages 2450--2462, 2018.

\bibitem{hafner2019learning}
Danijar Hafner, Timothy Lillicrap, Ian Fischer, Ruben Villegas, David Ha,
  Honglak Lee, and James Davidson.
\newblock Learning latent dynamics for planning from pixels.
\newblock In {\em International Conference on Machine Learning (ICML)}, pages
  2555--2565, 2019.

\bibitem{nakano2019neural}
Reiichiro Nakano.
\newblock Neural painters: A learned differentiable constraint for generating
  brushstroke paintings.
\newblock {\em arXiv preprint arXiv:1904.08410}, 2019.

\bibitem{lecun2004learning}
Yann LeCun, Fu~Jie Huang, Leon Bottou, et~al.
\newblock Learning methods for generic object recognition with invariance to
  pose and lighting.
\newblock In {\em Proceedings of the IEEE conference on computer vision and
  pattern recognition (CVPR)}, pages 97--104, 2004.

\bibitem{krizhevsky2012imagenet}
Alex Krizhevsky, Ilya Sutskever, and Geoffrey~E Hinton.
\newblock Imagenet classification with deep convolutional neural networks.
\newblock In {\em Advances in neural information processing systems (NIPS)},
  pages 1097--1105, 2012.

\end{thebibliography}

\appendix

\section{Appendix}

In this section, we discuss additional parameters that can futher fine-tune the channel stroke and the CPIA result.

\subsection{Stroke Sensitivity} \label{sec:disc-match}

\begin{figure}[!th]
    \centering
    \begin{subfigure}[]{\textwidth}
        \centering
        \includegraphics[width=0.137\textwidth]{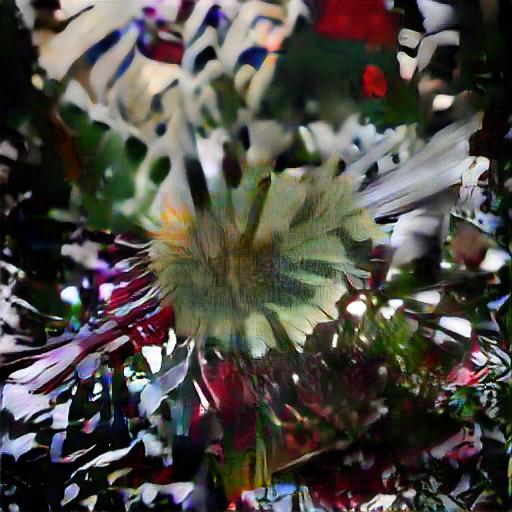}
        \hspace{-6em} \includegraphics[width=0.05\textwidth]{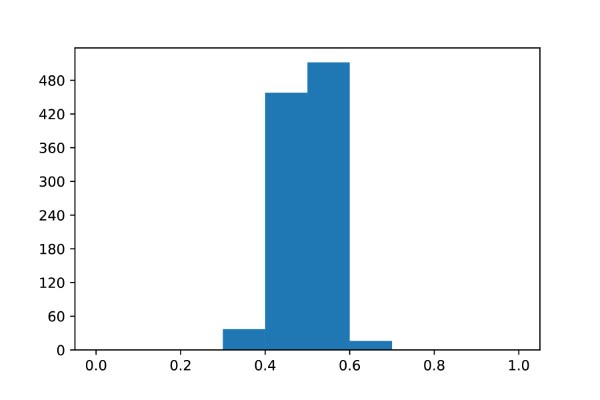}
        \hspace{3.2em} \includegraphics[width=0.137\textwidth]{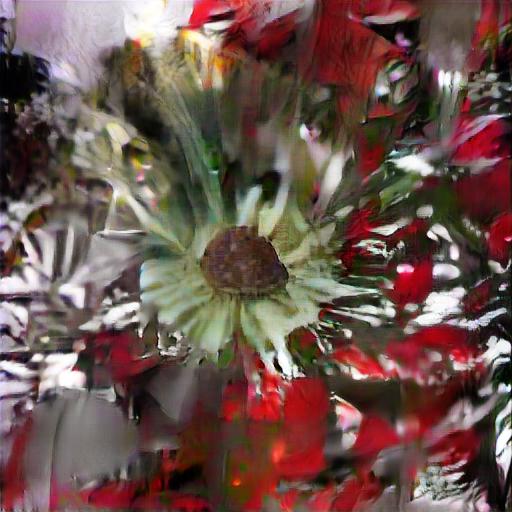}
        \hspace{-6em} \includegraphics[width=0.05\textwidth]{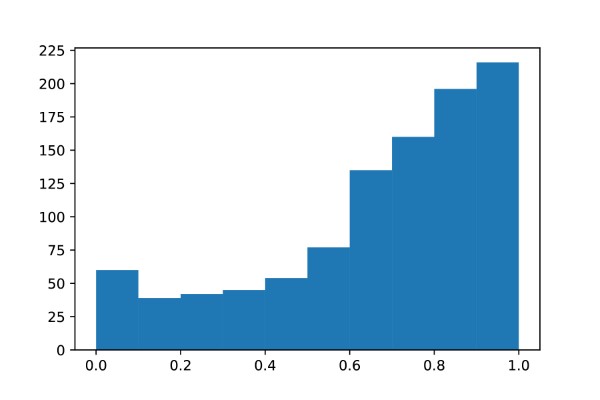}
        \hspace{3.2em} \includegraphics[width=0.137\textwidth]{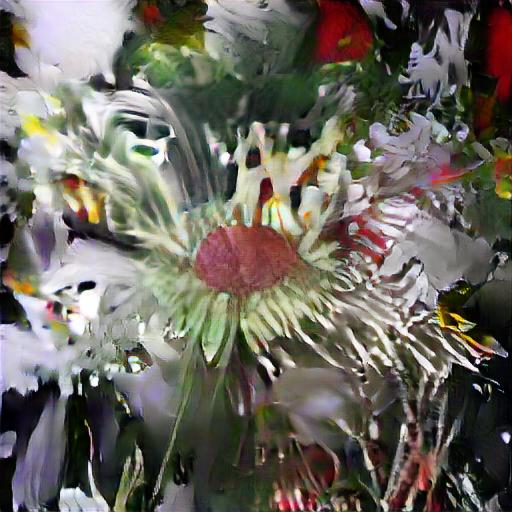}
        \hspace{-6em} \includegraphics[width=0.05\textwidth]{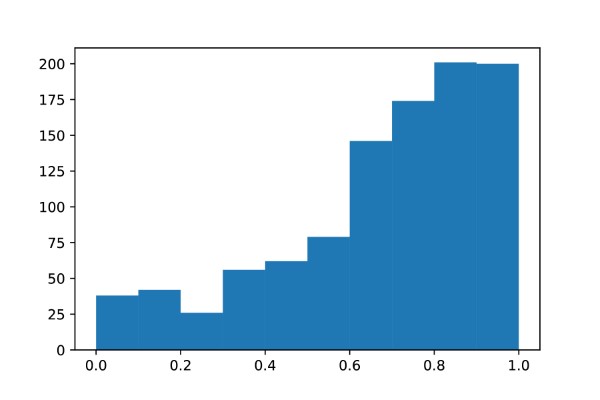}
        \hspace{3.2em} \includegraphics[width=0.137\textwidth]{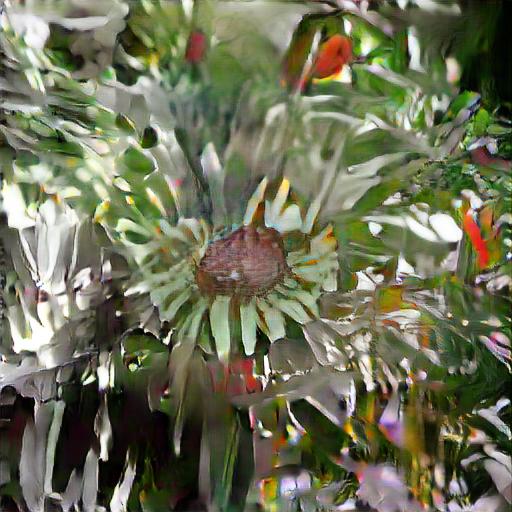}
        \hspace{-6em} \includegraphics[width=0.05\textwidth]{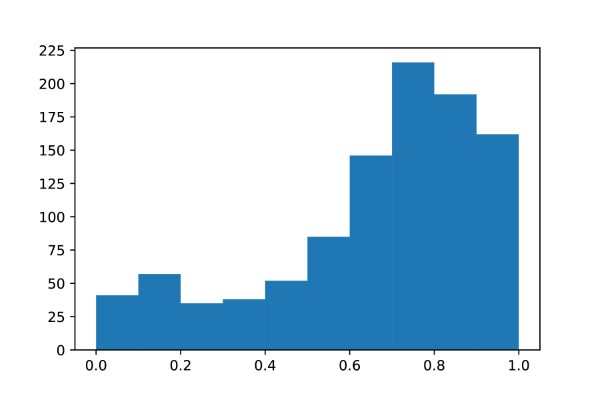}
        \hspace{3.2em} \includegraphics[width=0.137\textwidth]{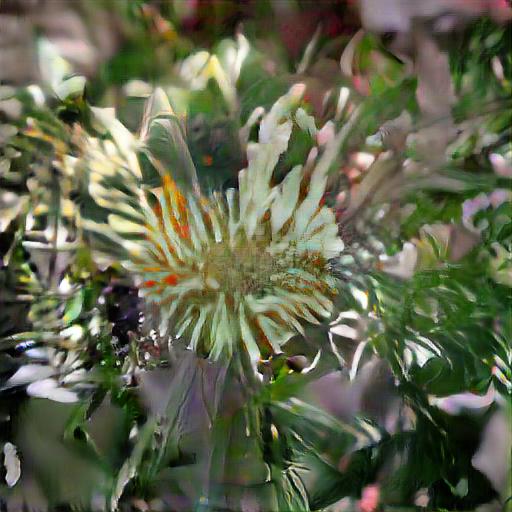}
        \hspace{-6em} \includegraphics[width=0.05\textwidth]{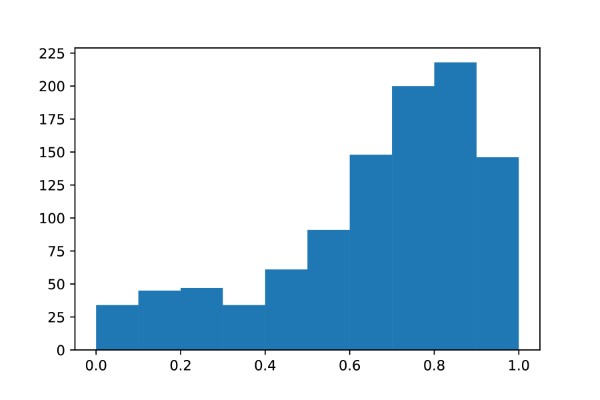}
        \hspace{3.2em} \includegraphics[width=0.137\textwidth]{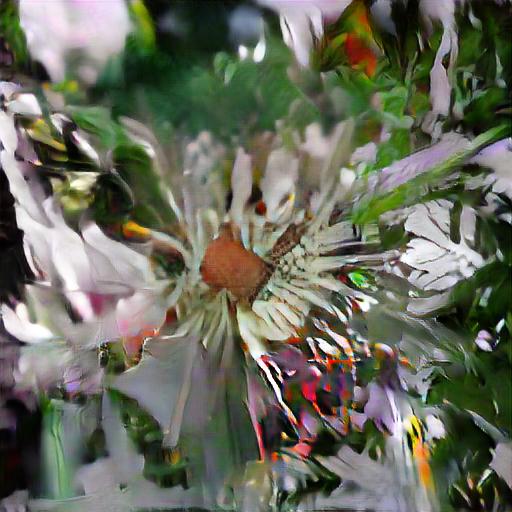}
        \hspace{-6em} \includegraphics[width=0.05\textwidth]{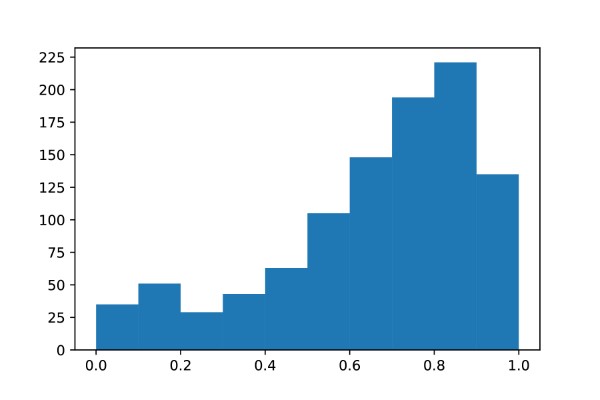}
        \hspace{3.2em} \includegraphics[width=0.137\textwidth]{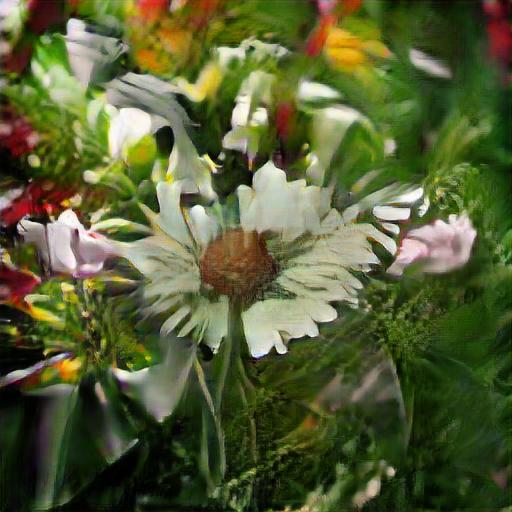}
    \end{subfigure}
    \begin{subfigure}[]{\textwidth}
        \centering
        \includegraphics[width=0.137\textwidth]{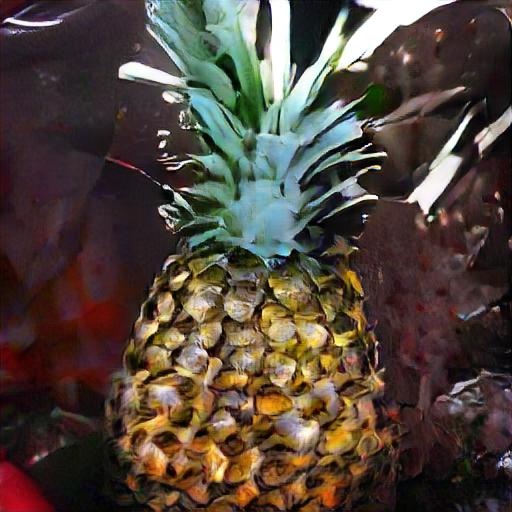}
        \hspace{-6em} \includegraphics[width=0.05\textwidth]{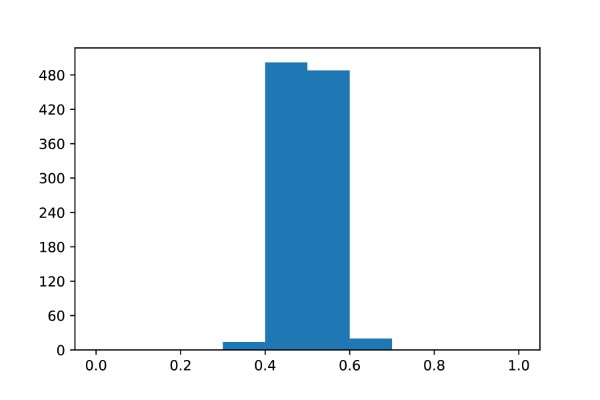}
        \hspace{3.2em} \includegraphics[width=0.137\textwidth]{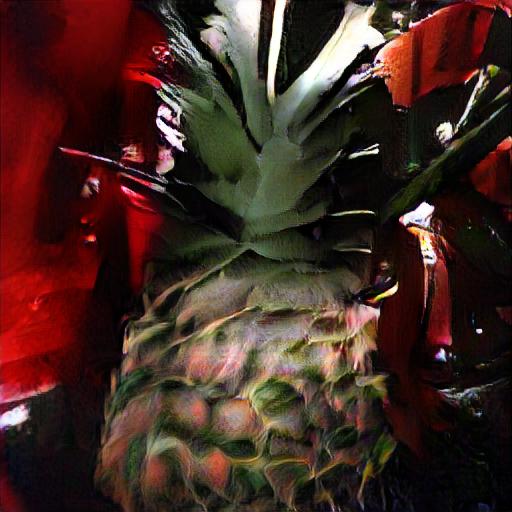}
        \hspace{-6em} \includegraphics[width=0.05\textwidth]{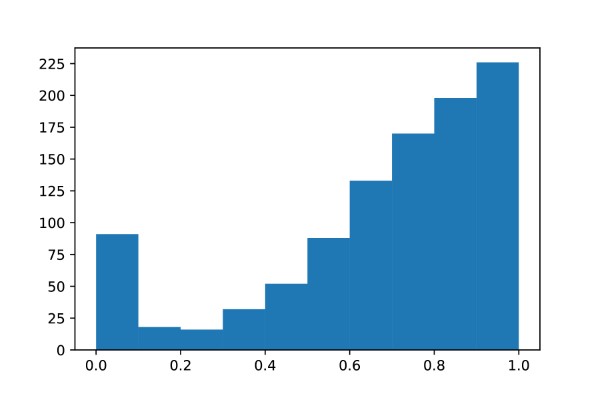}
        \hspace{3.2em} \includegraphics[width=0.137\textwidth]{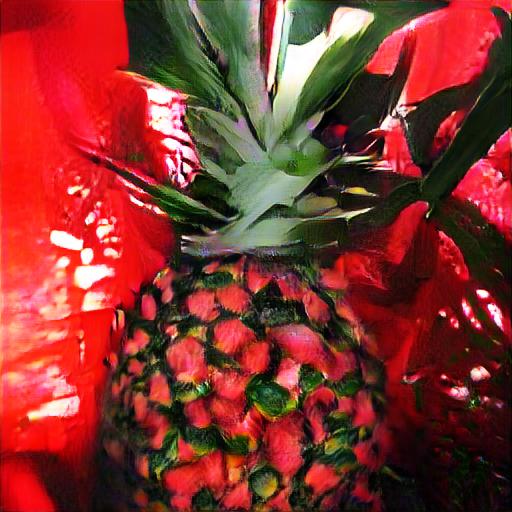}
        \hspace{-6em} \includegraphics[width=0.05\textwidth]{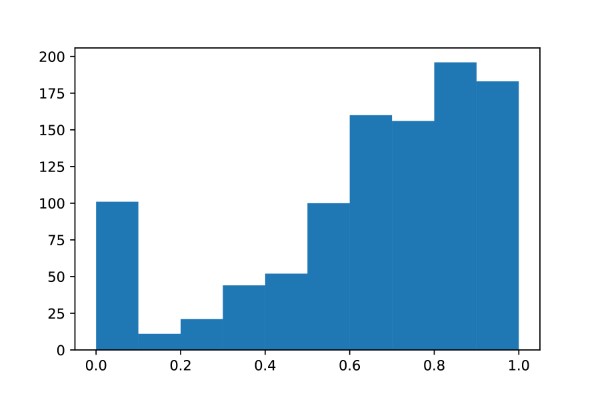}
        \hspace{3.2em} \includegraphics[width=0.137\textwidth]{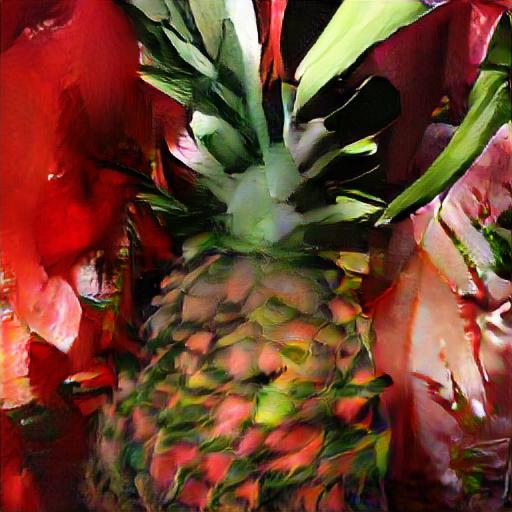}
        \hspace{-6em} \includegraphics[width=0.05\textwidth]{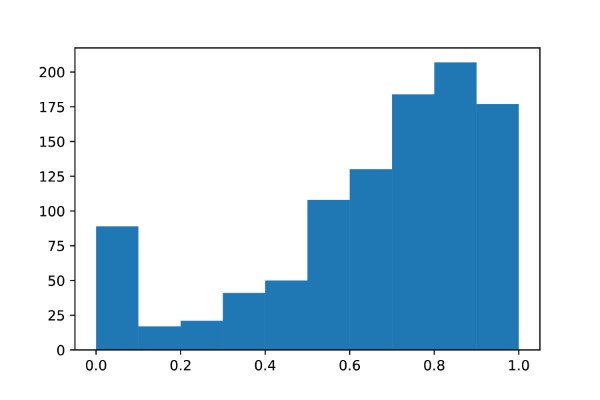}
        \hspace{3.2em} \includegraphics[width=0.137\textwidth]{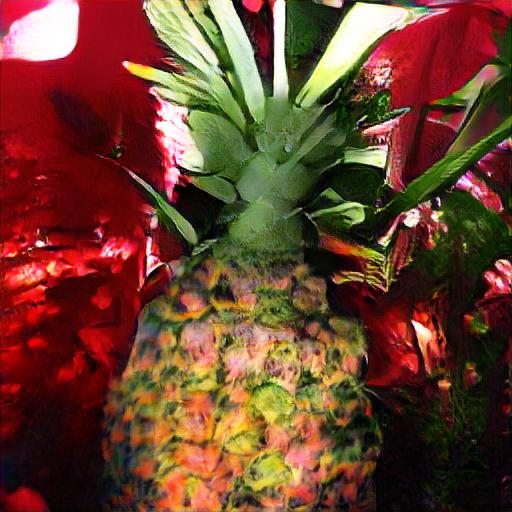}
        \hspace{-6em} \includegraphics[width=0.05\textwidth]{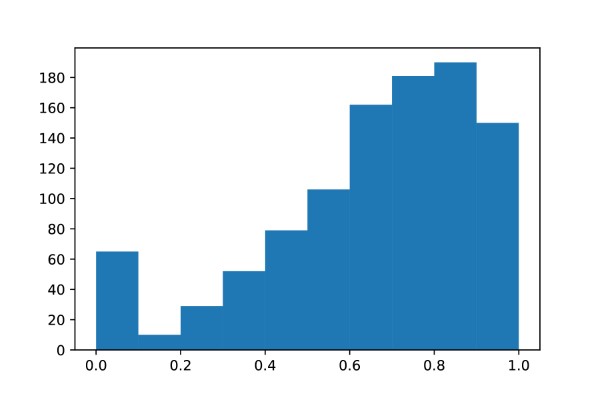}
        \hspace{3.2em} \includegraphics[width=0.137\textwidth]{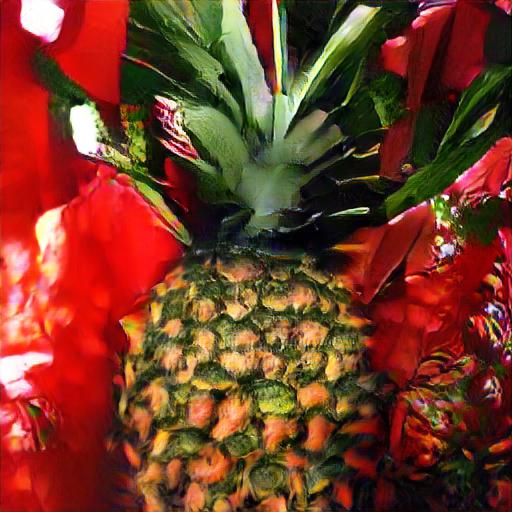}
        \hspace{-6em} \includegraphics[width=0.05\textwidth]{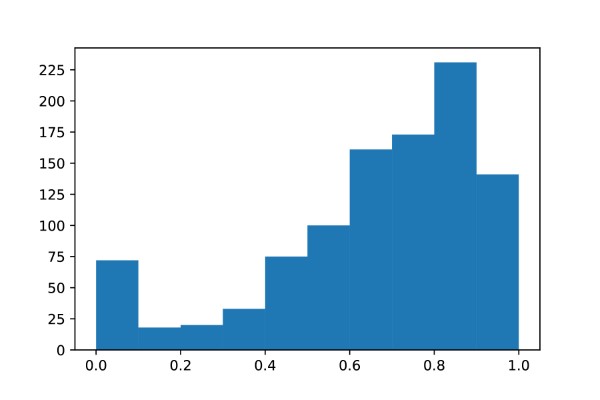}
        \hspace{3.2em} \includegraphics[width=0.137\textwidth]{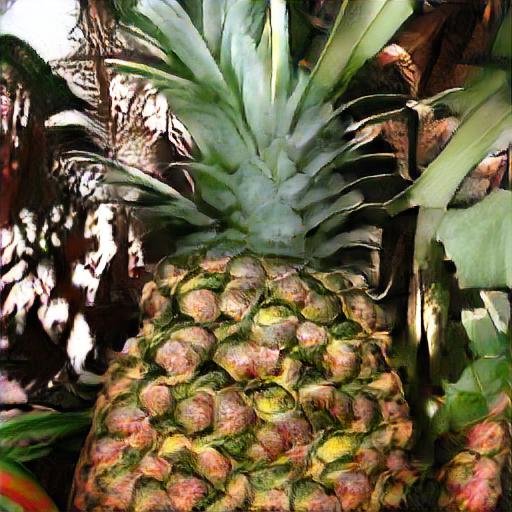}
    \end{subfigure}
    \begin{subfigure}[]{\textwidth}
        \centering
        \includegraphics[width=0.137\textwidth]{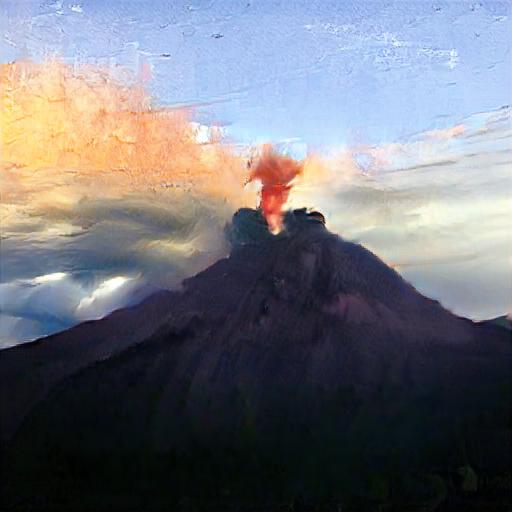}
        \hspace{-6em} \includegraphics[width=0.05\textwidth]{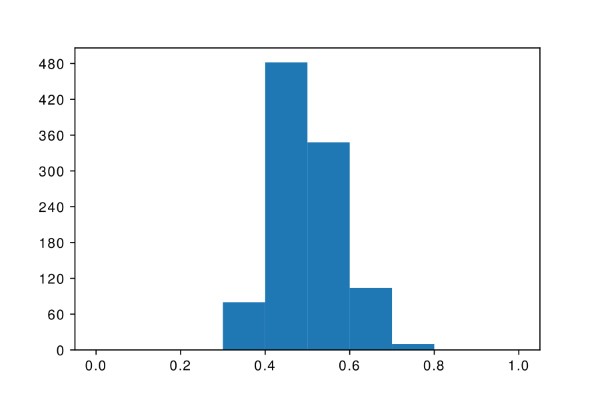}
        \hspace{3.2em} \includegraphics[width=0.137\textwidth]{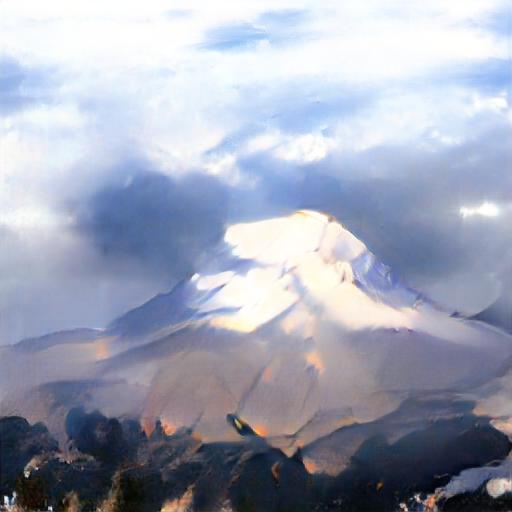}
        \hspace{-6em} \includegraphics[width=0.05\textwidth]{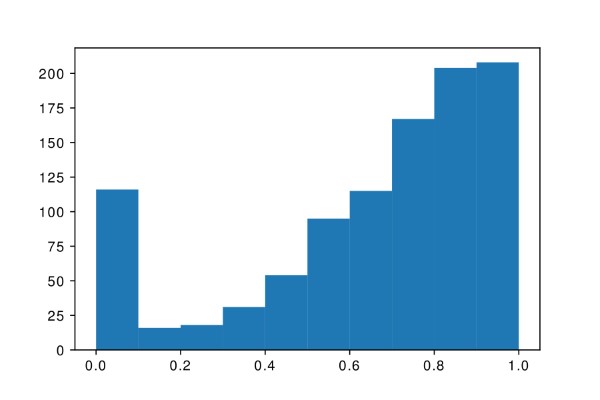}
        \hspace{3.2em} \includegraphics[width=0.137\textwidth]{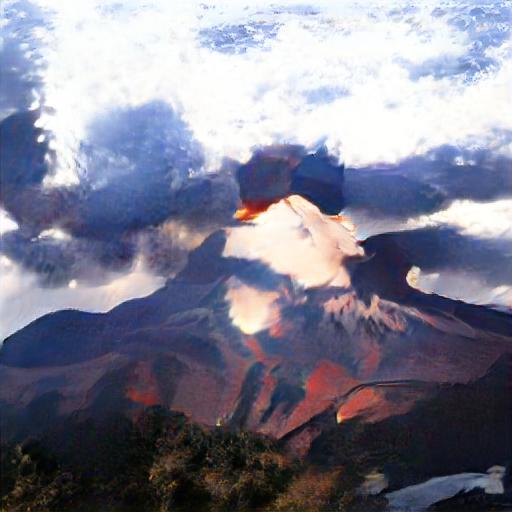}
        \hspace{-6em} \includegraphics[width=0.05\textwidth]{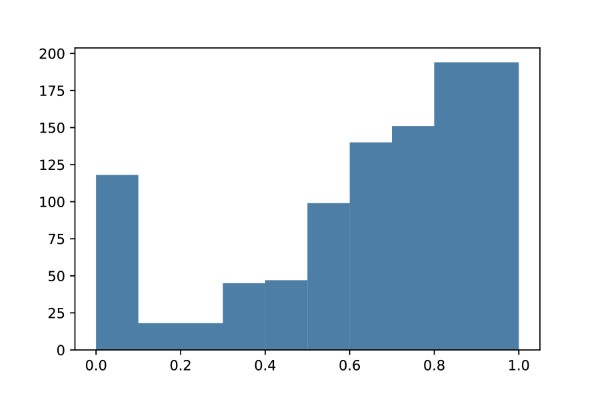}
        \hspace{3.2em} \includegraphics[width=0.137\textwidth]{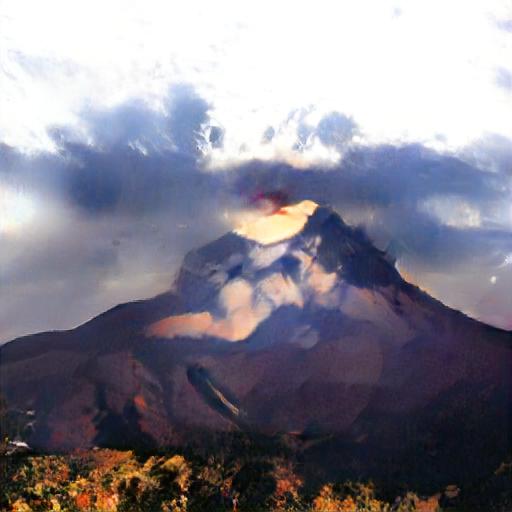}
        \hspace{-6em} \includegraphics[width=0.05\textwidth]{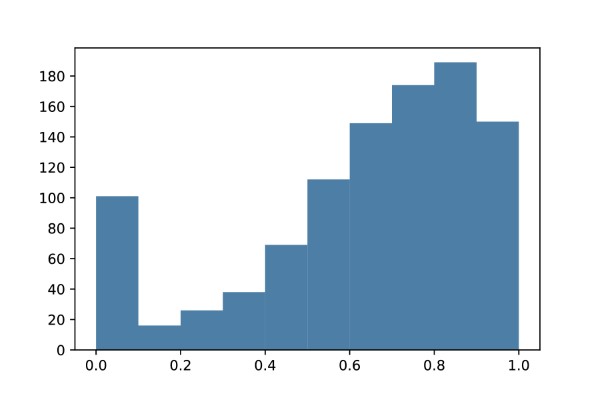}
        \hspace{3.2em} \includegraphics[width=0.137\textwidth]{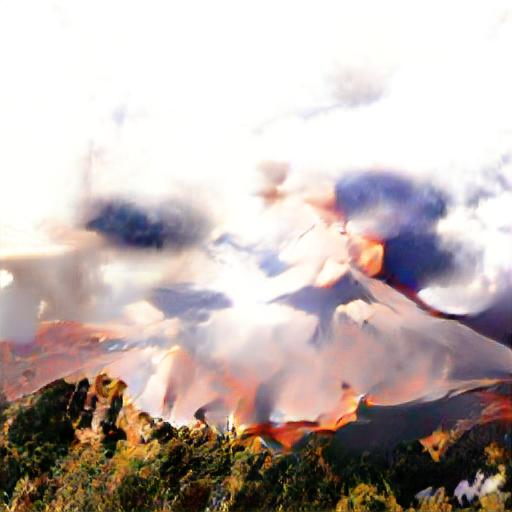}
        \hspace{-6em} \includegraphics[width=0.05\textwidth]{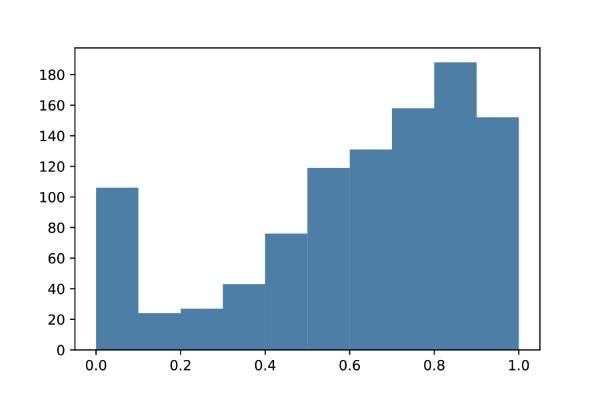}
        \hspace{3.2em} \includegraphics[width=0.137\textwidth]{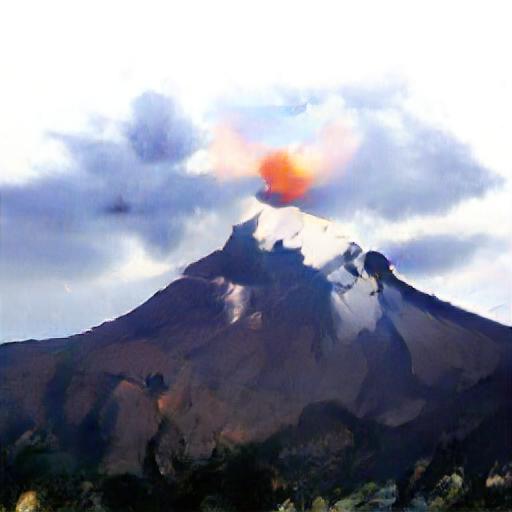}
        \hspace{-6em} \includegraphics[width=0.05\textwidth]{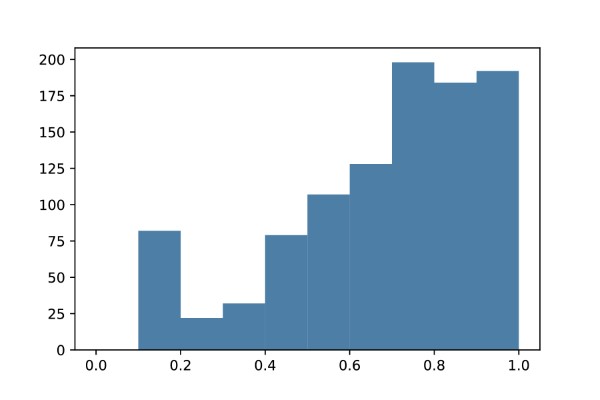}
        \hspace{3.2em} \includegraphics[width=0.137\textwidth]{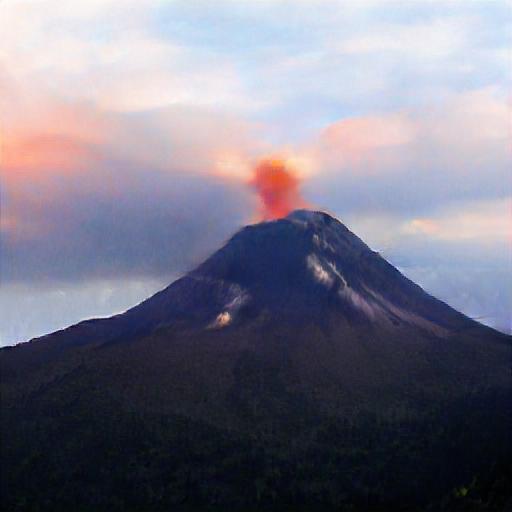}
    \end{subfigure}
    \begin{subfigure}[]{\textwidth}
        \centering
        \includegraphics[width=0.137\textwidth]{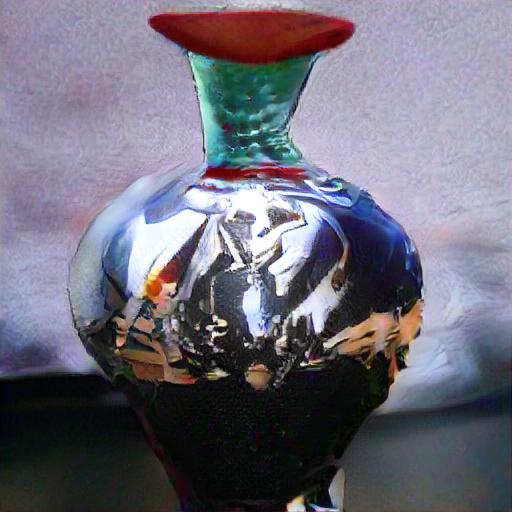}
        \hspace{-6em} \includegraphics[width=0.05\textwidth]{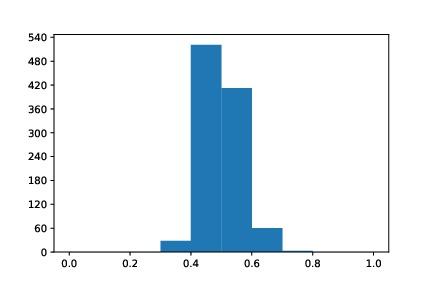}
        \hspace{3.2em} \includegraphics[width=0.137\textwidth]{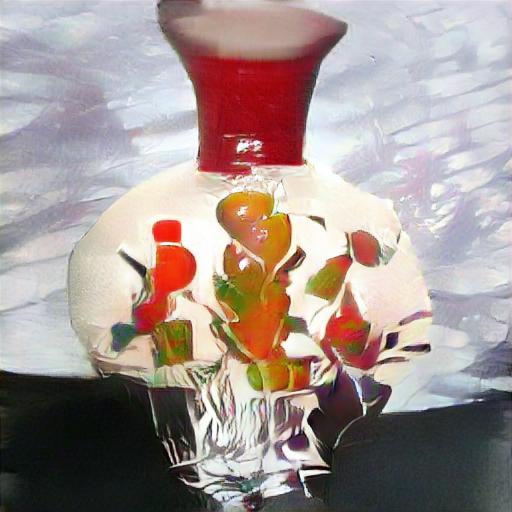}
        \hspace{-6em} \includegraphics[width=0.05\textwidth]{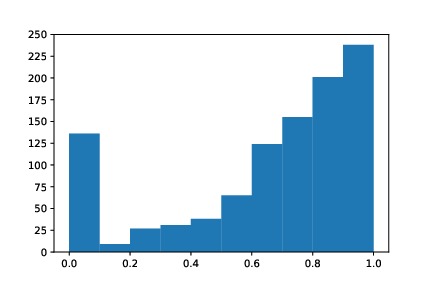}
        \hspace{3.2em} \includegraphics[width=0.137\textwidth]{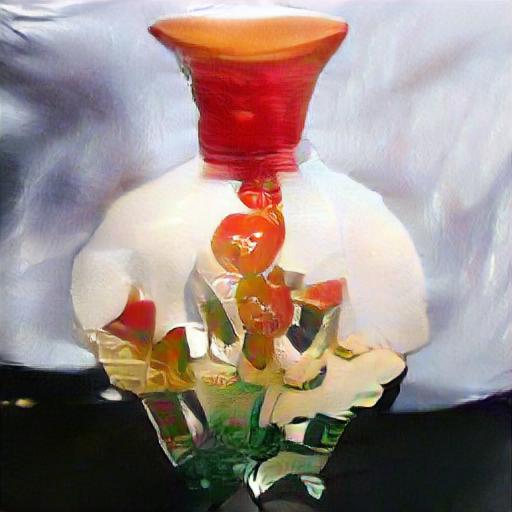}
        \hspace{-6em} \includegraphics[width=0.05\textwidth]{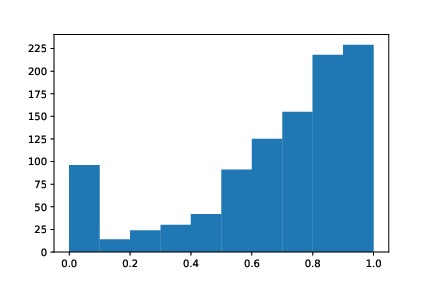}
        \hspace{3.2em} \includegraphics[width=0.137\textwidth]{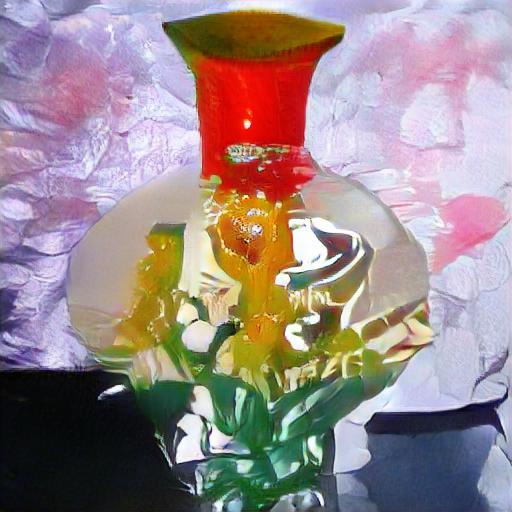}
        \hspace{-6em} \includegraphics[width=0.05\textwidth]{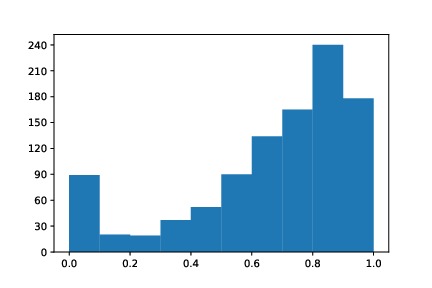}
        \hspace{3.2em} \includegraphics[width=0.137\textwidth]{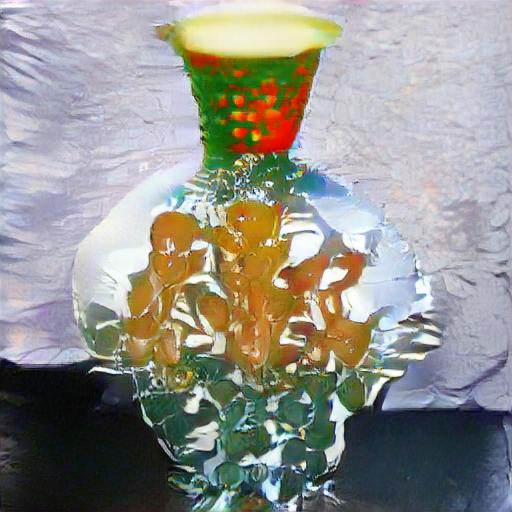}
        \hspace{-6em} \includegraphics[width=0.05\textwidth]{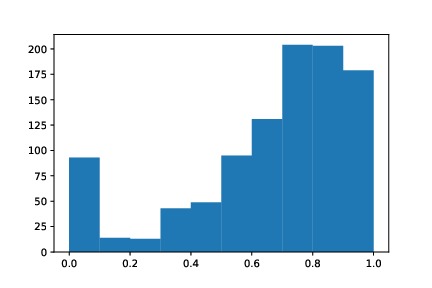}
        \hspace{3.2em} \includegraphics[width=0.137\textwidth]{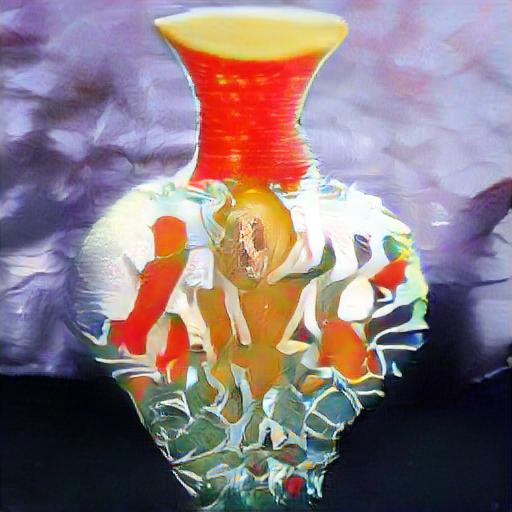}
        \hspace{-6em} \includegraphics[width=0.05\textwidth]{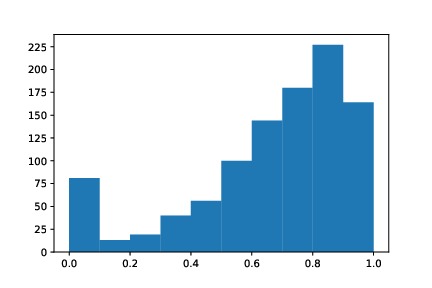}
        \hspace{3.2em} \includegraphics[width=0.137\textwidth]{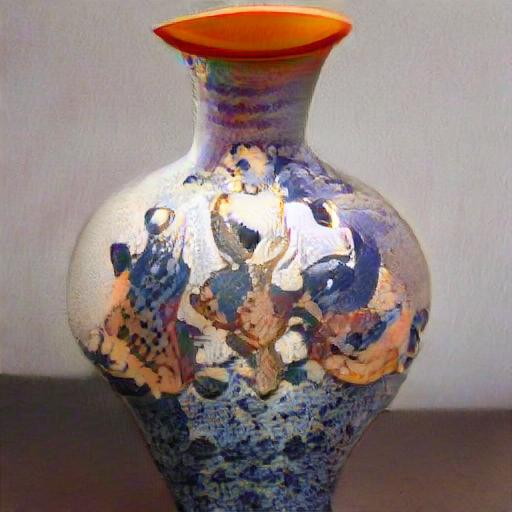}
    \end{subfigure}
    \begin{subfigure}[]{0.137\textwidth}
        \centering \caption*{channel flush}
    \end{subfigure}
    \begin{subfigure}[]{0.137\textwidth}
        \centering \caption*{$m=0.5$}
    \end{subfigure}
    \begin{subfigure}[]{0.137\textwidth}
        \centering \caption*{$m=0.6$}
    \end{subfigure}
    \begin{subfigure}[]{0.137\textwidth}
        \centering \caption*{$m=0.7$}
    \end{subfigure}
    \begin{subfigure}[]{0.137\textwidth}
        \centering \caption*{$m=0.8$}
    \end{subfigure}
    \begin{subfigure}[]{0.137\textwidth}
        \centering \caption*{$m=0.9$}
    \end{subfigure}
    \begin{subfigure}[]{0.137\textwidth}
        \centering \caption*{raw output}
    \end{subfigure}
     
    \caption{Comparison on various stroke sensitivities $m$ over the BigGAN\cite{brock2019} outputs. The channel operations are on layer $l=5$ with $\tau=512$ (out of 1024 available channels).}
    \label{fig:stroke_match}
\end{figure}

In the channel stroke (Alg.~\ref{alg:cs}), the parameter $m$ is used to qualify which of the neighboring pixels can be turned on during the current channel stroke. Therefore, both the scope of neighborhood $\mathcal{N}$ and the sensitivity $m$ decides the shape of the current stroke. In Fig.~\ref{fig:stroke_match}, different $m$'s are compared over the generated images from the BigGAN~\cite{brock2019} network.

Picking a smaller $m$ can further extend the current stroke into its neighboring pixels and create a wider stroke shape. For any pixel at the operation layer, being turned on at one channel reduces the chance of being turned on at other channels. The channel coverage histograms also confirms that smaller $m$'s cause more channels not being turned on. The channels not being turned on are the less responsive channels across the operation layer. Thus the output image from the smaller $m$'s focuses more on the globally responsive channels.

\begin{figure}[!th]
    \centering
    \begin{subfigure}[]{\textwidth}
        \centering
        \includegraphics[width=0.12\textwidth]{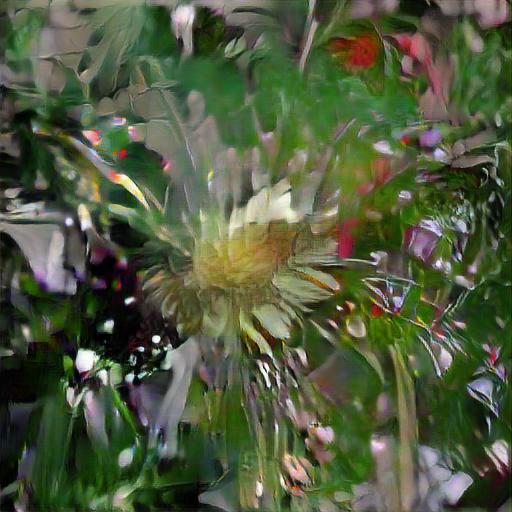}
        \hspace{-5.3em} \includegraphics[width=0.04\textwidth]{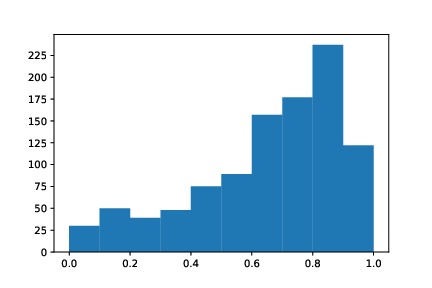}
        \hspace{2.9em} \includegraphics[width=0.12\textwidth]{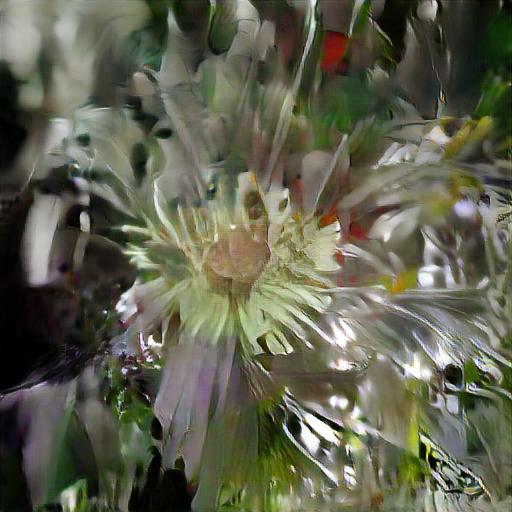}
        \hspace{-5.3em} \includegraphics[width=0.04\textwidth]{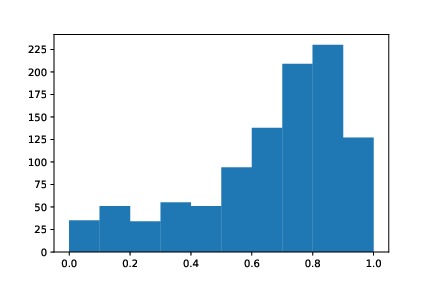}
        \hspace{2.9em} \includegraphics[width=0.12\textwidth]{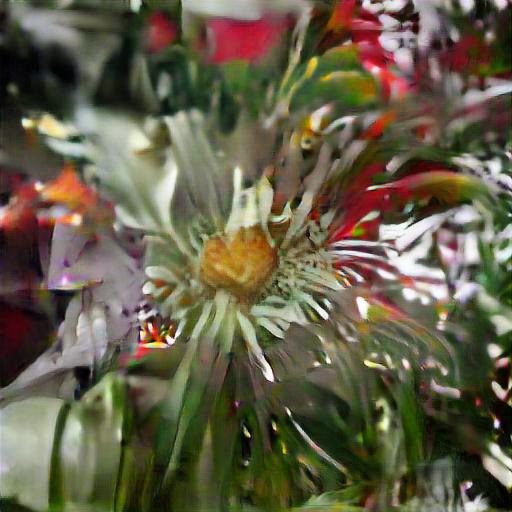}
        \hspace{-5.3em} \includegraphics[width=0.04\textwidth]{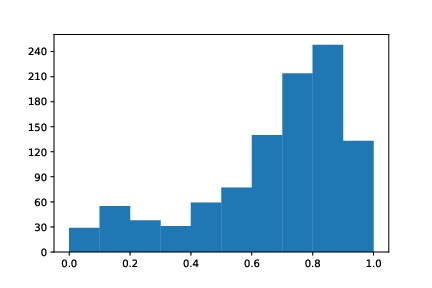}
        \hspace{2.9em} \includegraphics[width=0.12\textwidth]{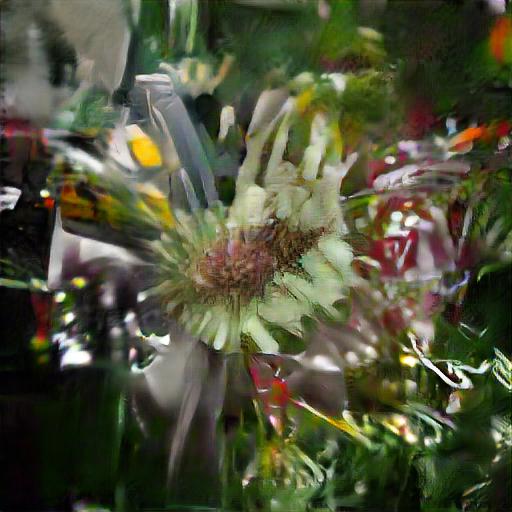}
        \hspace{-5.3em} \includegraphics[width=0.04\textwidth]{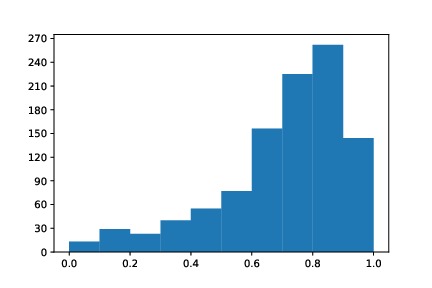}
        \hspace{2.9em} \includegraphics[width=0.12\textwidth]{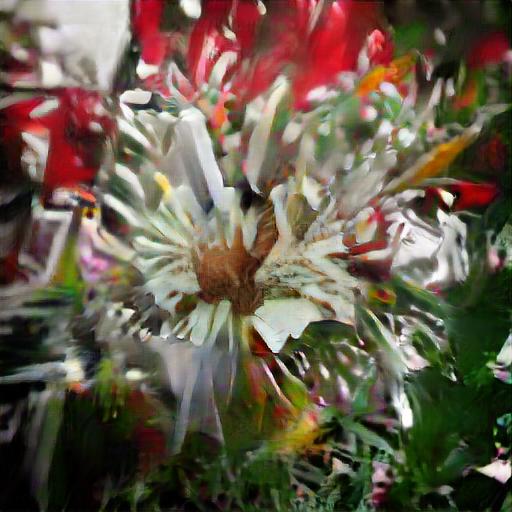}
        \hspace{-5.3em} \includegraphics[width=0.04\textwidth]{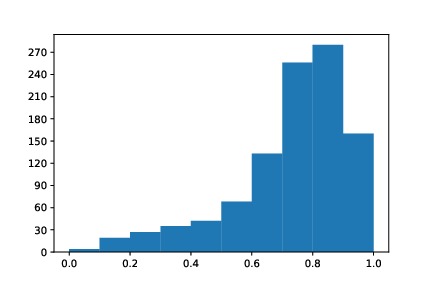}
        \hspace{2.9em} \includegraphics[width=0.12\textwidth]{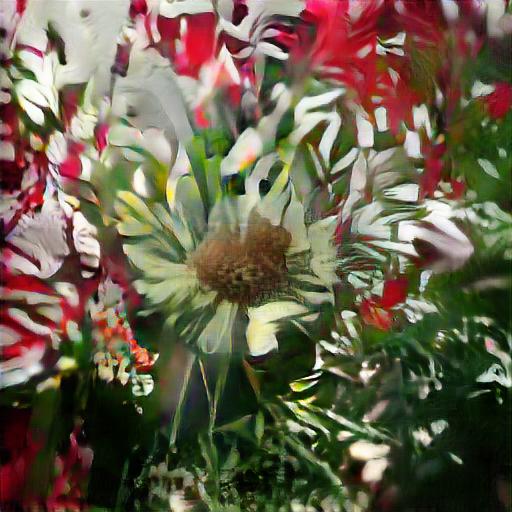}
        \hspace{-5.3em} \includegraphics[width=0.04\textwidth]{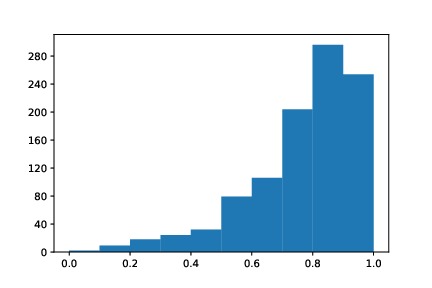}
        \hspace{2.9em} \includegraphics[width=0.12\textwidth]{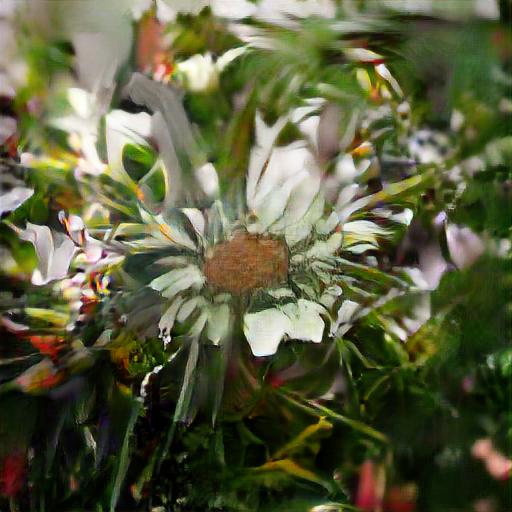}
        \hspace{-5.3em} \includegraphics[width=0.04\textwidth]{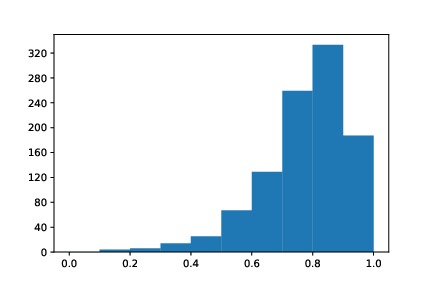}
        \hspace{2.9em} \includegraphics[width=0.12\textwidth]{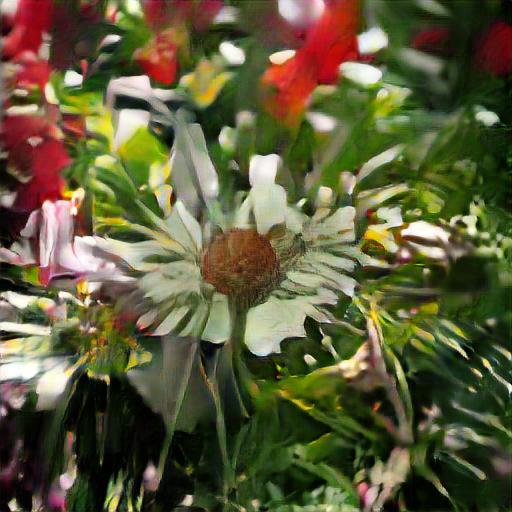}
        \hspace{-5.3em} \includegraphics[width=0.04\textwidth]{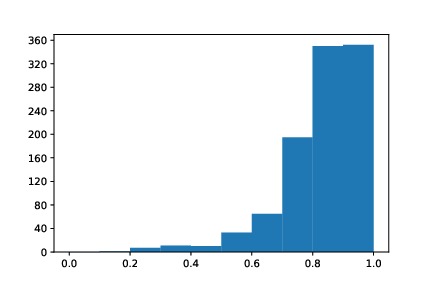}
        \hspace{2.9em} 
    \end{subfigure}

    \begin{subfigure}[]{\textwidth}
        \centering
        \includegraphics[width=0.12\textwidth]{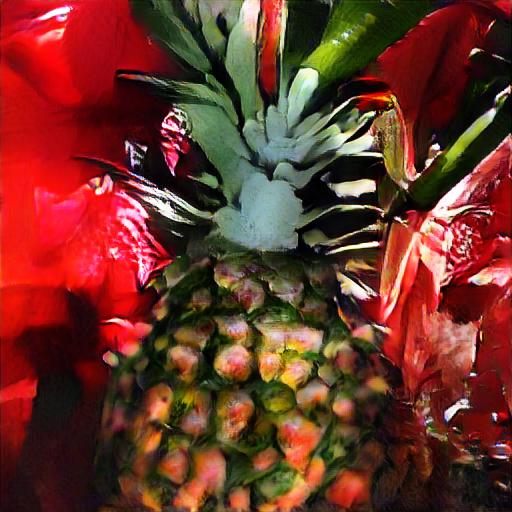}
        \hspace{-5.3em} \includegraphics[width=0.04\textwidth]{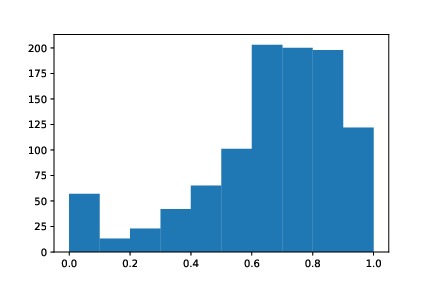}
        \hspace{2.9em} \includegraphics[width=0.12\textwidth]{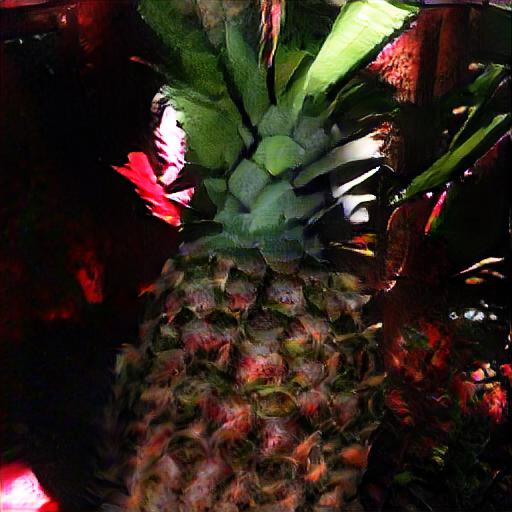}
        \hspace{-5.3em} \includegraphics[width=0.04\textwidth]{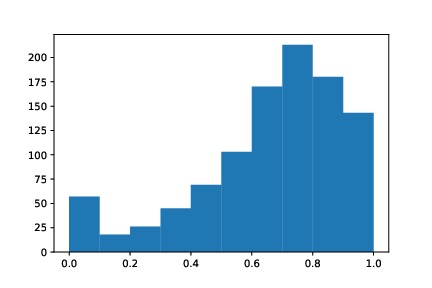}
        \hspace{2.9em} \includegraphics[width=0.12\textwidth]{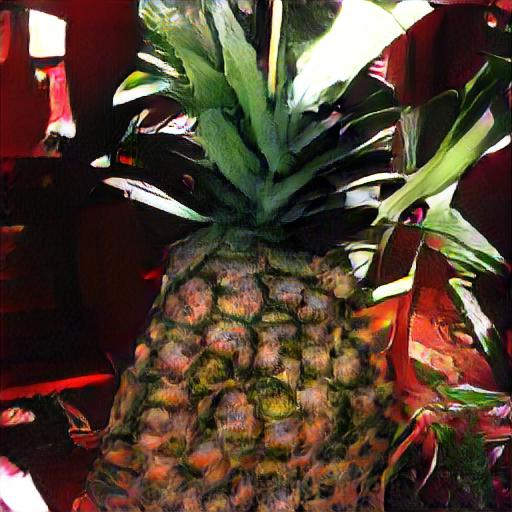}
        \hspace{-5.3em} \includegraphics[width=0.04\textwidth]{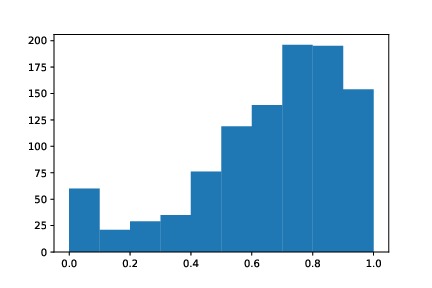}
        \hspace{2.9em} \includegraphics[width=0.12\textwidth]{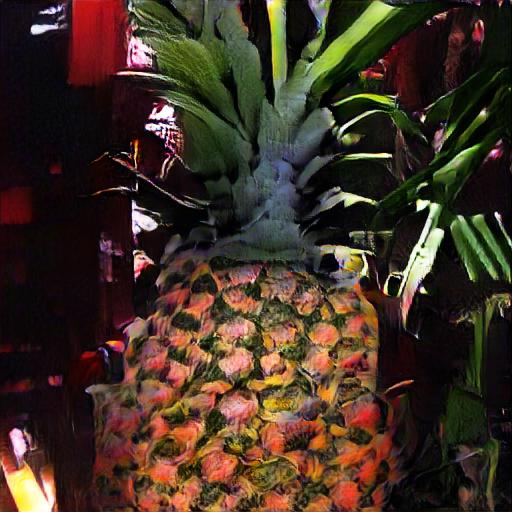}
        \hspace{-5.3em} \includegraphics[width=0.04\textwidth]{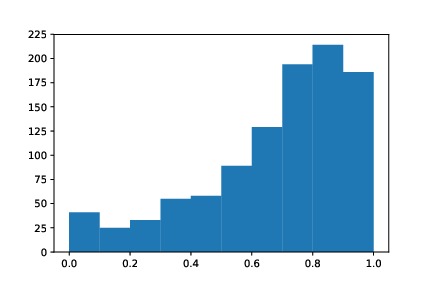}
        \hspace{2.9em} \includegraphics[width=0.12\textwidth]{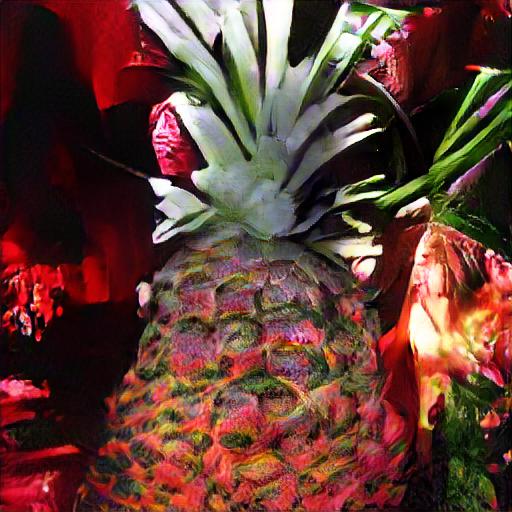}
        \hspace{-5.3em} \includegraphics[width=0.04\textwidth]{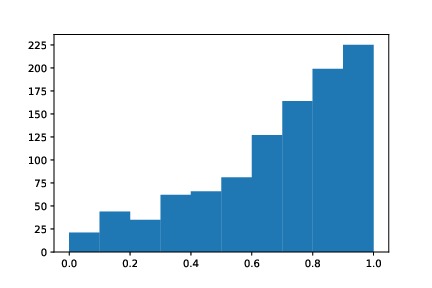}
        \hspace{2.9em} \includegraphics[width=0.12\textwidth]{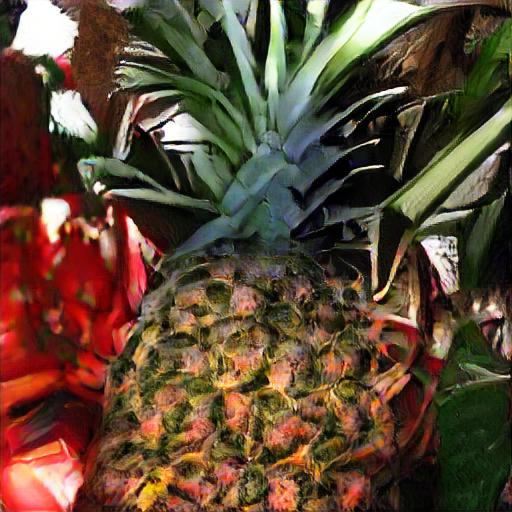}
        \hspace{-5.3em} \includegraphics[width=0.04\textwidth]{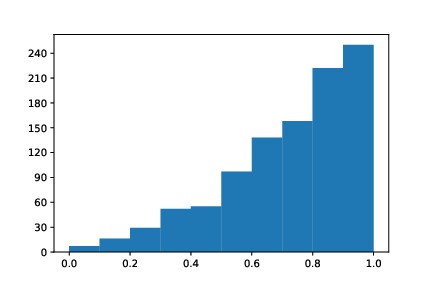}
        \hspace{2.9em} \includegraphics[width=0.12\textwidth]{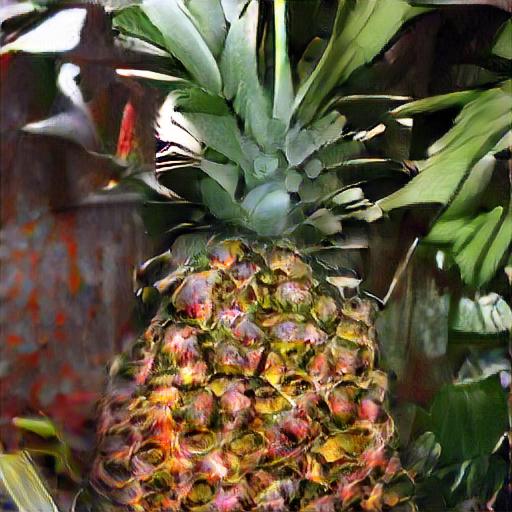}
        \hspace{-5.3em} \includegraphics[width=0.04\textwidth]{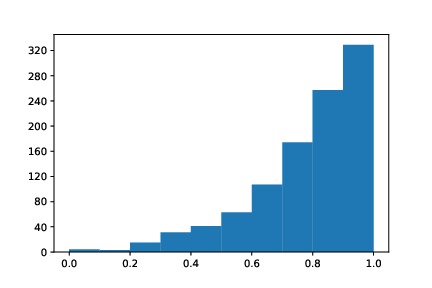}
        \hspace{2.9em} \includegraphics[width=0.12\textwidth]{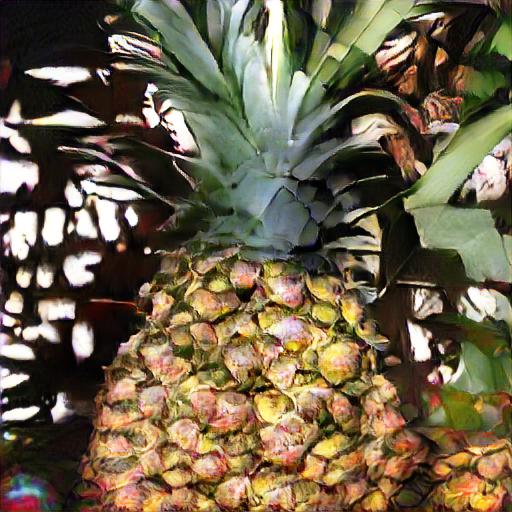}
        \hspace{-5.3em} \includegraphics[width=0.04\textwidth]{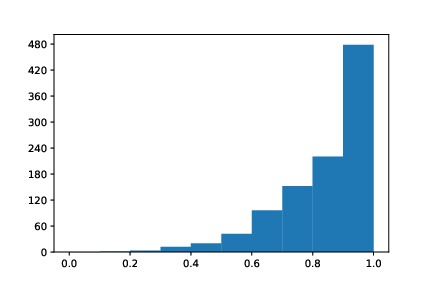}
        \hspace{2.9em} 
    \end{subfigure}

    \begin{subfigure}[]{\textwidth}
        \centering
        \includegraphics[width=0.12\textwidth]{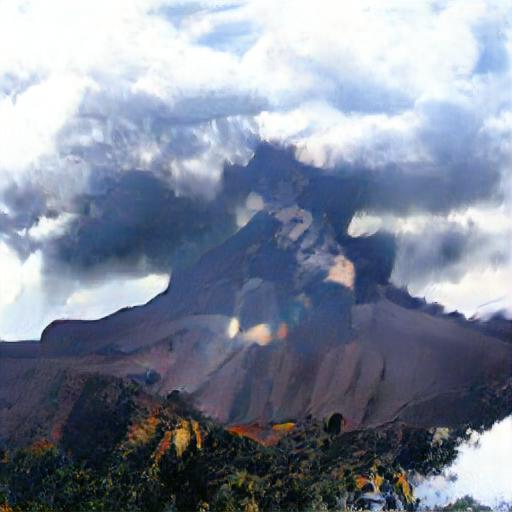}
        \hspace{-5.3em} \includegraphics[width=0.04\textwidth]{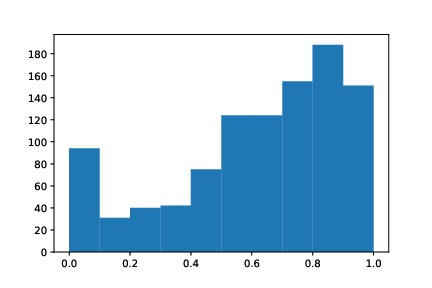}
        \hspace{2.9em} \includegraphics[width=0.12\textwidth]{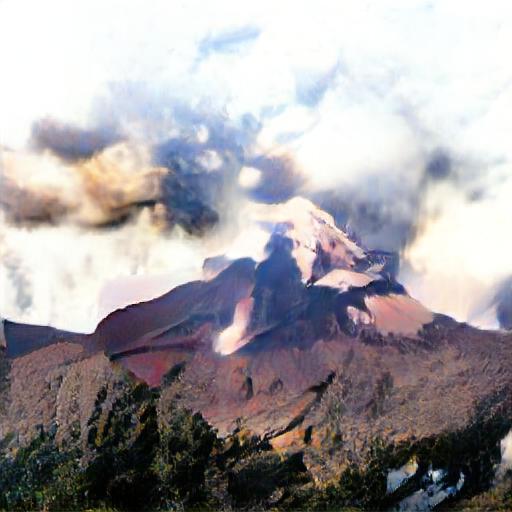}
        \hspace{-5.3em} \includegraphics[width=0.04\textwidth]{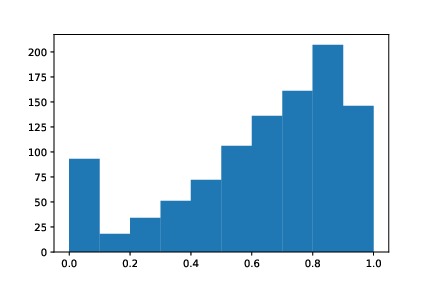}
        \hspace{2.9em} \includegraphics[width=0.12\textwidth]{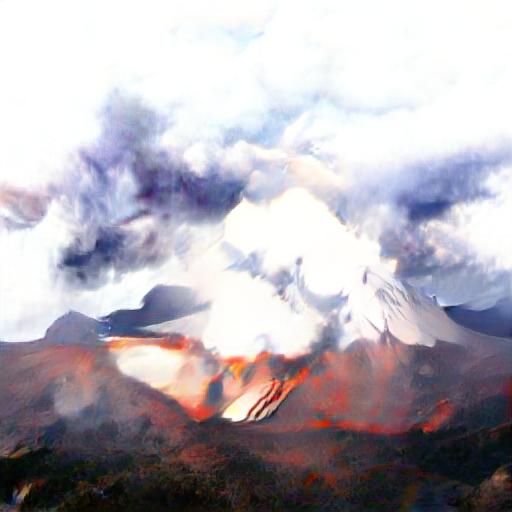}
        \hspace{-5.3em} \includegraphics[width=0.04\textwidth]{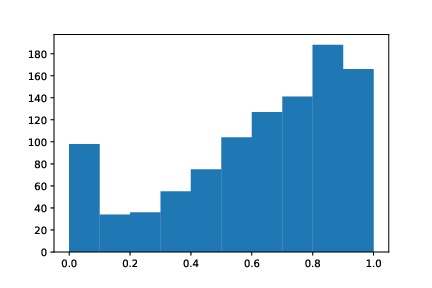}
        \hspace{2.9em} \includegraphics[width=0.12\textwidth]{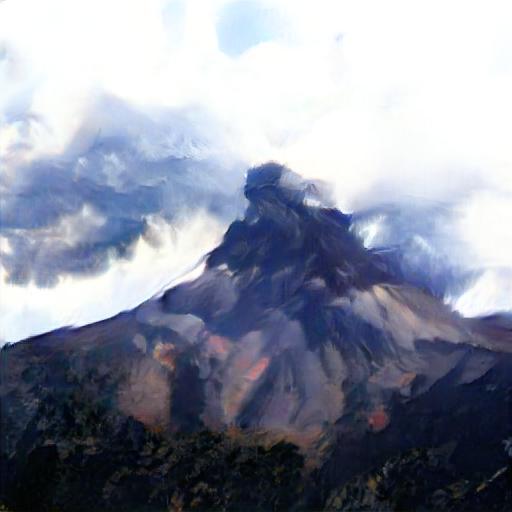}
        \hspace{-5.3em} \includegraphics[width=0.04\textwidth]{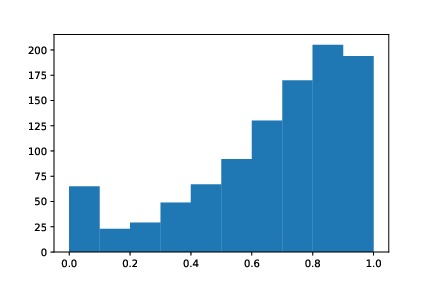}
        \hspace{2.9em} \includegraphics[width=0.12\textwidth]{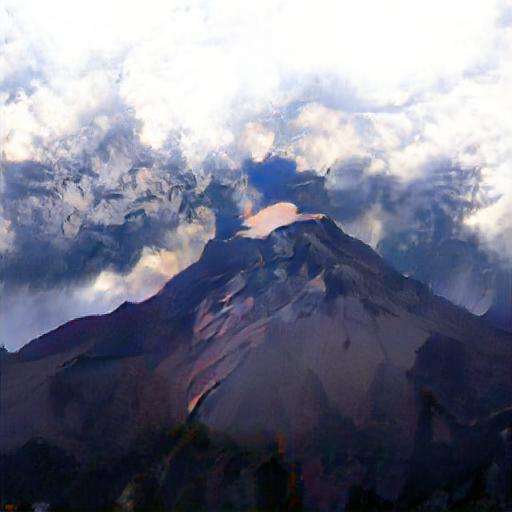}
        \hspace{-5.3em} \includegraphics[width=0.04\textwidth]{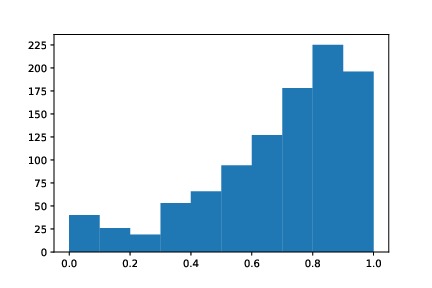}
        \hspace{2.9em} \includegraphics[width=0.12\textwidth]{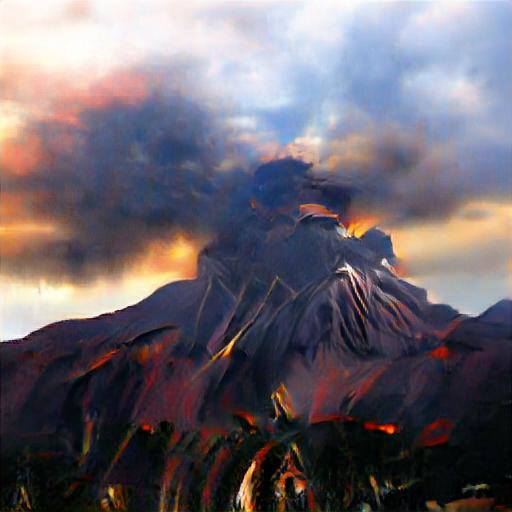}
        \hspace{-5.3em} \includegraphics[width=0.04\textwidth]{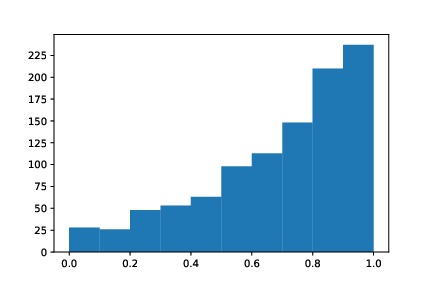}
        \hspace{2.9em} \includegraphics[width=0.12\textwidth]{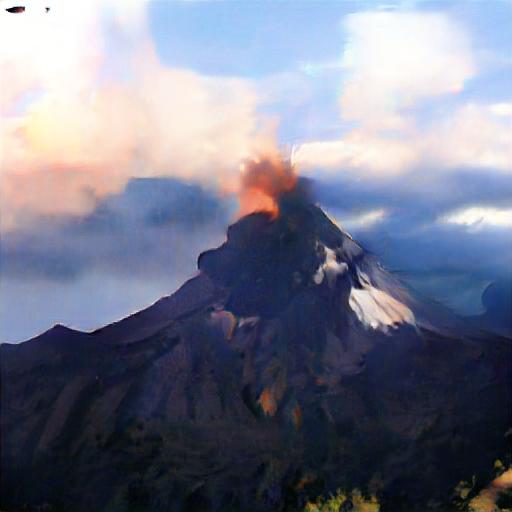}
        \hspace{-5.3em} \includegraphics[width=0.04\textwidth]{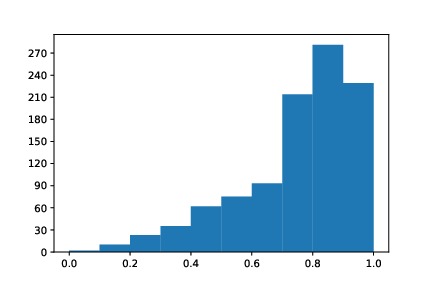}
        \hspace{2.9em} \includegraphics[width=0.12\textwidth]{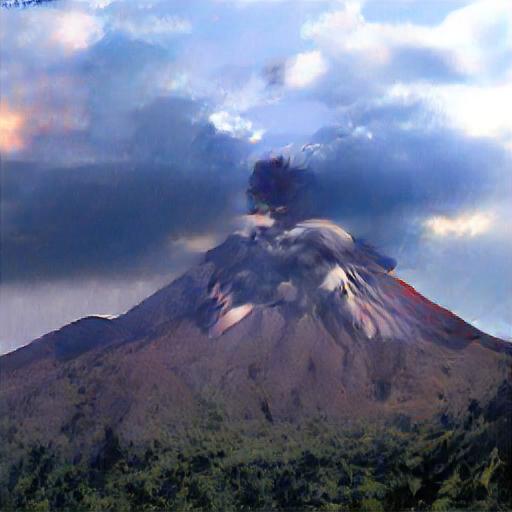}
        \hspace{-5.3em} \includegraphics[width=0.04\textwidth]{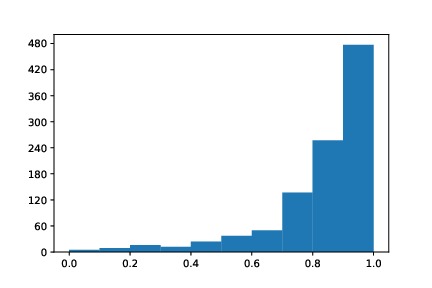}
        \hspace{2.9em} 
    \end{subfigure}

    \begin{subfigure}[]{\textwidth}
        \centering
        \includegraphics[width=0.12\textwidth]{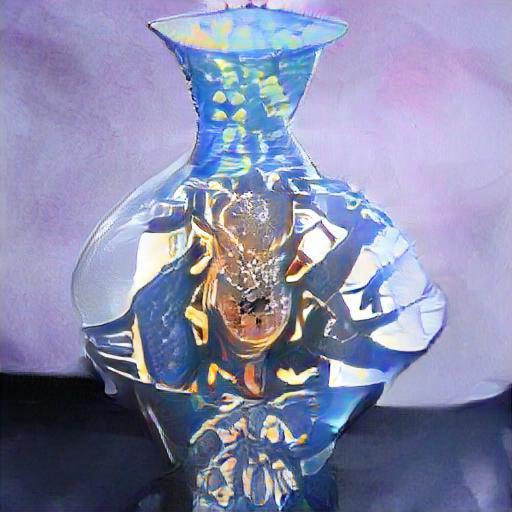}
        \hspace{-5.3em} \includegraphics[width=0.04\textwidth]{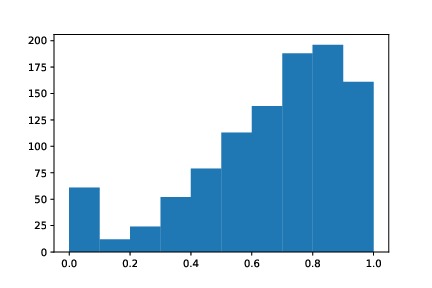}
        \hspace{2.9em} \includegraphics[width=0.12\textwidth]{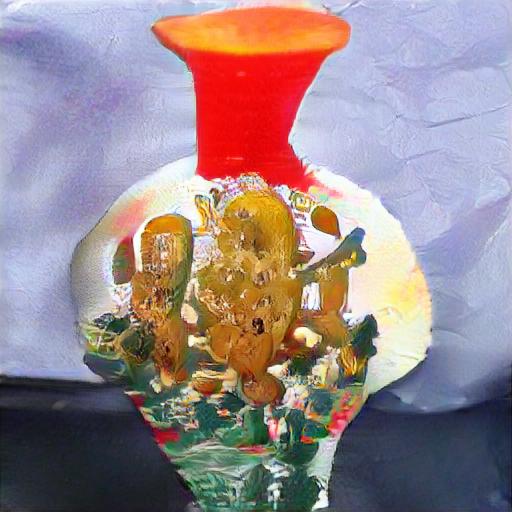}
        \hspace{-5.3em} \includegraphics[width=0.04\textwidth]{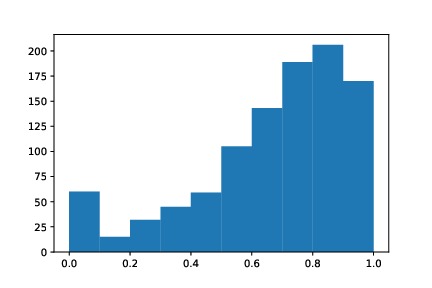}
        \hspace{2.9em} \includegraphics[width=0.12\textwidth]{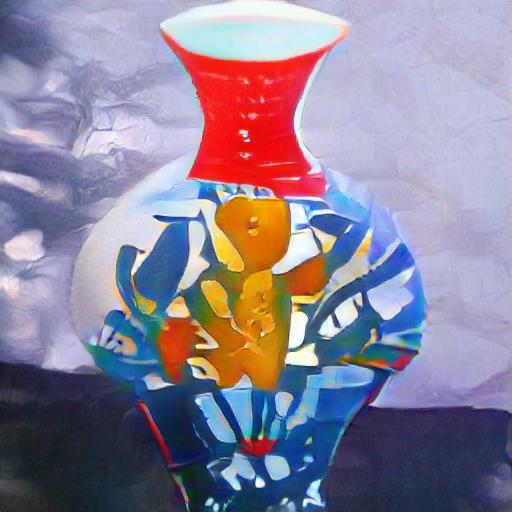}
        \hspace{-5.3em} \includegraphics[width=0.04\textwidth]{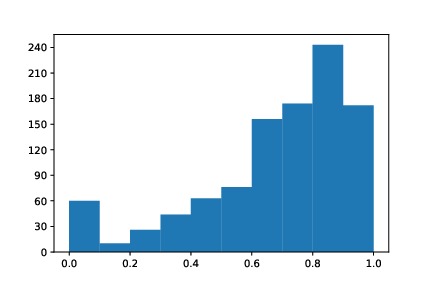}
        \hspace{2.9em} \includegraphics[width=0.12\textwidth]{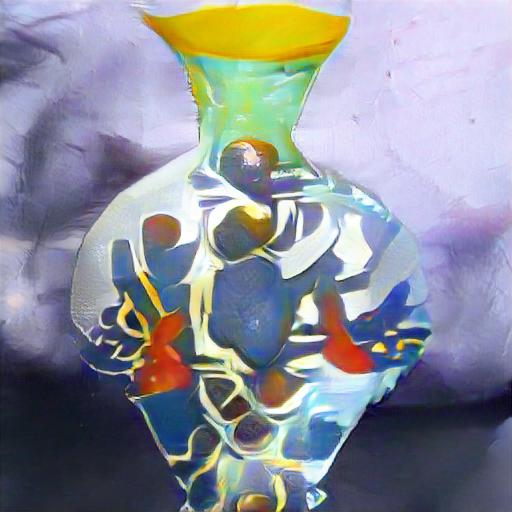}
        \hspace{-5.3em} \includegraphics[width=0.04\textwidth]{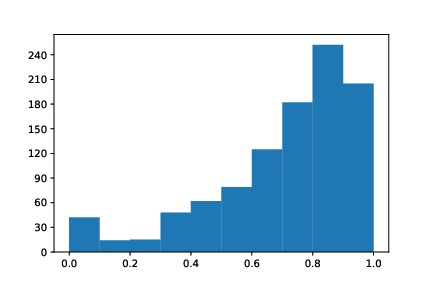}
        \hspace{2.9em} \includegraphics[width=0.12\textwidth]{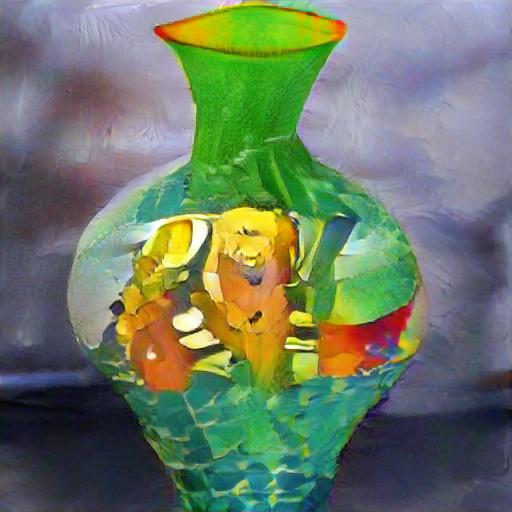}
        \hspace{-5.3em} \includegraphics[width=0.04\textwidth]{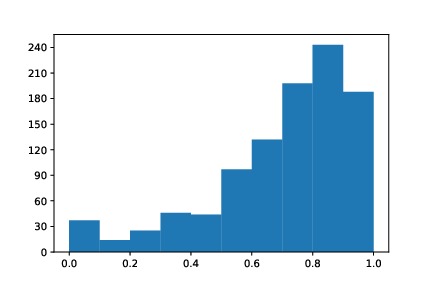}
        \hspace{2.9em} \includegraphics[width=0.12\textwidth]{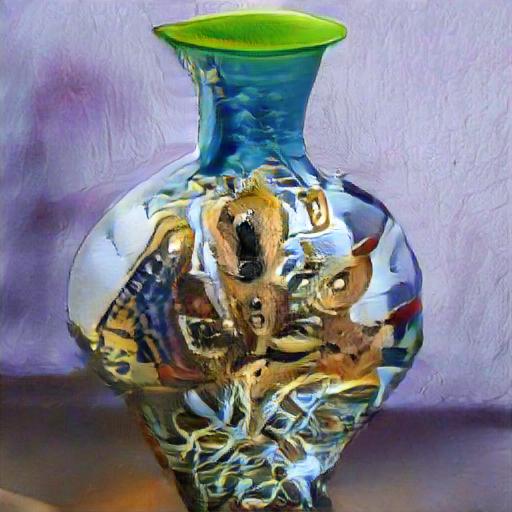}
        \hspace{-5.3em} \includegraphics[width=0.04\textwidth]{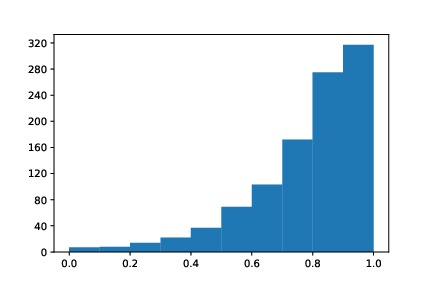}
        \hspace{2.9em} \includegraphics[width=0.12\textwidth]{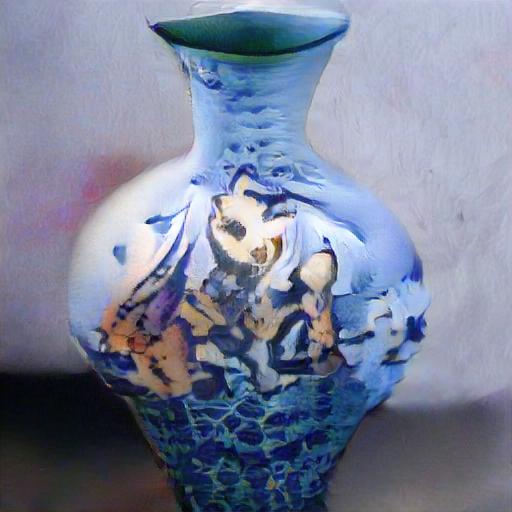}
        \hspace{-5.3em} \includegraphics[width=0.04\textwidth]{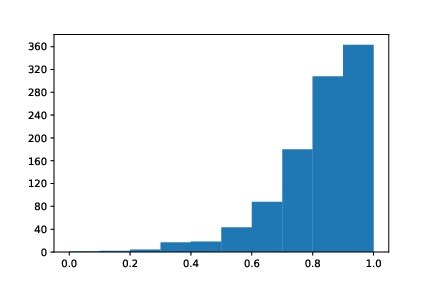}
        \hspace{2.9em} \includegraphics[width=0.12\textwidth]{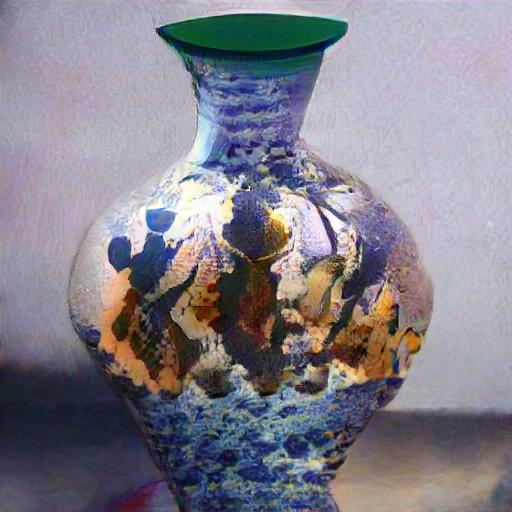}
        \hspace{-5.3em} \includegraphics[width=0.04\textwidth]{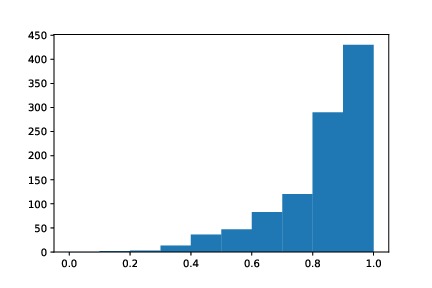}
        \hspace{2.9em} 
    \end{subfigure}

    \begin{subfigure}[]{\textwidth}
        \centering 
        \caption*{\hspace{1.5em} $p=1$ \hfill $p=2$ \hfill $p=4$ \hfill $p=8$ \hfill $p=16$ \hfill $p=32$ \hfill $p=64$ \hfill $p=128$ \hspace{0.3em} }
    \end{subfigure}

    \caption{Comparison on various stroke penetration $p$ over the BigGAN\cite{brock2019} outputs. The channel operations are on layer $l=5$ with $\tau=512$ (out of 1024 available channels).}
    \label{fig:stroke_pen}
\end{figure}

\subsection{Stroke Penetration} \label{sec:disc-pen}

A stroke typically affects multiple channels. For example, the last layer in the generator consists RGB channels. A stroke in yellow color at least impacts the red and the green channels. Reconsidering the channel strokes, another option is to enable the stroke to penetrate through several channels. The penetration happens when there exists several channels having similarly high response as the stroking channel $c$ at location $h,w$.

When stroking at the pixel $(c,h,w)$, other $p-1$ top response channels at the same location are also turned on, conditioned on they are previously muffled. The penetrating stroke then follow Alg.\ref{alg:cs} to complete the stroking process. The set of the penetrating channels decides the color and the pattern of the stroke. Depending on the operation layer, there can be more than a thousand paint-able channels. The depth of the layer gives room for the stroke penetration. We evaluate the stroke penetration in Fig.~\ref{fig:stroke_pen}, where the operation layer has $1024$ channels. 

The passing channels are less coupled on lower stroke penetration. This induces the variety in channel coverage, and generates high contrast bold results. On the other hand, the higher stroke penetration brings the output closer to the original full channel output. This can be verified with the histograms of channel coverage. While stroke sensitivity $m$ maintains the spatial continuity, the channel continuity is controlled by the stroke penetration $p$. Note that in Fig.~\ref{fig:stroke_match} the penetration is fixed at $p=2$, and in Fig.~\ref{fig:stroke_pen} the sensitivity is fixed at $m=0.95$.

\end{document}